\documentclass[a4paper,onecolumn]{IEEEtran}

\usepackage{enumerate}
\usepackage{multirow,array}
\usepackage{cite}
\usepackage{graphicx}
\usepackage{psfrag}
\usepackage{tcolorbox}

\usepackage{url}
\usepackage{amsmath}
\usepackage{amsthm}
\usepackage{subcaption}
\usepackage{array}

\usepackage{amssymb}
\usepackage{amsfonts}
\usepackage{float}
\usepackage{algorithm}
\usepackage{algpseudocode}

\usepackage{color, colortbl}
\definecolor{Gray}{gray}{0.9}
\usepackage{multirow}
\usepackage{multicol}
\usepackage{booktabs,ragged2e}

\newcolumntype{g}{>{\columncolor{Gray}}c}

\usepackage{setspace}

\makeatletter
\renewcommand{\boxed}[1]{\text{\fboxsep=.2em\fbox{\m@th$\displaystyle#1$}}}
\makeatother
\newcommand{\C}{\cal}

\newcommand{\B}{\boldsymbol}
\newcommand{\fq}{{\mathbb F}_q}
\newcommand{\gbinom}[2]{\begin{bmatrix}#1\\#2\end{bmatrix}_q}

\newtheorem*{conjecture*}{Conjecture}
\newtheorem{lemma}{Lemma}

\newtheorem{corollary}{Corollary}
\newtheorem{remark}{Remark}
\newtheorem{definition}{Definition}
\newtheorem{theorem}{Theorem}
\newtheorem{example}{Example}

\title{Subexponential and Linear Subpacketization Coded Caching via Projective Geometry}
\begin{document}

\author{
\IEEEauthorblockN{Hari Hara Suthan Chittoor, Prasad Krishnan,  K V Sushena Sree, and Bhavana MVN\\}
\vspace{-1cm}
}

\date{\today}
\maketitle
\thispagestyle{empty}	
\pagestyle{empty}
\let\thefootnote\relax\footnotetext
{



This work was presented in Parts at the \textit{2018 IEEE Information Theory Workshop} \cite{Prasad_ITW_2018}, at the \textit{2019 IEEE International Symposium on Information Theory} \cite{Hari_ISIT_2019}, at the \textit{2019 IEEE International Symposium on Communications and Information Technologies} \cite{Hari_ISCIT_2019}, and at the \textit{2019 IEEE Global
Communications Conference} \cite{Hari_GLOBECOM_2019}. 
This work was supported by the PhD fellowship given to Hari Hara Suthan Chittoor under Visvesvaraya PhD Scheme for Electronics \& IT (Govt.of India). Prasad Krishnan acknowledges support from SERB-DST projects MTR/2017/000827 and CRG/2019/005572.

H.H.S. Chittoor, P. Krishnan,  KVS. Sree, and Bhavana MVN are with the Signal Processing and Communication Research
Center, International Institute of Information Technology, Hyderabad 500032, India. Email: \{hari.hara@research., prasad.krishnan@, sushena.sree@research., bhavana.mvn@research.\}iiit.ac.in


}


\begin{abstract}
    Large gains in the rate of cache-aided broadcast communication are obtained using coded caching, but to obtain this most existing centralized coded caching schemes require that the files at the server be divisible into a large number of parts (this number is called subpacketization). In fact, most schemes require the subpacketization to be growing asymptotically as exponential in $\sqrt[\leftroot{-1}\uproot{1}r]{K}$ for some positive integer $r$ and $K$ being the number of users. On the other extreme, few schemes having subpacketization linear in $K$ are known; however, they require large number of users to exist, or they offer only little gain in the rate. In this work, we propose two new centralized coded caching schemes with low subpacketization and moderate rate gains utilizing projective geometries over finite fields. Both the schemes achieve the same asymptotic subpacketization, which is exponential in $O((\log K)^2)$ (thus improving on the $\sqrt[\leftroot{-1}\uproot{1}r]{K}$ exponent). The first scheme has a larger cache requirement but has at most a constant rate (with increasing $K$), while the second has small cache requirement but has a larger rate. As a special case of our second scheme, we get a new linear subpacketization scheme, which has a more flexible range of parameters than the existing linear subpacketization schemes. Extending our techniques, we also obtain low subpacketization schemes for other multi-receiver settings such as distributed computing and the cache-aided interference channel. We validate the performance of all our schemes via extensive numerical comparisons. For a special class of symmetric caching schemes with a given subpacketization level, we propose two new information theoretic lower bounds on the optimal rate of coded caching.
\end{abstract}



\vspace{-0.5cm}

\section{Introduction}

Present and future wireless communication systems (4G,5G and beyond) are becoming more content-centric. The majority of this content is video, which is generated well ahead of transmission. Further, it is predicted in \cite{Cis} that by 2022, four-fifths of all the internet traffic will be video. Therefore, we require new strategies to manage the data-heavy communication systems while ensuring quality of service.

Caching has been in vogue to lay off the traffic during the peak times in the network by storing part of the information demanded by the users (clients) in local storage known as \textit{caches}. In this way, during the peak hours, the server can transmit only the non-cached information, thus reducing the traffic. 
\textit{Coded caching} was proposed in a landmark paper by Ali-Niesen \cite{MaN} to exploit this aspect of cached content to reduce the network congestion in the peak traffic time by prefetching part of the content in cost-effective cache available at the users during the off-peak time, while using coded transmissions during the peak times. 


In \cite{MaN}, the authors considered an error-free broadcast channel with a server containing $N$ files of the same size and $K$ users (clients) each having a cache capable of storing $M$ files, where $M\leq N$. According to the scheme presented in \cite{MaN} the system operates in two phases. During the \textit{caching phase} (happens in the off-peak time) each file in the server is divided into $F$ equal-sized subfiles ($F$ is known as the \textit{subpacketization} parameter), and placed in the caches of the clients. The caching phase occurs well ahead of the appearance of receiver demands, and thus the caching phase has to be designed in a demand-oblivious manner. During the \textit{delivery phase} (happens in the peak time) each user demands a file from the server. Based on the demands and the cache contents of the users, the server makes multiple coded transmissions. The goal is to design the caching and delivery phase so that the demands of all the users are satisfied. Since the caching phase is also designed centrally, this framework for coded caching is known as the \textit{centralized coded caching}. Decentralized coded caching was introduced in \cite{decentralizedcodedcaching}, in which a coded delivery scheme is shown to achieve large gains in the rate, under a random or decentralized caching phase. Other important variations of this setting include coded caching in a popularity-based caching setting \cite{multi_level_popularity_Zhang}, online coded caching \cite{online_coded_caching}, and hierarchical coded caching \cite{Hierarchical_Karamchandani}.  We assume the basic centralized coded caching framework in the present work.  

The delivery scheme in \cite{MaN} serves $\gamma=1+\frac{MK}{N}$ users per transmission. The parameter $\gamma$ is known as the \textit{global caching gain} and the \textit{rate} of the scheme is given as $R=\frac{K\left(1-\frac{M}{N}\right)}{\gamma}$, which has a $\Theta(K)$ gain over the uncoded delivery rate $\big($for constant $\frac{M}{N} \big)$ which is $K\left(1-\frac{M}{N}\right)$. The rate achieved by Ali-Niesen scheme \cite{MaN} was shown to be optimal for a given cache size $M$ in \cite{WTP}, under the assumption of uncoded cache placement and $N\geq K$. Further, for large $K$, this scheme surprisingly achieves (approximately) a rate that is independent of $K$. However, it suffers from the problem of large subpacketization, which we now describe. 
 
The achievability of the scheme shown in \cite{MaN} is ensured by splitting each file into $F=\binom{K}{MK/N}$ equal-sized subfiles, where $F$ is known as the \textit{subpacketization level} or simply, subpacketization. 
It was noticed in \cite{user_grouping_Shanmugam} that for this scheme, the subpacketization required grows exponentially in $K$ for constant $\frac{M}{N}$ as $K$ grows large (as $\binom{K}{Kp}\approx 2^{KH(p)}$ for constant $0<p<1$, where $H(p)$ is the binary entropy). For instance, with $K=25$ users and with the capability to store one-fifth of the file library in the cache of each user $\left(\frac{M}{N}=\frac{1}{5}\right)$, the subpacketization becomes $\binom{25}{5}$, which is $53130$. A high subpacketization requirement poses multiple issues in the implementation of coded caching. A straightforward issue is that of the size of the file itself; the file-size has to be at least as large as the product of subpacketization $F$ and the size of any accessible file-segment. Assuming a file-segment size of about $512$ KB, the file-size has to be at least $27$ GB for a subpacketization level of $53130$, which is  prohibitive in practice. Higher subpacketization levels also mean higher indexing overheads to identify the subfiles. Also, a large subpacketization level implies smaller chunks of the file, which in turn means higher normalized read times from the storage media, along with search-and-read overheads. Having longer indexing overheads to identify the subfiles is also a factor to consider when there are a large number of subfiles in any file. Because of these reasons, coded caching schemes with low-subpacketization and with good caching gains are preferable for practical applications.


\begin{table*}
\centering
\captionsetup{justification=centering}
\caption{\small Parameters of some known coded caching schemes.\\ The asymptotic nature of the subpacketization and the rate with large $K$ and constant cache fraction $\frac{M}{N}$ are shown wherever possible.\\ For simplicity, we ignore the constant multiplicative factors in the exponents of asymptotics ($e$ is the base of the natural logarithm)}. 
\label{table known results}
\setlength{\tabcolsep}{6pt} 
\renewcommand{\arraystretch}{1.7}

\scriptsize
\begin{tabular}{|c|c|c|c|c|c|}

\hline

\textbf{Scheme} & \textbf{Number of Users}  & \textbf{Cache Fraction } & \textbf{Subpacketization}    & \multicolumn{2}{|c|}{\textbf{Rate } $\mathbf{R}$} \\

\cline{5-6}

& $\mathbf{K}$ & $\mathbf{\frac{M}{N}}$ & $\mathbf{F}$ & Expression & Asymptotics \\


\hline

Ali-Niesen \cite{MaN} & any $K$ & $\frac{M}{N}$ for $M<N$ & $O\left(2^{K}\right)$ & $\frac{K\left(1-\frac{M}{N}\right)}{\frac{MK}{N}+1}$ & $O(1)$ \\

& & $\frac{MK}{N}\in {\mathbb Z}^{+}$ & & & \\

\hline

Ali-Niesen scheme with & $K$ & Same as \cite{MaN} & $O\left(e^{g}\right)$ & $\frac{K}{g+1}\left(1-\frac{1}{\lceil\frac{N}{M}\rceil}\right)$, where & \\

Grouping \cite{user_grouping_Shanmugam} & & & & $g\in {\mathbb Z}^{+}$ such that $\frac{K}{g\lceil\frac{N}{M}\rceil}\in{\mathbb Z}^{+}$. &  $O\left(\frac{K}{g}\right)$\\

\hline

Yan et al. \cite{YCTCPDA} based on & Any $K$ & $1-\frac{1}{q}$ or $\frac{1}{q}$  & $O\left(e^{K}\right)$ & $\dfrac{K\left(1-\frac{M}{N}\right)}{\frac{MK}{N}}$ & $O(1)$ \\

Placement Delivery Arrays (PDAs)& & &  & & \\
\hline
Shangguan et al. \cite{SZG} &	& & & $R\approx (2q-1)^2$, such that $q=\frac{\lambda}{2}$, &  \\

(PDAs based on hypergraphs) & Specific choices & $1-\frac{1}{q}$ or $\frac{1}{q}$ & $O\left(e^{\sqrt{K}}\right)$ &where $\lambda$ is such that $\frac{M}{N}=\frac{2\lambda-1}{\lambda^2}$ & $O(1)$ \\ 

\hline

Yan et al. \cite{strongedgecoloringofbipartitegraphs} (for integers $0<a,b<m$  & & & & & \\

and $\lambda <min\left\{a,b\right\}$ based on & $\binom{m}{a}$ & $\frac{\binom{a}{\lambda}\binom{m-a}{b-\lambda}}{\binom{m}{a}}$ & $\binom{m}{b}$ & $\dfrac{\binom{m}{a+b-2\lambda}\binom{a+b-2\lambda}{a-\lambda}}{\binom{m}{b}}$ & \textemdash \\

strong edge coloring of bipartite graph) & & & & & \\

\hline

Tang et al. \cite{TaR} based & $nq$ & &  & & \\

on resolvable designs using & (for some  & $1-\frac{1}{q}$ or $\frac{1}{q}$ & $O\left(q^{K}\right)$ & $\dfrac{K\left(1-\frac{M}{N}\right)}{\frac{cK}{q}+1}$ & $O(1)$ \\

$n$ length linear block code of rate $c$ & constant $q$) & &  & & \\

\hline

Scheme from \cite{STD} based on & $K$ &  &  & & \\

induced matchings of a & (necessarily & $ K^{-\epsilon}$ & $K$ 	&$K^\delta$  &  $K^{\delta}$\\

Ruzsa Szemeredi graph &  extremely large) & (some small $\epsilon$) & & (some small $\delta$) & \\

\hline

PDA scheme $P_1$ from Cheng et al. \cite{cheng2017coded} & \normalsize $\binom{k}{t+1}$ & $1-\frac{t+1}{\binom{k}{t}}$ & \normalsize $\binom{k}{t}$ & \normalsize $\frac{k}{\binom{k}{t}}$ & \textemdash \\

$k,t \in \mathbb{Z}^{+}$ &  &  &  &  &  \\

\hline


Two PDA Schemes from \cite{CJYT} & $\binom{m}{t}q^t$ and & & & $\left((q-z)/{\big\lfloor \frac{q-1}{q-z}\big\rfloor}\right)^t$ &  \\

$q,z,m,t \in \mathbb{Z}^{+}$, $q\geq2, z<q, t<m$ & $(m+1)q$ & $1-\left(\frac{q-z}{q}\right)^t$ and $\frac{z}{q}$ &  $O\left(q^{\sqrt[\leftroot{-1}\uproot{1}t]{K}}\right)$ & and $(q-z)/{\big\lfloor \frac{q-1}{q-z}\big\rfloor}$ & $O(1)$ \\

\hline

\end{tabular}

\end{table*}


Since this problem was identified, a number of papers, for instance \cite{user_grouping_Shanmugam,strongedgecoloringofbipartitegraphs,YCTCPDA,SZG,STD,TaR,cheng2017coded,CJYT,SLB} have presented new schemes for the coded caching which use smaller subpacketization than \cite{MaN} at the cost of having increased rate for the same cache requirement compared to \cite{MaN}. The summary of some important known schemes is given in Table \ref{table known results}. The third column lists the cache fraction of any file $\big($a fraction $\frac{M}{N}$ of each file is cached by a user$\big)$. Many of the schemes presented in Table \ref{table known results} require exponential subpacketization (in $K$, for large $K$), as shown in the fourth column of Table \ref{table known results} to achieve a constant rate (shown in the fifth column). The asymptotics of the rate are presented in the last column of Table \ref{table known results}. A user-grouping technique combined with the scheme of \cite{MaN} was used in \cite{user_grouping_Shanmugam} to reduce the subpacketization at the cost of rate. A variety of techniques and structures from combinatorics, coding theory, and graph theory, have also been employed to obtain many of these constructions. For instance, in \cite{YCTCPDA}, a special combinatorial structure called \textit{placement delivery arrays (PDA)} was presented and used to construct coded caching schemes with reduced subpacketization than \cite{MaN} (though it remained still exponential in $K$). The work \cite{TaR} used resolvable designs obtained from linear block codes for the same. The papers \cite{SZG,strongedgecoloringofbipartitegraphs} used hypergraphs and bipartite graphs respectively, and showed schemes with subpacketization subexponential in $K$. The subpacketization of particular schemes of \cite{CJYT} have been shown to be subexponential, while some schemes of \cite{cheng2017coded} have subpacketization that is linear or polynomial (in $K$) at the cost of either requiring much larger cache $M$ or much larger rate compared to \cite{MaN}. Interestingly, a linear subpacketization scheme ($F=K$) was shown in \cite{STD} using a graph theoretic construction with near constant rate and small memory requirement. However, the construction in \cite{STD} holds for very large values of $K$ only. In \cite{SZG}, it was shown that the schemes with subpacketization linear in $K$ is impossible if we require constant rate. In \cite{SLB}, the authors consider caching schemes without file splitting, i.e., the scenario when $F=1$. Constructions of low-subpacketization coded caching schemes is an active area of research, with many other recent works such as \cite{cheng2018linear,Michel2019PlacementDA,Song2019SomeNC,Cheng2019AUF,SSKCombidesigns} presenting new or modifying existing constructions, using a variety of techniques including combinatorial designs, bipartite graphs, orthogonal arrays, covering arrays, etc. 

The coded caching scheme proposed in \cite{MaN} was extended to a variety of other settings, including device-to-device communication (D2D) networks \cite{D2D}, distributed computing \cite{distributed_computing_maddah_ali}, and interference management in wireless interference channels \cite{interferencemanagement}. Every one of these settings can be modelled as a multi-client communication scenario with one or more transmitters, with the clients (receivers), and in some situations the transmitters as well, having cache. This enables \textit{coded} transmissions, which generates rate advantages in all such situations. Because the fundamental scheme of \cite{MaN} is adapted to each of these settings, the subpacketization issue continues to affect the adapted schemes as well. In fact, the problem is sometimes exacerbated because the special features of the setting requires a further division of the subfiles into smaller packets (for instance, the subpacketization in the D2D scheme of \cite{D2D} is $\frac{MK}{N}\binom{K}{MK/N}$, thus having a multiplicative factor of $\frac{MK}{N}$ over that of the subpacketization of \cite{MaN}). 

In this work, we construct low-subpacketization schemes for coded caching utilizing ideas from graph theory and projective geometry over finite fields. We give the summary of the contributions of this paper in the next section.


\section{Summary of Main Contributions} \label{section summary of results}

The contributions and organization of this work are as follows. In Section \ref{section system model, examples}, we review the formal system model for coded caching on broadcast networks from \cite{MaN}. In Section \ref{subsection bipartite model}, we present the bipartite graph model for coded caching given in \cite{strongedgecoloringofbipartitegraphs}. For \textit{symmetric} coded caching schemes in which the caching is file-index invariant, the bipartite graph model captures the caching scheme and a specific class of delivery schemes in the form of a bipartite graph and its subgraphs.  

The central contribution of this work is the construction of coded caching schemes which have subpacketization subexponential in the number of users $K$ (for large $K$). Using the bipartite graph framework, and utilizing some basic ideas from the projective geometries over finite fields, we construct two new coded caching schemes, given in Section \ref{section scheme 1} (Scheme A) and Section \ref{section scheme 2} (Scheme B). We briefly describe the salient features of these schemes. Scheme A, which we present in Section \ref{section scheme 1}, has subexponential subpacketization, but uses high cache size, to achieve a rate upper bounded by a constant. When compared to Scheme A, Scheme B presented in Section \ref{section scheme 2} has equivalent subpacketization (as an asymptotic function of $K$), while having lower cache size but a higher rate. Because of the low-cache requirement which is relevant in practice, we consider Scheme B to be the more important among the two schemes presented. For ranges of users from 10s-1000s and cache-fraction in the range $0.05$ to $0.35$, we show via numerical examples that we obtain the practical values of subpacketization in the range of $10^2-10^4$, achieving the rates in the range $10-100$, (serving number of users in the range of $3-15$ per transmission). These numerical comparisons of Scheme A and Scheme B with existing schemes are shown in Table \ref{table for construction 1} (Section \ref{section scheme 1}), and Tables \ref{table for construction 2 with Ali-Niesen scheme with grouping},\ref{table for construction 2 with PDA basic paper} and \ref{table for construction 2 with ramamoorthy} (Section \ref{section scheme 2}) respectively.


\begin{table*}

\caption{Summary of asymptotic behaviour (as $K$ grows large) of the coded caching schemes for the broadcast channel presented in this paper. (where $n,m \in \mathbb{Z}^+$ and $q$ is prime power.)}

\label{table summary of schemes}

\setlength{\tabcolsep}{5.9pt} 
\renewcommand{\arraystretch}{1.5}

\centering

\begin{tabular}{|m{8em}||c|c|c||p{6em}|}

\hline
\textbf{Scheme name and} & \textbf{Cache fraction $\left(\frac{M}{N}\right)$} & \textbf{Subpacketization $(F)$} & \textbf{Rate $(R)$} & \textbf{Results, examples,} \\

\textbf{Characteristics} & & & & \textbf{numerical comparisons} \\

\hline

\multirow{5}{9em}{\textbf{Low (subexponential)
subpacketization scheme with large cache fraction}} &  $\leq$ $constant$ ($\geq 0.5$) & $O(poly(K))$ & $\Theta(K)$ &  
\multirow{1}{9em}{Section \ref{subsection scheme 1}}
\multirow{1}{9em}{Theorem \ref{theorem construction 1}}

\\ \cline{2-4}

& & & & \multirow{1}{9em}{Table \ref{table for construction 1}}
\\

&  $1-\Theta\left(\frac{1}{\sqrt{K}}\right)$ & $q^{O\left((\log_q{K})^2\right)}$ &  $O(1)$ & 
\multirow{1}{9em}{Examples \ref{example of construction 1},\ref{example of construction 1 continuation}}

\\

(\textbf{Scheme A}) &  & (\textbf{subexponential in $K$}) &  & 
\multirow{1}{9em}{Appendix \ref{appendix asymptotics of scheme 1} }

\\ 

\hline
\hline

& & & & 
\multirow{1}{9em}{Section \ref{subsection scheme 2}}
\\

\multirow{3}{9em}{\textbf{Low (subexponential) subpacketization scheme with small cache fraction}} &  $\leq$ $constant$ & $q^{O\left((\log_q{K})^2\right)}$ & $\Theta\left(\frac{K}{(\log_{q}K)^n}\right)$ &  
\multirow{1}{9em}{Theorem \ref{theorem construction 2}} 
\multirow{1}{9em}{Table \ref{table for construction 2 with Ali-Niesen scheme with grouping}}
\multirow{1}{9em}{Table \ref{table for construction 2 with PDA basic paper}}
\multirow{1}{9em}{Table \ref{table for construction 2 with ramamoorthy}}
\multirow{1}{9em}{Example \ref{example of construction 2}}

\\

(\textbf{Scheme B}) &  & (\textbf{subexponential in $K$}) &  & \multirow{1}{9em}{Appendix \ref{appendix asymptotics of scheme 2}}

\\

\hline
\hline

\multirow{2}{9em}{\textbf{Linear subpacketization $(F=K)$ scheme with small cache fraction }} & & & &

\\ 

&  $\leq$ $constant$
& $=K$ & $\Theta\left(\frac{K}{(\log_{q}K)^n}\right)$ & Section \ref{subsection linear scheme} Corollary \ref{corollary linear scheme})\\

\vspace{0.2cm}

(\textbf{Scheme C}) & & (\textbf{linear in $K$}) & & (Theorem \ref{theorem construction 2}

with $n=m$) \\

\hline

\end{tabular}

\end{table*}



\begin{table*}
    \caption{Summary of the subpacketization dependent lower bounds on the rate presented in this paper. Here $D=F(1-\frac{M}{N})$ and $d=K(1-\frac{M}{N})$.}
    \label{Table summary of lower bounds}
    \centering
    \begin{tabular}{|p{5cm}|c|}
    \hline
        Corollary \ref{corollaryfirstlowerbound}  ~~ (Section \ref{section lowerbound})  ~~ Table \ref{table lower bound}  & \\
        (Symmetric caching with every user caching the same number of subfiles) &  $R^{*}F\geq (K+F)\left(1-\frac{M}{N}\right)-1$ \\
    \hline \hline
         Theorem \ref{theorem lower bound 2} ~~~ (Section \ref{section lowerbound}) ~~ Table \ref{table lower bound} & \\
        (Symmetric caching with every user caching the same number of subfiles and every subfile cached in the same number of users) & $R^*F  \geq D+ \left\lceil{\tiny\frac{(d-1)D}{K-1}}\right\rceil +  \cdots + \left\lceil{\frac{1}{\frac{KM}{N}+1}\left\lceil {\frac{2}{\frac{KM}{N}+2}\left\lceil\cdots\left\lceil \frac{d-2}{K-2}\left\lceil{\frac{(d-1)D}{K-1}}\right\rceil\right\rceil\cdots\right\rceil}\right\rceil}\right\rceil$ \\
    \hline
    \end{tabular}

\end{table*}


Observing the subpacketization and the rate of any given scheme, as the number of users grow, gives us an understanding of the performance of the scheme, and is done in the prior literature also (see the subpacketization column of Table \ref{table known results}, for instance). It also enables comparison with the existing schemes, in situations when parameters cannot be matched accurately. Asymptotically in $K$, both the schemes A and B achieve subpacketization $F=q^{O\left((\log_q{K})^2\right)}$, Scheme A achieves this $F$ when $R=O(1)$ and $\frac{M}{N}=1-\Theta\left(\frac{1}{\sqrt{K}}\right)$, whereas Scheme B achieves this $F$ when $R=\Theta \left(\frac{K}{(\log_q{K})^{n}}\right)$ and $\frac{M}{N}$ is smaller than some constant (where $n$ is an integer constant and $q$ is some prime power). So clearly, Scheme A requires a large cache size that grows with $K$, whereas in scheme $B$ the cache fraction can be maintained constant. In the regime when $\frac{M}{N}$ is smaller than some constant, Scheme A achieves subpacketization $F=O(poly(K))$ when $R=\Theta(K)$ (thus the gain of coded caching is small and therefore this regime is not very interesting). These asymptotics are proved in respective sections, and also captured in Table \ref{table summary of schemes}. The last column of Table \ref{table summary of schemes} provides the location of the related results, numerical comparisons and examples in the paper.

For specific values of the scheme parameters, we get a new linear subpacketization scheme (Scheme C) in Section \ref{subsection linear scheme}, parametrized with two parameters, $q$ a prime power and $\lambda\in (0,1)$. For this scheme, we get the number of users $K\leq \frac{q^{2\lambda^2 q^2}}{(\lambda q)!}$, the subpacketization level $F=K$, the cache fraction $\frac{M}{N}\leq \lambda,$ and the rate being $\frac{K(1-M/N)}{\gamma}$, where $\gamma\geq \frac{4^{\lambda q}}{2\sqrt{\lambda q}}$. The asymptotics of Scheme C and location of its results in the paper are captured in Table \ref{table summary of schemes}.  A generalized version of Scheme B, with one more tunable parameter, is presented in Section \ref{subsection scheme 3}.

In Section \ref{section extensions}, we extend our  low-subpacketization low-cache Scheme B to some other settings explored in the literature where coded caching helps. In particular, we extend our Scheme B to the distributed computing in Section \ref{subsection distributed computing} (Theorem \ref{theorem distributed computing scheme}), and to the cache-aided interference channel setting in Section \ref{subsection interference channel} (Theorem \ref{theorem interference channel scheme}). In each of these settings, our extended scheme is compared with the existing schemes numerically to illustrate our low subpacketization advantage (Tables  \ref{table distributed computing}, \ref{table interferen channel}) in Section \ref{section extensions}.

Utilizing the perspective that is given by the bipartite graph model in Section \ref{subsection bipartite model}, we obtain two information-theoretic lower bounds for the rate of coded caching schemes with some fixed finite subpacketization level in Section \ref{section lowerbound}. Prior literature, for instance, in \cite{WTP,cheng2017coded}, has such lower bounds on the rate. However, not all of them take into account the finite-subpacketization constraint. The two lower bounds on the rate we obtain for given parameters $K, F, M$ and $N$, are given in Table \ref{Table summary of lower bounds} in this section. In this table, the parameter $D\triangleq F\left(1-\frac{M}{N}\right)$ and $d\triangleq K\left(1-\frac{M}{N}\right)$. Using numerical examples, we also compare the performance of these lower bounds with the bounds from \cite{WTP,cheng2017coded} in Table \ref{table lower bound} (Section \ref{section lowerbound}).
We conclude the paper with discussions regarding future work in Section \ref{section conclusion}. 

\textbf{\textit{Notations and terminology:}} $\mathbb{Z}^{+}$ denotes the set of positive integers. We denote the set $\{1,\hdots,n\}$ by $[n]$ for some $n \in \mathbb{Z}^+$. The empty set is denoted by $\phi$. For sets $A,B$, the set of elements in $A$ but not in $B$ is denoted by $A\backslash B$. The set $A\setminus a$ denotes $A\setminus \{a\}$. The set of all $b$ sized subsets of $A$ is denoted by $\binom{A}{b}$. For $a,b\in \mathbb{Z}^+$ such that $1\leq a \leq b$, $\binom{a}{b}$ represents the binomial coefficient. A set $\{A_1,A_2,\cdots,A_n\}$ is said to partition the set $A$ if $\bigcup\limits_{i=1}^{n} A_i =A$ (where the union is a disjoint union). The finite field with $q$ elements is $\fq$. The $k$-dim (dimensional) vector space over $\fq$ is represented as $\fq^k$. The dimension of a vector space $V$ over $\fq$ is given as $dim(V)$. The zero vector is represented as $\mathbf{0}$. For two subspaces $V,W$, their subspace sum is denoted by $V+W$. Note that $V+W=V\oplus W$ (the direct sum) if $V\cap W=\{\mathbf{0}\}$. The subspaces $V_1,\cdots,V_n$ (all are subspaces of a vector space) are said to be linearly independent if $v_1+\cdots +v_n = \mathbf{0}$ $\big($for each $v_i \in V_i, i\in [n]\big)$ holds only for $v_1=v_2=\cdots =v_n= \mathbf{0}$, and linearly dependent otherwise. A graph $G$ is defined by its vertex set $V(G)$ and edge set $E(G) \subseteq \binom{V(G)}{2}$. A graph $H$ is said to be a subgraph of $G$ if $V(H) \subseteq V(G)$ and $E(H) \subseteq E(G)$ such that the vertices of each edge in $E(H)$ are in $V(H)$. Further, $H$ is said to be an induced subgraph of $G$ if $E(H)$ consists of all those edges of $G$ both vertices of which are in $V(H)$. For a graph $G$, the neighbourhood of a vertex $u\in V(G)$ is defined as $\mathcal{N}(u) = \{v \in V(G) : \{u,v\} \in E(G)\}$ and $|\mathcal{N}(u)|$ denotes the degree of $u$. A graph $G$ is said to be a bipartite graph if there exist $A,B\subseteq V(G)$ such that $A\cup B =V(G)$ (where the union is a disjoint union) and every edge in $E(G)$ connects a vertex in $A$ to a vertex in $B$. Further, $A$ and $B$ are called as the set of left and right vertices respectively. A bipartite graph $G$ is said to be left-regular (right-regular) if the degree of every left (right) vertex is the same. A bipartite graph $G$ is said to be bi-regular if it is both left and right regular. A graph $G$ is said to be regular if the degree of every vertex is the same. WLOG stands for ``Without loss of generality".

\section{System Model}
\label{section system model, examples}

In this section, we present the classical coded caching setup as given by Ali-Niesen in \cite{MaN}. We then discuss the bipartite graph model for coded caching as given in \cite{strongedgecoloringofbipartitegraphs}. 
\subsection{System Model}
\label{subsection system model}

\begin{figure}
    \centering
    \includegraphics[height=2in]{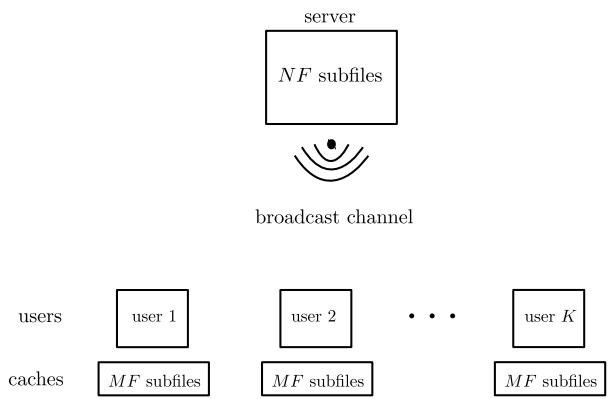}
    \caption{Broadcast coded caching setup.}
    \label{fig:broadcastchannel}
    \hrule
\end{figure}

Consider a broadcast coded caching setup as shown in Fig. \ref{fig:broadcastchannel}. Let $\cal K$ be the set of users (clients) ($|{\cal K}|=K$) in a system consisting of one server having a library of $N$ files, $\{W_i:i\in[N]\}$, connected to the clients via an error-free broadcast channel. We assume $K\leq N$, i.e., the number of files is larger than the number of users. This is typically the case in most research works in the coded caching literature (some exceptions exist, for instance \cite{more_users_than_files_KaiWan}). It is also true in many practical situations, as the library (set of all files) has possibly many more files than the number of receivers in the network. Further, if the number of receivers is large, then they may be grouped into multiple groups, and coded caching may be applied to the modified system, in which each group is considered like a single user (see, for instance, \cite{user_grouping_Shanmugam}); in which case the setting of $K\leq N$ is more relevant.


Let $F$ be the subpacketization level, i.e., we assume each file is composed of $F$ subfiles, each taking values according to a uniform distribution from some finite Abelian group ${\cal A}$. The subfiles of file $W_i$ are denoted as $W_{i,f}:f\in {\cal F}$ for some set ${\cal F}$ of size $F$. Each user can store $M$ files (equivalently, $MF$ subfiles) in its cache. A \textit{coded caching scheme }consists of two sub schemes (as in \cite{MaN}), a \textit{caching scheme} according to which subfiles of the files are placed in the user caches during periods when the traffic is low (the caching phase), and a \textit{transmission scheme} or a \textit{delivery scheme} that consists of  broadcast transmissions from the server satisfying the demands of the clients appearing during the demand phase. We assume \textit{symmetric caching} throughout the paper, i.e., for any $f\in {\cal F}$, any user either caches $W_{i,f}$ $\forall i\in [N]$ or does not cache $W_{i,f}$ $\forall i\in [N]$. Most schemes in the literature, including those for instance in \cite{strongedgecoloringofbipartitegraphs,user_grouping_Shanmugam,YCTCPDA,SZG,STD,TaR,cheng2017coded,CJYT}, employ symmetric caching. The parameters $F$ and $\frac{M}{N}$ in the symmetric caching will lead to the quantity $\frac{MF}{N}$ being an integer, indicating the number of subfiles of any particular file stored in a user's cache. 

During the demand phase, each client demands one file from the library. In the delivery scheme, the transmissions must be done so that the demands of the clients are all satisfied. The worst-case demand scenario corresponds to that situation in which each receiver demands a unique file. As in \cite{MaN}, the rate $R$ (for the worst-case demands) of such a coded caching scheme is defined as,

\[
\text{Rate}~R \triangleq  \small \frac{\stackrel{\text{\small{Number of bits transmitted in the transmission}}}{\text{scheme considering worst-case demands}} }{\text{Number of bits in a file}}.
\]
The delivery scheme typically involves multiple transmission rounds. Under the assumption that the size of each transmission is the same as that of the subfile size, the rate expression can be simplified as,

\begin{equation}
\label{eqn rate definition}
R= \small \frac{\stackrel{\text{\small{Number of transmission rounds in the}}}{\text{transmission scheme for worst-case demands}} }{\text{Number of subfiles in a file $(F)$}}.    
\end{equation}

\vspace{0.1cm}

In an achievable scheme in which the delivery scheme consists of uncoded transmissions, note that the rate must be $K\left(1-\frac{M}{N}\right)$, as the uncached fraction of each demanded file consisting of $F\left(1-\frac{M}{N}\right)$ subfiles are sent uncoded. Note that in this scheme, each transmitted subfile is intended for one particular user only.

The ratio  of the rate of the uncoded scheme to the rate $R$ of a coded caching scheme is defined as the \textit{global caching gain} $(\gamma)$ of the coded caching scheme. Thus, \begin{equation}
\label{globalcachinggaineqn}
\gamma=\frac{K(1-M/N)}{R}.
\end{equation}
The global caching gain $\gamma$ of a coded caching scheme also represents the average number of users served per transmission in the coded caching scheme. 
We are interested in designing coded caching schemes with low rate (or high gain) and low subpacketization level, which also uses low cache size at the users. 

\vspace{-0.2cm}

\subsection{Bipartite Graph based Coded Caching and Delivery based on \cite{strongedgecoloringofbipartitegraphs}}
\label{subsection bipartite model}


We can visualize the symmetric caching scheme (with fully populated caches) using a bipartite graph, following \cite{strongedgecoloringofbipartitegraphs}. We shall use this bipartite coded caching picture to obtain our coded caching schemes in Section \ref{section scheme 1} and Section \ref{section scheme 2}, as well as to obtain lower bounds on the rate of coded caching in Section \ref{section lowerbound}.

Consider a bipartite graph $B$ with ${\cal K}$ being the left (user) vertices and the right (subfile) vertices being ${\cal F}$. We then define the edges of the bipartite graph to denote the uncached subfiles of the files, i.e., for $k\in {\cal K}, f\in {\cal F}$, an edge $\{k,f\}\in E(B)$ exists if and only if the user $k$ does \textit{not} contain in its cache the subfile $W_{i,f}, \forall i\in [N]$. Clearly, this bipartite graph is left-regular, with $F\left(1-\frac{M}{N}\right)$ being the degree of any user vertex. Indeed any left-regular bipartite graph defines a caching scheme, which we formalize below. 
\begin{definition}[Bipartite Caching Scheme, Bipartite Caching Graph]
\label{definition bipartite caching scheme,graph}
Given a bipartite $D$-left-regular graph with $K$ left vertices and $F$ right vertices denoted by $B(K,F,D)$ (or in short, $B$), the symmetric caching scheme defined on $K$ users with subpacketization $F$ with the edges of $B$ indicating the uncached subfiles at the users, is called the $(K,F,D)$ bipartite caching scheme associated with the bipartite graph $B$. Further, $B(K,F,D)$ is known as the bipartite caching graph.
\end{definition}
\begin{remark}
We observe that the bipartite caching scheme associated with the graph $B(K,F,D)$ has the cache fraction $\frac{M}{N}=1-\frac{D}{F}$.
\end{remark}


\begin{figure}
    \centering
    \includegraphics[height=1.7in]{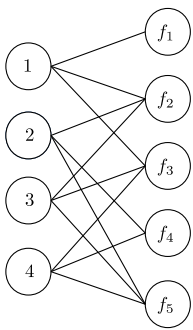}
    \caption{The figure is a bipartite caching graph $B(K=4,F=5,D=3)$ and represents a symmetric caching scheme with $4$ users, $5$ subfiles and cache fraction $\frac{M}{N}=\frac{2}{5}$. Edges indicate the missed subfiles.}
    \label{fig: bipartite caching graph}
    \hrule \vspace{-0.3cm}
\end{figure}


\vspace{-0.2cm}

\begin{example}
\label{example of bipartite caching}
Fig. \ref{fig: bipartite caching graph} shows a graph describing a $(K=4,F=5,D=3)$ bipartite caching scheme. The cache-fraction is $\frac{M}{N}=\frac{2}{5},$ meaning that each receiver caches $2$ out of the $5$ subfiles in each file. For instance, the user $2$ caches the subfiles $W_{i,f_1}, W_{i,f_3}$ and does not cache $W_{i,f_2},W_{i,f_4},W_{i,f_5}$, $\forall i\in [N]$. Similarly, the subfile $W_{i,f_1}$ is cached in the users $2,3,4$, and not at the user $1$, $\forall i\in [N]$ (where, $N$ represents the number of files in the library). 
\end{example}

Most schemes in the literature can be captured via the bipartite caching model. The delivery scheme of such schemes are the so-called `all-but-one' delivery schemes, in which each transmission is simply a sum of subfiles (one for each client in some subset of clients), with each summand subfile being demanded by a unique client in the subset, while the other summands are available in the unique clients cache. It turns out that this all-but-one delivery scheme can be captured in a graph-theoretic sense, as given in \cite{strongedgecoloringofbipartitegraphs}, an equivalent version of which we now briefly discuss. 

A matching of a graph $G$ is a subset of edges with no common vertices between any two distinct edges in the subset. An \textit{induced matching} $\C C$ of a graph $G$ is a matching such that the induced subgraph of the vertices of $\C C$ is $\C C$ itself.  The formal definition is as follows:

\begin{definition}[Induced Matching, Induced Matching Cover] \label{defintion induced matching&cover}
    Consider a bipartite graph $B$. A set of edges $\mathcal{C} \subseteq E(B)$ is called an induced matching of $B$, if every $\{k_1,f_1\},\{k_2,f_2\} \in \mathcal{C}$ satisfies $k_1\neq k_2, f_1\neq f_2$ and $\{k_1,f_2\},\{k_2,f_1\}\notin E(B)$. A set of induced matchings $\{\mathcal{C}_i \colon i\in [S]\}$ is called an induced matching cover of $B$ if it partitions $E(B)$.
\end{definition}

Now, an induced matching cover $\{\mathcal{C}_i : i\in [S]\}$ of $B$ is equivalent to the set of color classes of an $S$-\textit{strong-edge-coloring} of $B$. The work \cite{strongedgecoloringofbipartitegraphs} connects the strong edge coloring of bipartite graphs to coded caching. For more details on the definition of strong edge coloring we refer the reader to \cite{strongedgecoloringofbipartitegraphs}.

Let $W_{d_k}$ denote the demanded file of the user $k\in {\cal K}$, in the demand phase.
For an induced matching $\C C$ of the bipartite caching graph $B$ consisting of the edges $\left\{\{k_i,f_i\}:i\in[g]\right\}$ (where $g$ represents the number of edges in the induced matching $\C C$), consider the associated transmission
\begin{equation}
\label{eqn1}
Y_{\C C}=\sum_{i=1}^g W_{d_{k_i},f_i}.
\end{equation}
As ${\C C}$ is an induced matching, $ W_{d_{k_i},f_i}$ is a subfile unavailable but demanded at the user $k_i$. By the same reason, each user $k_i$ has all the subfiles in (\ref{eqn1}) in its cache except for $W_{d_{k_i},f_i}$, hence user $k_i$ can decode $W_{d_{k_i},f_i}, \forall i\in[g].$ 
If $\{ \mathcal{C}_i, i\in[S]\}$ is an induced matching cover then it is easy to see that the transmissions $Y_{\mathcal{C}_i}:i\in[S]$ (constructed as in (\ref{eqn1})) corresponding to $\mathcal{C}_i:i\in[S]$ satisfy the demands of all the users, as all the edges of the bipartite caching graph are `covered' by the induced matching cover. Therefore, we have obtained a valid delivery scheme. We refer to this delivery scheme as the \textit{induced-matching based delivery scheme}. Here, the parameter $S$ represents the number of induced matchings in the induced matching cover or equivalently the number of transmissions in the associated delivery scheme. By (\ref{eqn rate definition}), the rate of this transmission scheme is $\frac{S}{F}$. Further, if $|{\C C}_i|=g, \forall i \in [S]$ then the rate will be $\frac{S}{F}=\frac{K(1-M/N)}{g}$ and the coded caching gain is $g$. Thus, each transmission corresponding to each induced matching will serve exactly $g$ users.

The above discussion of the bipartite caching scheme and the induced-matching based delivery scheme is summarized into the following theorem, which will be used to derive the coded caching parameters from the bipartite caching graphs that we construct in Section \ref{section scheme 1} and Section \ref{section scheme 2}. The equivalent description of this theorem can be found in \cite{strongedgecoloringofbipartitegraphs} in the language of strong edge coloring.

\begin{tcolorbox}
\begin{theorem}
\label{theorem bipartite coded caching}
Consider a bipartite caching graph $B(K,F,D)$ with an induced matching cover $\{{\C C}_i : i\in [S]\}$ such that $|{\C C}_i|=g, \forall i \in [S]$.
Then there is a coded caching scheme for a broadcast system with $K$ users, each with cache size $M$, with the number of files $N\geq K$, consisting of the caching scheme defined by $B(K,F,D)$ with subpacketization $F$, cache fraction $\frac{M}{N}= 1-\frac{D}{F}$, and the associated delivery scheme based on the induced matching cover $\{{\C C}_i : i\in [S]\}$ having rate $R=\frac{S}{F}$ and global caching gain $\gamma =g$.
\end{theorem}
\end{tcolorbox}

The following example illustrates an induced matching cover for the bipartite caching scheme presented in Example \ref{example of bipartite caching}.
\begin{example}[Continuation of Example \ref{example of bipartite caching}]
\label{example of induced matching}
Consider the following subsets of edges of the bipartite caching graph presented in Fig. \ref{fig: bipartite caching graph}. ${\C C}_1 = \left\{ \{1,f_1\},\{3,f_5\} \right\}$ , ${\C C}_2 = \left\{ \{1,f_2\},\{4,f_5\} \right\}$ , ${\C C}_3 = \left\{ \{1,f_3\},\{2,f_5\} \right\}$ , ${\C C}_4 = \left\{ \{2,f_2\},\{4,f_3\} \right\}$ , ${\C C}_5 = \left\{ \{2,f_4\},\{3,f_3\} \right\}$ , ${\C C}_6 = \left\{ \{3,f_2\},\{4,f_4\} \right\}$. Further $\bigcup\limits_{i=1}^{6} \mathcal{C}_i =E(B)$, (where the union is a disjoint union). It is easy to see that each $\mathcal{C}_i, \forall i\in [6]$ is an induced matching and $\{\mathcal{C}_i: i\in [6]\}$ is an induced matching cover by Definition \ref{defintion induced matching&cover}. Therefore, by Theorem \ref{theorem bipartite coded caching}, we have a delivery scheme corresponding to the induced matchings of $B$, with rate $R=\frac{6}{5}$ and global caching gain $\gamma = 2$. In Example \ref{example of lower bound}, we show that this rate is optimal for the symmetric caching scheme defined by $B(4,5,3)$.
\end{example}




\section{A New Low Subpacketization Scheme with Large Cache Fraction (Scheme A)}
\label{section scheme 1}

In this section and in Section \ref{section scheme 2}, we present new coded caching schemes using the bipartite graph framework, which we have recollected in Section \ref{subsection bipartite model}, via tools from the projective geometry. Intuitively, these schemes combine ideas from the base-line coded caching scheme of \cite{MaN}, which is based on sets and set-containment, with ideas from the projective geometry, i.e., subspaces and subspace-containment (containment or the lack thereof of smaller subspaces within larger ones).  

In this section we present a construction of coded caching scheme (Scheme A) which achieves a subpacketization which is subexponential in $K$, with cache fraction $(\frac{M}{N})$ $\geq 0.5$.  Example \ref{example of construction 1} is a motivating example corresponding to the construction in Scheme A. 

The construction of Scheme A (also Scheme B in Section \ref{section scheme 2}) is based on the bipartite graph approach recollected in Section \ref{subsection bipartite model}. We construct the bipartite caching graph by giving user vertices and subfile vertices, and then identify an induced matching cover in it. Then by using Theorem \ref{theorem bipartite coded caching}, we obtain the parameters of the coded caching scheme.

As our constructions use some simple results from the projective geometry, we first review some basic concepts from the projective geometry over finite fields and develop some mathematical terminology.

\subsection{Review of the Projective Geometries over Finite Fields \cite{hirschfeld1998projective}}
\label{subsection review of projective geometry}
Let $k,q\in \mathbb{Z}^+$ such that $q$ is a prime power. Let $\fq^k$ be a $k$-dim vector space over a finite field $\fq$. Let `$\mathbf{0}$' represent the zero vector of $\fq^k$. Consider the set of equivalence classes of $\fq^k\setminus \{\boldsymbol{0}\}$ under the equivalence relation $\sim$ defined by $x \sim y$ if there is a nonzero element $ \alpha \in \fq$ such that $x = \alpha y$.
The $(k-1)$-dim \textit{projective space} over $\fq$ is denoted by $PG_q(k-1)$ and is defined as the set of these equivalence classes. For $m\in [k]$, let $PG_q(k-1,m-1)$ denote the set of all $m$-dim subspaces of $\fq^k$.
From Chapter $3$ in \cite{hirschfeld1998projective} it is known that  $|PG_q(k-1,m-1)|$ is equal to the \textit{Gaussian(}or \textit{q)-binomial coefficient} $\gbinom{k}{m}$ where,
\[\gbinom{k}{m}
\triangleq \frac{(q^k-1)\hdots(q^{k-m+1}-1)}{(q^m-1)\hdots(q-1)}. \text{  (where }k\geq m \text{) }\]
In fact, $\gbinom{k}{m}$ gives the number of $m$-dim subspaces of any $k$-dim vector space over $\fq$. Further, by definition, $\gbinom{k}{0}=1.$
In any Gaussian binomial coefficient $\gbinom{a}{b}$ given in this paper we assume that $a,b\in \mathbb{Z}^+$ and $1\leq b\leq a.$

The following known results from \cite{hirschfeld1998projective} are used to describe our schemes (Scheme A and Scheme B).

\begin{lemma}\cite{hirschfeld1998projective} 
\label{lemma gaussiancoeff}
Consider a $k$-dim vector space $\mathbb{F}_q^k$. Let $1\leq r,s,l <k$.
\begin{itemize}
\item[A1:] $\gbinom{k}{r}=\gbinom{k}{k-r}$.
\item[A2:] The number of distinct $r$-dim subspaces of $\mathbb{F}_q^k$ containing a fixed $l$-dim subspace is $\gbinom{k-l}{r-l}$.
\item[A3:] The number of distinct $r$-dim subspaces of $\mathbb{F}_q^k$ intersecting a fixed $s$-dim subspace in some $l$-dim subspace is

$$q^{(r-l)(s-l)}\gbinom{k-s}{r-l} \gbinom{s}{l}.$$
\end{itemize}
\end{lemma}

    
    




We are essentially interested in finding out some asymptotic results of the schemes which we develop in next sections. For this reason, we use the following simple upper and lower bounds on Gaussian binomial coefficients and their relationships. 
\begin{lemma}
\label{lemma bounds on gaussian binomial coefficients}
For non-negative integers $a,b,f$, for $q$ being some prime power, 

\begin{align}
\label{eqn31}
q^{(a-b)b} & \leq \gbinom{a}{b} \leq  q^{(a-b+1)b}
\end{align}

\begin{align}
\label{eqn32}
q^{(a-f-1)b} & \leq \frac{\gbinom{a}{b}}{\gbinom{f}{b}} \leq  q^{(a-f+1)b}
\end{align}

\begin{align}
\label{eqn33}
q^{(a-f-b-1)\delta} &\leq \frac{\gbinom{a}{b}}{\gbinom{a}{f}} \leq  q^{(a-f-b+1)\delta}.
\end{align}
where, $\delta=max(b,f)-min(b,f)$. 
\end{lemma}
\begin{IEEEproof}
The first lower bound for $\gbinom{a}{b}$ is well known from the combinatorics literature (see, for instance, \cite{Tak}). All the other bounds are proved by definition of the Gaussian binomial coefficient and by noting that $q^a-1\geq q^{a-1}$ (since $q\geq 2$), and $q^a-1\leq q^a$. This completes the proof.
\end{IEEEproof}

\subsection{An Illustrative Example}
We now show an example which we shall see illustrates the idea behind the construction in Section \ref{subsection scheme 1}.
\begin{example}
\label{example of construction 1}

Consider a caching system with $K=7, F=7, \frac{M}{N}=\frac{4}{7}$. To present this system, we need to provide the indexing for the users $(\mathcal{K})$ and the subfiles $(\mathcal{F})$. For this purpose, we consider some quantities from the projective geometry over finite fields. Consider a $3$-dim vector space $(\mathbb{F}_{2}^{3})$ over a binary field $\mathbb{F}_{2}$. Consider the $1$-dim subspaces of $\mathbb{F}_{2}^{3}$ which are listed as follows,


\begin{align*}
V_1 &= span\{(0,0,1)\}  \\
V_2 &= span\{(0,1,0)\}  \\
V_3 &= span\{(1,0,0)\}  \\
V_4 &= span\{(1,1,0)\}  \\
V_5 &= span\{(1,0,1)\}  \\
V_6 &= span\{(0,1,1)\}  \\
V_7 &= span\{(1,1,1)\}. 
\end{align*}

Let $\mathbb{V} = \{V_1,V_2,V_3,V_4,V_5,V_6,V_7\}$. Consider the $2$-dim subspaces of $\mathbb{F}_{2}^{3}$ which are listed as follows,

\begin{align*}
    X_1 &= \{(0,0,1),(0,1,0),(0,1,1),(0,0,0)\} \\
    X_2 &= \{(0,0,1),(1,0,0),(1,0,1),(0,0,0)\} \\
    X_3 &= \{(0,1,0),(1,1,0),(1,0,0),(0,0,0)\} \\
    X_4 &= \{(0,0,1),(1,1,0),(1,1,1),(0,0,0)\} \\
    X_5 &= \{(0,1,0),(1,0,1),(1,1,1),(0,0,0)\} \\
    X_6 &= \{(1,0,0),(0,1,1),(1,1,1),(0,0,0)\} \\
    X_7 &= \{(1,1,0),(0,1,1),(1,0,1),(0,0,0)\}.
\end{align*}

Let $\mathbb{X}=\{X_1,X_2,X_3,X_4,X_5,X_6,X_7\}$. We now proceed to describe the caching phase and the delivery phase.

\textbf{\underline{Caching phase:}} Let \underline{$\mathcal{K}=\mathbb{V}$} and \underline{$\mathcal{F}=\mathbb{X}$}. During the caching phase, every file $(W_i,i\in [N])$ is divided into $F=7$ subfiles. The subfiles of $W_i$ are denoted as $W_{i,X}, \forall X\in \mathcal{F}$. The caching scheme is,

\begin{itemize}
    \item For each $i\in[N]$, the user $V_l \in \mathcal{K}$ caches the subfile $W_{i,X}$ if $V_l$ is not a subspace of $X$.  
\end{itemize}

Following this rule, we have the cached and uncached subfile indices of each user as shown in Table \ref{table example scheme 1}. The first two columns of Table \ref{table example scheme 1}, provides users and indices of cached subfiles respectively. Therefore, every user caches $4$ out of the $7$ subfiles in every file. Hence, the cache fraction $\frac{M}{N}= \frac{4}{7}$.


\begin{table}[htbp]

\caption{\small Indices of the cached and uncached subfiles for the caching scheme presented in Example \ref{example of construction 1}.}
\label{table example scheme 1}

\setlength{\tabcolsep}{7pt} 
\renewcommand{\arraystretch}{1.6}

\centering
\begin{tabular}{|c|c|c|}
\hline
\textbf{Users} & \textbf{Indices of} & \textbf{Indices of uncached} \\

& \textbf{cached subfiles} & \textbf{(equivalently demanded) subfiles} \\

\hline

$V_1$ & $X_3,X_5,X_6,X_7$ & $X_1,X_2,X_4$ \\

\hline

$V_2$ & $X_2,X_4,X_6,X_7$ & $X_1,X_3,X_5$ \\

\hline

$V_3$ & $X_1,X_4,X_5,X_7$ & $X_2,X_3,X_6$ \\

\hline

$V_4$ & $X_1,X_2,X_5,X_6$ & $X_3,X_4,X_7$ \\

\hline

$V_5$ & $X_1,X_3,X_4,X_6$ & $X_2,X_5,X_7$ \\

\hline

$V_6$ & $X_2,X_3,X_4,X_5$ & $X_1,X_6,X_7$ \\

\hline

$V_7$ & $X_1,X_2,X_3,X_7$ & $X_4,X_5,X_6$ \\

\hline
\end{tabular}

\end{table}


\textbf{\underline{Delivery phase:}}
Let demand of an arbitrary user $V_l \in \mathcal{K}$ be $W_{d_{V_l}}$. The demanded (uncached) subfile indices corresponding to each user are given in the last column of Table \ref{table example scheme 1}. To satisfy the demands of the users, the server transmits the following $7$ transmissions.

\begin{align*}
    & W_{d_{V_6},X_1}+W_{d_{V_5},X_2}+W_{d_{V_7},X_4} \\
    & W_{d_{V_1},X_1}+W_{d_{V_4},X_3}+W_{d_{V_7},X_5}\\
    & W_{d_{V_1},X_2}+W_{d_{V_2},X_3}+W_{d_{V_7},X_6} \\
    & W_{d_{V_3},X_3}+W_{d_{V_1},X_4}+W_{d_{V_6},X_7}\\
    & W_{d_{V_3},X_2}+W_{d_{V_2},X_5}+W_{d_{V_4},X_7}\\
    & W_{d_{V_2},X_1}+W_{d_{V_3},X_6}+W_{d_{V_5},X_7}\\
    & W_{d_{V_4},X_4}+W_{d_{V_5},X_5}+W_{d_{V_6},X_6}.
\end{align*}

Each transmission is a linear combination of $3$ demanded subfiles. It is easy to see (using Table \ref{table example scheme 1}) that any user decodes their demanded subfiles. For instance, the user $V_1$ decodes $W_{d_{V_1},X_1}$ from the second transmission as it contains $W_{d_{V_4},X_3}$ and $W_{d_{V_7},X_5}$ in its cache. Similarly, the user $V_1$ decodes $W_{d_{V_1},X_2}$ from the third transmission and $W_{d_{V_1},X_4}$ from the fourth transmission. Thus, each transmission serves $3$ users. Hence, the global caching gain is $\gamma=3$. By using (\ref{eqn rate definition}), the rate of the scheme is $R=\frac{7}{7}=1$.
\end{example}

\begin{remark}
\label{remarkbelowexampleschemeA}
The caching scheme of the Example \ref{example of construction 1} is obtained based on the idea of `subspace containment', which we shall see motivates the caching scheme of our new construction in this section.  We shall show in Example \ref{example of construction 1 continuation} in Section \ref{subsection scheme 1} that the delivery scheme of Example \ref{example of construction 1} is  obtained from an induced matching cover of  the bipartite caching graph of the new construction.
\end{remark}

We are now ready to construct Scheme A, which is the main result of this section.

\subsection{Construction of Scheme A}
\label{subsection scheme 1}

We will construct Scheme A by constructing a bipartite caching graph $B(K,F,D)$ and identify an induced matching cover in it. Then by using Theorem \ref{theorem bipartite coded caching}, we obtain the coded caching parameters $(K,F,\frac{M}{N},R,\gamma)$ of Scheme A.

Let $k,m,t,q$ be positive integers such that $m+t\leq k$ and $q$ is some prime power. Consider a $k$-dim vector space $\fq^k$. Consider the following sets of subspaces which are used to index our user vertices, subfile vertices and the induced matchings of the bipartite caching graph $B$ which we construct.

\begin{align*}
    \mathbb{V} &\triangleq  PG_q(k-1,t-1). \text{  (set of all $t$-dim subspaces)}\\
    \mathbb{X} &\triangleq  PG_q(k-1,m+t-1). \text{\small{(set of all $(m+t)$-dim subspaces)}}\\
    \mathbb{T} & \triangleq  PG_q(k-1,m-1). \text{  (set of all $m$-dim subspaces)}
\end{align*}

Construct a bipartite graph $B$ with left (user) vertex set \underline{$\mathcal{K}=\mathbb{V}$} and right (subfile) vertex set \underline{$\mathcal{F}=\mathbb{X}$}. Define the edge set of $B$ as,
\[
E(B) \triangleq \{\{V,X\} : V\in \mathbb{V} , X\in \mathbb{X}, V\subseteq X\}.
\]
We now find the values of $K,F$ and the degree of any user vertex (left degree) $D$.

\begin{lemma}
\label{lemma K,F,D,c expressions of scheme 1}
The following relationships hold for the construction we have presented. 
\[K = |\mathbb{V}|= \gbinom{k}{t}, ~~~~~~
F = |\mathbb{X}| = \gbinom{k}{m+t},\]
\[D \triangleq |\mathcal{N}(V)| =\gbinom{k-t}{m},\]
where the last relationship holds for any $V\in \mathbb{V}$.
\end{lemma}

\begin{IEEEproof}
By using the ideas presented in Section \ref{subsection review of projective geometry} we can write,
\[ K = |\mathbb{V}| = |PG_q(k-1,t-1)| = \gbinom{k}{t}.\]
\[ F = |\mathbb{X}| = |PG_q(k-1,m+t-1)| = \gbinom{k}{m+t}.\]
Now, we will find $|\mathcal{N}(V)|$. Consider an arbitrary $V\in \mathbb{V}$. 
It is easy to see that $\mathcal{N}(V) = \{X \in \mathbb{X} : V \subseteq X\}$. Therefore, finding $|{\C N}(V)|$ is equivalent to counting the number of $(m+t)$-dim subspaces of $\fq^k$ containing the fixed $t$-dim subspace $V$. By applying A2 of Lemma \ref{lemma gaussiancoeff} we get,
\[D=|{\C N}(V)|=\gbinom{k-t}{(m+t)-t}=\gbinom{k-t}{m}.\]
This completes the proof.
\end{IEEEproof}

Note that, by Lemma \ref{lemma K,F,D,c expressions of scheme 1}, $B$ is a $D$-left regular bipartite graph with $K$ left vertices and $F$ right vertices. Therefore, by Definition \ref{definition bipartite caching scheme,graph}, $B(K,F,D)$ is a valid bipartite caching graph.

\begin{remark}
\label{remark on bi-regularity of graph in scheme A}
Similar to left degree $D$, it is easy to see that the degree of any right (subfile) vertex $X\in \mathbb{X}$ is $|\mathcal{N}(X)| = |PG_q(m+t-1,t-1)| = \gbinom{m+t}{t}$. Therefore, $B$ is a bi-regular bipartite caching graph.
\end{remark}

We now show that $B$ has an induced matching cover
$\{{\C C}_i : i\in [S]\}$ such that $|{\C C}_i|=g, \forall i \in [S]$, for some $g,S \in \mathbb{Z}^+$. 


\textbf{\underline{Induced matching cover:}} The induced matchings of $B$, that we wish to obtain, is based on a relabeling of the edges $B$ based on $m$-dim subspaces of $\fq^k$. Towards that end, we first require the following lemmas (Lemma \ref{lemma perfectmatching} and Lemma \ref{lemma valid alternate labeling}) using which we can find `matching' labels to the $t$-dim and $m$-dim subspaces of some $X \in \mathbb{X}$. Subsequently, using Lemma \ref{lemma transmission clique in scheme 1} and Lemma \ref{lemma transmission clique cover in scheme 1}, we show the induced matching cover of $B$.

 
\begin{lemma}
\label{lemma perfectmatching}
Consider some element $X\in \mathbb{X}$. Let $\left\{ V_i, i=1,\hdots,\gbinom{m+t}{t}\right\}$ denote the $t$-dim subspaces of $X$ taken in some fixed order.  Then the set of $m$-dim subspaces of $X$ can be written as an indexed set as $\left\{T_{V_i,X},~ i=1,\hdots,\gbinom{m+t}{m}\right\}$ such that $T_{V_i,X} \oplus V_i=X, \forall i$ (where $\oplus$ denotes direct sum). Moreover, such an indexed set can be found in operations polynomial in $\gbinom{m+t}{t}$.
\end{lemma}
\begin{IEEEproof}
See Appendix \ref{appendix proof perfect matching}.
\end{IEEEproof}
For a $t$-dim subspace $V_i$ contained in  an $(m+t)$-dim subspace $X,$ let $T_{V_i,X}$ (the $m$-dim subspace as obtained in Lemma \ref{lemma perfectmatching} such that $T_{V_i,X}\oplus V_i=X$) be called \textit{the matching subspace of $V_i$ in $X$}. Using these matching subspaces, we can obtain an alternate labeling scheme for the edges of our bipartite caching graph $B$. The alternate labels are given as follows:
\begin{itemize}
\item Let the alternate label for $\{V,X\}$ be $\{V,T_{V,X}\}$, where $T_{V,X}$ is the $m$-dim matching subspace of $V$ in $X$ obtained using Lemma \ref{lemma perfectmatching}.
\end{itemize}
The following lemma ensures that the alternative labeling given above is indeed a valid labelling, i.e., it uniquely identifies the edges of $B$. 
\begin{lemma}
\label{lemma valid alternate labeling}
No two edges of $B$ have the same alternate label.
\end{lemma}
\begin{IEEEproof}
If $\{V_1,X_1\},\{V_2,X_2\}\in E(B)$ have the same alternate label $\{V,T_{V,X}\}$, then clearly $V_1=V_2=V$. Moreover, we should also, by definition of the alternate labels, have that $X_1=T_{V,X}\oplus V=X_2$. Therefore $\{V_1,X_1\}=\{V_2,X_2\}$. This completes the proof.
\end{IEEEproof}
We are now in a position to present the induced matching cover of $B$. Our induced matchings (defined in Lemma \ref{lemma transmission clique in scheme 1}) are represented in terms of the alternate labels given to the edges of $B$. We first show the structure of one such induced matching.
\begin{lemma}
\label{lemma transmission clique in scheme 1}
For an $m$-dim subspace $T \in \mathbb{T}$, consider the subset of $E(B)$ (identified by their alternate labels) as follows: 
\[
{\cal C}_T \triangleq \{ \{V,T\} \in  E(B): V\in \mathbb{V}\}.
\]
Then ${\C C}_T$ is a $\gbinom{k-m}{t}$-sized induced matching of $B$. 
\end{lemma}
\begin{IEEEproof}
Firstly, we observe that ${\C C}_T$ is a well-defined set because the $T$ is an $m$-dim subspace of precisely $\gbinom{k-m}{(m+t)-m}$  subspaces of dimension $(m+t)$ by A2 of Lemma \ref{lemma gaussiancoeff}. 
Note that $\{V,T\}$ is the alternate label for $\{V,T\oplus V\}\in E(B)$. Also, we can observe that for distinct $\{V_1,T\}, \{V_2,T\}\in {\C C}_T$, we must have $V_1\oplus T\neq V_2\oplus T$. This is due to the fact that each $m$-dim subspace within an $(m+t)$-dim subspace $X$ is matched to a unique $t$-dim subspace of $X$. Hence, by Lemma \ref{lemma perfectmatching} and our alternate labeling scheme, we should have $|{\C C}_T|=\gbinom{k-m}{(m+t)-m}=\gbinom{k-m}{t}$. 

\vspace{0.2cm}

We now show that ${\C C}_T$ forms an induced matching of $B$. We do this using Definition \ref{defintion induced matching&cover}.  Consider two distinct and arbitrary edges $\{V_1,T\},\{V_2,T\}\in {\C C}_T$. These are the alternate labels for $\{V_1,T\oplus V_1\},\{V_2,T\oplus V_2\}$ respectively. We have that $V_1\neq V_2$, and have already checked that $T\oplus V_1\neq T\oplus V_2.$ Further, note that $V_1\not\subset T\oplus V_2$. This is because if $V_1\subset T\oplus V_2$, then, $T\oplus V_1=T\oplus V_2,$ which is not true by the definition of ${\cal C}_T$. Thus $\{V_1,T\oplus V_2\}\notin E(B)$. Similarly, $\{V_2,T\oplus V_1\}\notin E(B)$. By invoking the Definition \ref{defintion induced matching&cover}, it is clear that $\mathcal{C}_T$ is an induced matching of $B$. This completes the proof. 
\end{IEEEproof}
We now show that the set of induced matchings $\{{\C C}_T : T\in \mathbb{T}\}$ partition $E(B)$. 
\begin{lemma}
\label{lemma transmission clique cover in scheme 1}
\[
\bigcup\limits_{T\in \mathbb{T}}{\C C}_T=E(B),
\]
where the above union is a disjoint union (the induced matchings $\mathcal{C}_T$ are as defined in Lemma \ref{lemma transmission clique in scheme 1}).
\end{lemma}
\begin{IEEEproof}
It should be clear from our alternate labeling scheme and the definition of ${\C C}_{T}$ that any edge $\{V,X\}\in E(B)$ $\big($which gets some alternate label $\{V,T_{V,X}\}\big)$ appears in the induced matching of $B$ given by ${\C C}_{T_{V,X}}.$ Furthermore, by definition ${\C C}_{T_1}$ and ${\C C}_{T_2}$ are disjoint for any distinct $T_1$ and $T_2$ in $\mathbb T$. This completes the proof.
\end{IEEEproof}

Therefore, $\{\mathcal{C}_T : T\in \mathbb{T} \}$ is an induced matching cover of $B$. In the light of the construction of the bipartite caching graph $B$ of this section and the induced matching cover of $B$, we are now ready to present our coded caching scheme, Scheme A, in the following theorem.

\begin{tcolorbox}
\begin{theorem}
\label{theorem construction 1}
\textbf{(Scheme A)} Let $k,m,t$ be positive integers such that $m+t\leq k$ and $q$ be any prime power. The bipartite graph $B$ constructed in Section \ref{subsection scheme 1} is a $B(K,F,D)$ bipartite caching graph with an induced matching cover having $S=\gbinom{k}{m}$ induced matchings, each having $g=\gbinom{k-m}{t}$ edges and defines a coded caching scheme with,
\[ K=\gbinom{k}{t}, ~~~~~~~~~~~~~~~~~~~~~ F=\gbinom{k}{m+t},\] \[\frac{M}{N}=1-\frac{\gbinom{k-t}{m}}{\gbinom{k}{m+t}}, ~~~~~~~~~~ R=\frac{\gbinom{k}{m}}{\gbinom{k}{m+t}},\]
\[\text{Global caching gain } \gamma = \gbinom{k-m}{t}. \]
\end{theorem}
\end{tcolorbox}

\begin{IEEEproof}
From Lemma \ref{lemma K,F,D,c expressions of scheme 1}, we get the expressions of $K, F$ and $D$. By Lemma
\ref{lemma transmission clique in scheme 1} and Lemma \ref{lemma transmission clique cover in scheme 1}, the size of the induced matchings of  $B$ is $g= |\mathcal{C}_T|= \gbinom{k-m}{t}$ for any $T\in \mathbb{T}$ and they partition the edge set $E(B)$. Further, by Lemma \ref{lemma transmission clique in scheme 1}, the number of induced matchings in the induced matching cover is $S=|\mathbb{T}| = |PG_q(k-1,m-1)| = \gbinom{k}{m}$. Hence,  the bipartite graph $B$ satisfies all the conditions in Theorem \ref{theorem bipartite coded caching}. Therefore, there exists a coded caching scheme with $K$ users, subpacketization $F$,
\[\frac{M}{N}=1-\frac{D}{F}=1-\frac{\gbinom{k-t}{m}}{\gbinom{k}{m+t}},\]
\[R = \frac{S}{F} = \frac{\gbinom{k}{m}}{\gbinom{k}{m+t}},\]
\[\text{Global caching gain }\gamma =g = \gbinom{k-m}{t}.\]
This completes the proof.
\end{IEEEproof}

\begin{figure}
  \centering
        \includegraphics[height=1in]{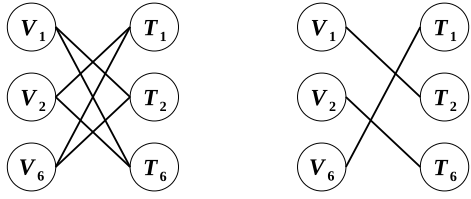}
\caption{Illustration of the relabeling procedure presented in Example \ref{example of construction 1 continuation}: The left figure is a bi-regular bipartite graph $(B_1)$ with the left and right vertices being the $t$-dim and $m$-dim subspaces of $X_1$ respectively. Note that $\{V_i,T_j\} \in E(B)$ if $V_i \oplus T_j =X_1$ where $i,j \in \{1,2,6\}$. The right figure is a perfect matching of $B_1$. Hence, the relabels of $\{V_1,X_1\},\{V_2,X_1\},\{V_6,X_1\}$ are $\{V_1,T_2\},\{V_2,T_6\},\{V_6,T_1\}$ respectively.}
\label{fig:perfect_matching}
\hrule
\end{figure}

\begin{example}{(Continuation of Example \ref{example of construction 1})}
\label{example of construction 1 continuation}
Example \ref{example of construction 1} gives us an illustration of our Scheme A for the values of $t=1,m=1,k=3$ and $q=2$. It is not difficult to see that the bipartite caching graph $B(K,F,D)$ obtained according to Scheme A gives us the caching scheme as given in Example \ref{example of construction 1}. We now illustrate how we obtained the delivery scheme as shown in Example \ref{example of construction 1}, using the induced matching cover of $B$ obtained through perfect matchings of the bipartite graph constructed in Appendix \ref{appendix proof perfect matching} (proof of Lemma \ref{lemma perfectmatching}) and our alternate labelling scheme. 

In the bipartite graph terminology we have recollected, the edge set of the bipartite caching graph $B$ can be inferred from Table \ref{table example scheme 1}. For instance, the edges incident at the user vertex $V_1$ are $\{V_1,X_1\},\{V_1,X_2\},\{V_1,X_4\}$ and the edges incident at the subfile vertex $X_1$ are $\{V_1,X_1\},\{V_2,X_1\},\{V_6,X_1\}$. First, we relabel the edges of $B$ as per Lemma \ref{lemma perfectmatching}. Let $\mathbb{T}$ be the set of all $m$-dim subspaces of $\mathbb{F}_2^3$. Since $t=m=1$, we can consider $\mathbb{T}=\mathbb{V}$ with $T_i=V_i, i\in [7]$. Consider an $(m+t)$-dim subspace $X_1$. This can be written as $X_1=V_1\oplus T_2=V_1\oplus T_6=V_2\oplus T_1=V_2\oplus T_6=V_6\oplus T_1=V_6\oplus T_2$. 

The relabeling procedure is illustrated in Fig. \ref{fig:perfect_matching}. The left figure represents a bipartite graph with the $t$-dim ($m$-dim) subspaces of $X_1$ as the left (right) vertices. The edges denote the subspaces in direct sum. This bipartite graph is not the caching graph, but it corresponds to the one defined in Appendix \ref{appendix proof perfect matching} used to find the relabeling. The right figure of Fig. \ref{fig:perfect_matching} denotes a perfect matching of the bipartite graph. In the same way, for each $(m+t)$-dim subspace $X\in{\mathbb X}$, a perfect matching is obtained. Following the edges of the perfect matchings gives us the new labels for the elements of $E(B)$, which are presented in Table \ref{Table perfect matching}. 

Now we describe how we can get the delivery scheme using the induced matchings of the bipartite caching graph $B$, which are indexed using $\mathbb T$. For each $m$-dim subspace $\mathbb{T}$  we can define an induced matching of $B$ as per Lemma \ref{lemma transmission clique in scheme 1} using the alternate labels for the edges of $B$. For instance, the induced matching corresponding to $T_1$ is $\mathcal{C}_{T_1}=\{\{V_6,T_1\},\{V_5,T_1\},\{V_7,T_1\}\}=\{\{V_6,X_1\},\{V_5,X_2\},\{V_7,X_4\}\}$. Therefore, the transmission corresponding to $\mathcal{C}_{T_1}$ is $W_{d_{V_6},X_1}+W_{d_{V_5},X_2}+W_{d_{V_7},X_4}$ as mentioned in Example \ref{example of construction 1}. Similarly, other transmissions can be obtained from the induced matchings ${\cal C}_T:T\in{\mathbb T}$, corresponding to which the transmissions are shown in Example \ref{example of construction 1}.
\end{example}


\begin{table}[htbp]

\caption{Old and new labels of the edges in $E(B)$ through perfect matchings, for the bipartite caching graph $B$ presented in Example \ref{example of construction 1 continuation}.}

\label{Table perfect matching}

\setlength{\tabcolsep}{9pt} 
\renewcommand{\arraystretch}{1.75}

\centering
\begin{tabular}{|c|c||c|c|}

\hline

\textbf{Old labels} & \textbf{New labels} & \textbf{Old labels} & \textbf{New labels} \\

\hline
$\{V_1,X_1\}$ & $\{V_1,T_2\}$ & $\{V_7,X_4\}$ & $\{V_7,T_1\}$ \\

\hline \cline{3-4} \cline{3-4}
$\{V_2,X_1\}$ & $\{V_2,T_6\}$ & $\{V_2,X_5\}$  & $\{V_2,T_5\}$ \\

\hline
$\{V_6,X_1\}$ & $\{V_6,T_1\}$ & $\{V_5,X_5\}$ & $\{V_5,T_7\}$ \\

\hline \cline{1-2} \cline{1-2}
$\{V_1,X_2\}$ & $\{V_1,T_3\}$ & $\{V_7,X_5\}$ & $\{V_7,T_2\}$ 
\\

\hline \cline{3-4} \cline{3-4}
$\{V_3,X_2\}$ & $\{V_3,T_5\}$ & $\{V_3,X_6\}$ & $\{V_3,T_6\}$ \\

\hline
$\{V_5,X_2\}$ & $\{V_5,T_1\}$ & $\{V_6,X_6\}$ & $\{V_6,T_7\}$ \\

\hline \cline{1-2} \cline{1-2}
$\{V_2,X_3\}$ & $\{V_2,T_3\}$ & $\{V_7,X_6\}$ & $\{V_7,T_3\}$ \\

\hline \cline{3-4} \cline{3-4}
$\{V_3,X_3\}$ & $\{V_3,T_4\}$ & $\{V_4,X_7\}$ & $\{V_4,T_5\}$ \\

\hline
$\{V_4,X_3\}$ & $\{V_4,T_2\}$ & $\{V_5,X_7\}$ & $\{V_5,T_6\}$ \\

\hline \cline{1-2} \cline{1-2}
$\{V_1,X_4\}$ & $\{V_1,T_4\}$ & $\{V_6,X_7\}$ & $\{V_6,T_4\}$ \\

\hline \cline{3-4} \cline{3-4}
$\{V_4,X_4\}$ & $\{V_4,T_7\}$ & & \\

\hline
\end{tabular}

\end{table}



\begin{table*}[htbp]

\caption{Comparison of the coded caching scheme in Theorem \ref{theorem construction 1} (Scheme A) with the scheme in \cite{user_grouping_Shanmugam} (Ali-Niesen scheme with grouping). We match the cache fraction, the gain, and number of users as closely as possible, and compare the subpacketization level. }
\label{table for construction 1}

\setlength{\tabcolsep}{6pt} 
\renewcommand{\arraystretch}{1.1}
\centering

\begin{tabular}{|c|c||c||c||g|c|}
\hline

\multicolumn{2}{|c||}{\textbf{Number}} & \textbf{Cache} & \textbf{Global} & \multicolumn{2}{|c|}{\textbf{Subpacketization}} \\

\multicolumn{2}{|c||}{\textbf{of users}} & \textbf{fraction} & \textbf{caching gain} & \multicolumn{2}{|c|}{} \\

\hline

$(k,m,t,q)$ & $(K^{'},l)$ & & & &  \\

$K_1$ & $K_2$ & $\frac{M}{N}$ & $\gamma$ & $F_1$ & $F_2$ \\

(Theorem \ref{theorem construction 1}) & \cite{user_grouping_Shanmugam} & & & (Theorem \ref{theorem construction 1}) & \cite{user_grouping_Shanmugam}  \\

\hline

$(8,3,1,2)$ & $(60,4)$ & & & & \\

$255$ & $240$ & $\dfrac{16}{17}$ & $31$ & $200787$ & $10^{17}$ \\

& & & & & \\
 
\hline

$(6,3,2,3)$ & $(24,459)$ & & & & \\

$11011$ & $11016$ & $\dfrac{81}{91}$ & $13$ & $364$ & $2704156$ \\

& & & & & \\

\hline

$(6,3,2,2)$ & (12,54) & & & &  \\

$651$ & $648$ & $\dfrac{16}{21}$ & $7$ & $63$ & $924$ \\

& & & & & \\

\hline





$(7,4,1,2)$ & $(12,10)$ & & & & \\

$127$ & $120$ & $\dfrac{96}{127}$ & $7$ & 2667 & $924$ \\

& & & & & \\

\hline





\end{tabular}

\end{table*}


\vspace{-0.5cm}

\subsection{Asymptotic Analysis and Numerical Comparisons of Scheme A with Ali-Niesen Scheme}
\label{subsection asymptotics & numerical comparisons for scheme 1}

In Appendix \ref{appendix asymptotics of scheme 1}, we provide the asymptotic analysis (as $K$ grows) for the scheme presented in Theorem \ref{theorem construction 1}. We have provided two cases. In case 1 of Appendix \ref{appendix asymptotics of scheme 1}, the cache fraction is upper bounded by a constant $(\frac{M}{N} \leq constant)$ and our scheme has subpacketization, $F = O(poly(K))$. The rate, however, increases linearly with $K$ i.e., $R=\Theta(K)$ (similar to the uncoded caching rate); hence this regime does not have much significance. In case 2 of Appendix \ref{appendix asymptotics of scheme 1}, we keep the rate upper bounded by a constant $(R =O(1))$ then our scheme has subexponential subpacketization, $F=q^{O\left((\log_q{K})^2\right)}$ and cache fraction $\frac{M}{N}=1-\Theta\left(\frac{1}{\sqrt{K}}\right)$. Hence, the drawback of Scheme A is that it requires higher cache fraction even though it has subexponential subpacketization (when rate is upper bounded by a constant). The asymptotics of Scheme A obtained in this section are summarized in the first $2$ rows of Table \ref{table summary of schemes} in Section \ref{section summary of results}.

In Table \ref{table for construction 1}, we compare numerically the scheme in Theorem \ref{theorem construction 1} with the scheme in \cite{user_grouping_Shanmugam} (Section V-A in \cite{user_grouping_Shanmugam}, this is also given in row 2 of Table \ref{table known results}), which is a modified version of the coded caching scheme in \cite{MaN} with user grouping. The scheme in \cite{user_grouping_Shanmugam} is parameterized by the cache fraction $\frac{M}{N}$, global caching gain $\gamma$ and number of user groups $l$, and gives a scheme with the number of users $K_2=K^{'}l$ and subpacketization $F_2=\binom{K^{'}}{\gamma -1}$, where $K^{'} =(\gamma -1) \lceil\frac{N}{M}\rceil$. The number of users and subpacketization corresponding to Theorem \ref{theorem construction 1}, are labelled as $K_1$ and $F_1$ respectively. From the table it is clear that Scheme A performs better than \cite{user_grouping_Shanmugam}, for most of the cases, in terms of the subpacketization.


We also observe from the construction of Scheme A as well as Table \ref{table for construction 1} that the cache fraction of our scheme is at least $0.5$ in general. To overcome the drawback of high cache requirement of Scheme A, we propose a new scheme (Scheme B) in Section \ref{section scheme 2}. Scheme B also uses ideas from the projective geometry.


\section{A New Low Subpacketization Scheme with Small Cache Fraction (Scheme B)}
\label{section scheme 2}

In this section, we present a coded caching scheme (Scheme B), which achieves subexponential (in $K$) subpacketization when memory is upper bounded by a constant. In Section \ref{subsection linear scheme}, we give Scheme C, which is a special case of Scheme B, and achieves linear subpacketization and is comparatively more flexible than the existing linear subpacketization schemes in the literature with non-trivial caching gain. Finally, in Section \ref{subsection scheme 3}, we show a generalized version of Scheme B which adds one more tunable parameter to the construction of Scheme B. 

The construction of Scheme B uses similar techniques (bipartite graph and projective geometry) as that of Scheme A. Before presenting the new construction of bipartite caching graph $B(K,F,D)$ (which gives Scheme B), we first present some simple results using the projective geometry over finite fields which are used in this section.

\subsection{An useful lemma about sets of subspaces}

Let $\mathbb{T} \triangleq PG_q(k-1,0)$. Let $\theta(k)$ denote the number of distinct $1$-dim subspaces of $\mathbb{F}_q^k$. Therefore,
\[\theta (k) = |\mathbb{T}|=\gbinom{k}{1}= \frac{q^k -1}{q-1}.\]

The following lemma will be used repeatedly in this section.
\begin{lemma} \label{lemma no of sets of LI 1D spaces}
Let $k,a,b \in \mathbb{Z}^+$ such that $1\leq a+b\leq k$. Consider a $k$-dim vector space $V$ over $\fq$ and a fixed $a$-dim subspace $A$ of $V$. The number of distinct (un-ordered) $b$-sized sets $\{T_1,T_2,\cdots,T_b\} \subseteq \mathbb{T}$ and $A\oplus T_1 \oplus T_2 \oplus \cdots \oplus T_b \in PG_q(k-1,a+b-1)$ is 
$\frac{1}{b!}\prod\limits_{i=0}^{b-1}(\theta(k)-\theta(a+i))$.
\end{lemma}
\begin{IEEEproof}
First we find the number of $T_1\in \mathbb{T}$ such that $A\oplus T_1$ is an $(a+1)$-dim subspace of $V$. To pick such a $T_1$ we define, $T_1= span(\mathbf{t_1})$ for some $\mathbf{t_1}\in V \setminus A$. Such a $\mathbf{t_1}$ can be picked in $(q^k-q^{a})$ ways. However, for one such fixed $\mathbf{t_1}$,
there exist $(q-1)$ number of $\mathbf{t_1'}\in V\setminus A 
$ such that $span(\mathbf{t_1})=span(\mathbf{t_1'})=T_1$. Thus, the required number of unique $T_1\in \mathbb{T}$ is $\frac{q^k-q^{a}}{q-1}=\theta(k)-\theta(a)$. Similarly, for every such $T_1$ we can select $T_2$ with the condition that $A\oplus T_1 \oplus T_2$ is $(a+2)$-dim subspace of $V$ in $(\theta(k)-\theta(a+1))$ ways. So the number of distinct ordered sets $\{T_1,T_2\}$ is $(\theta(k)-\theta(a))(\theta(k)-\theta(a+1))$. By induction the number of distinct ordered sets $\{T_1,T_2, \cdots, T_b\}$, such that $A \bigoplus\limits_{i=1}^{b}T_i$ is an $(a+b)$-dim subspace of $V$ is $\prod\limits_{i=0}^{b-1}(\theta(k)-\theta(a+i))$.
We know that the number of permutations of a $b$-sized set is $b!$. Therefore, the number of distinct (un-ordered) sets satisfying the required conditions is $\frac{1}{b!}\prod\limits_{i=0}^{b-1}(\theta(k)-\theta(a+i))$. This completes the proof.
\end{IEEEproof}

We also use the following corollary to Lemma \ref{lemma no of sets of LI 1D spaces}. The proof of this follows from Lemma \ref{lemma no of sets of LI 1D spaces} (by taking $k=a, b=1, a=a-1$ in Lemma \ref{lemma no of sets of LI 1D spaces}).

\begin{corollary} \label{no of 1D spaces outside a hper space}
Consider two subspaces $A',A$ of a $k$-dim vector space $V$ over $\fq$ such that $dim(A')=a-1, dim(A)=a$ and $A'\subset A$. The number of distinct $T\in \mathbb{T}$ such that $A'\oplus T=A$ is $\frac{1}{1!}\left(\theta(a)-\theta(a-1)\right)=\frac{q^{a}-q^{a-1}}{q-1}=q^{a-1}$.
\end{corollary}

We now give an example of another coded caching scheme, which we shall see in Section \ref{subsection scheme 2} to be illustrative of Scheme B of this paper. 


\subsection{An Illustrative Example}

\begin{example}
\label{example of construction 2}

Consider a caching system with $K=7, F=21, \frac{M}{N}=0.4285$. Similar to Example \ref{example of construction 1}, we need to provide the indexing for the users $(\mathcal{K})$ and the subfiles $(\mathcal{F})$. For this purpose, we consider some quantities from the projective geometry over finite fields. Consider a $3$-dim vector space $(\mathbb{F}_{2}^{3})$. Consider the $1$-dim subspaces of $\mathbb{F}_{2}^{3}$ which are listed as follows,

\begin{align*}
T_1 &= span\{(0,0,1)\}  \\
T_2 &= span\{(0,1,0)\}  \\
T_3 &= span\{(1,0,0)\}  \\
T_4 &= span\{(1,1,0)\}  \\
T_5 &= span\{(1,0,1)\}  \\
T_6 &= span\{(0,1,1)\}  \\
T_7 &= span\{(1,1,1)\}. 
\end{align*}

Let $\mathbb{X}=\{T_1,T_2,T_3,T_4,T_5,T_6,T_7\}$. We know that any two distinct $1$-dim subspaces are linearly independent. Let $\mathbb{Y}$ be the set of all $2$-sized sets of linearly independent $1$-dim subspaces of $\mathbb{F}_{2}^{3}$ i.e., $\mathbb{Y}=\{\{T_i,T_j\}, i,j\in \{1,2,\cdots,7\}, i\neq j\}$. Therefore $|\mathbb{Y}|=21$.

Let $\mathbb{Z}_1 $ be the set of all $3$-sized sets of linearly dependent $1$-dim subspaces of $\mathbb{F}_{2}^{3}$. It is easy to see that 
\begin{align*}
\mathbb{Z}_1 = \{ &\{T_1,T_2,T_6\},\{T_2,T_3,T_4\},\{T_1,T_3,T_5\},\{T_1,T_4,T_7\}, \\ &\{T_2,T_5,T_7\},\{T_3,T_6,T_7\},\{T_4,T_5,T_6\}\}.    
\end{align*}
Let $\mathbb{Z}$ be the set of all $3$-sized sets of linearly independent $1$-dim subspaces of $\mathbb{F}_{2}^{3}$. Therefore $\mathbb{Z} = \binom{\mathbb{X}}{3} \setminus \mathbb{Z}_1$. Therefore $|\mathbb{Z}|=28$. We now proceed to describe the caching phase and delivery phase.

\textbf{\underline{Caching phase:}} Let the user set be \underline{$\mathcal{K}=\mathbb{X}$} and the set of subfiles be \underline{$\mathcal{F}=\mathbb{Y}$}. During the caching phase, every file $(W_i,i\in [N])$ is divided into $F=|\mathbb{Y}|=21$ subfiles. The subfiles of $W_i$ are denoted as $W_{i,Y}, \forall Y\in \mathcal{F}$. The caching scheme is,

\begin{itemize}
    \item  For each $i \in [N]$, the user $T_l \in \mathcal{K}$ caches the subfile $W_{i,Y}$ if $\{T_l\} \cup Y \notin \mathbb{Z}$.
\end{itemize}

For instance, the user $T_1$ caches the subfiles $W_{i,\{T_1,T_2\}},$ $W_{i,\{T_1,T_3\}},W_{i,\{T_1,T_4\}},W_{i,\{T_1,T_5\}},W_{i,\{T_1,T_6\}},W_{i,\{T_1,T_7\}}$ and $W_{i,\{T_2,T_6\}},$ $W_{i,\{T_3,T_5\}},W_{i,\{T_4,T_7\}}$ for every $i\in [N]$. It is easy to see that, every user caches $9$ subfiles of every file. Hence, the cache fraction $\frac{M}{N}= \frac{9}{21} = 0.4285$.

\textbf{\underline{Delivery phase:}} Let demand of an arbitrary user $T_l \in \mathcal{K}$ be $W_{d_{T_l}}$. Note that the subfiles requested by the user $T_l$ are precisely $\{W_{{d_{T_l}},Z\backslash T_l}: \forall Z\in{\mathbb Z}\}$. The transmission scheme is,
    
\begin{itemize}
    \item For each $Z\in \mathbb{Z}$, the server makes the transmission $\bigoplus\limits_{T_l \in Z} W_{d_{T_l},Z\setminus T_l}$.
\end{itemize}
For instance, the transmission corresponding to $\{T_1,T_2,T_3\} \in \mathbb{Z}$ is $W_{d_{T_1},\{T_2,T_3\}} + W_{d_{T_2},\{T_1,T_3\}} + W_{d_{T_3},\{T_1,T_2\}}$. 
It is clear that, from this transmission, the user $T_1$ decodes $W_{d_{T_1},\{T_2,T_3\}}$, the user $T_2$ decodes $W_{d_{T_2},\{T_1,T_3\}}$, and the user $T_3$ decodes $W_{d_{T_3},\{T_1,T_2\}}$. Since one transmission is made for each $Z\in{\mathbb Z}$, all the user demands will be satisfied. As 3 users are served in any transmission, the global caching gain is $3$. By using (\ref{eqn rate definition}), the rate of the scheme is $R=\frac{|\mathbb{Z}|}{|\mathcal{F}|}=\frac{28}{21}=1.33$.

\end{example}

We are now ready to construct Scheme B, which is the main result of this section. 

\subsection{Construction of Scheme B}
\label{subsection scheme 2}

We now proceed to develop some notations which are used to label our user vertices, subfile vertices and induced matchings of the bipartite caching graph that we construct.

Let $k,m,n,q$ be positive integers such that $n+m\leq k$ and $q$ is some prime power. Consider a $k$-dim vector space $\fq^{k}$ and the following sets of subspaces.
\begin{align*}
    \mathbb{T} &\triangleq  PG_q(k-1,0). \text{  (set of all $1$-dim subspaces)}\\
    \mathbb{R} &\triangleq  PG_q(k-1,n-1). \text{  (set of all $n$-dim subspaces)}\\
    \mathbb{S} & \triangleq  PG_q(k-1,m-1). \text{  (set of all $m$-dim subspaces)}\\
    \mathbb{U} & \triangleq  PG_q(k-1,n+m-1). \text{   \footnotesize{(set of all $(n+m)$-dim subspaces)}}
\end{align*}

Now, consider the following sets, which are used to present our bipartite caching graph.

\begin{align}
    \label{eqn X}
    \mathbb{X} & \triangleq \left\{\{T_1,T_2,\cdots,T_{n}\}: \forall T_i \in \mathbb{T}, \sum\limits_{i=1}^{n} T_i\in \mathbb{R} \right\}. \\
    \label{eqn Y}
    \mathbb{Y} & \triangleq \left\{\{T_1,T_2,\cdots,T_{m}\}: \forall T_i \in \mathbb{T}, \sum\limits_{i=1}^{m}T_i \in \mathbb{S} \right\}. \\
    \label{eqn Z}
    \mathbb{Z} & \triangleq \left\{\{T_1,\cdots,T_{n+m}\}: \forall T_i \in \mathbb{T}, \sum\limits_{i=1}^{n+m}T_i \in \mathbb{U}\right\}.
\end{align}

Thus, $\mathbb{X}$ is the set of all $n$-sized sets of $1$-dim  subspaces such that their sum is an $n$-dim subspace. Intuitively, each $T_i,i\in [n]$ in $\{T_1,T_2,\cdots,T_{n}\}$ has a different extra dimension. So, the dimension of the subspace $\sum\limits_{i=1}^{n}T_i$ is $n$. Similarly, we have $\mathbb{Y}$ and $\mathbb{Z}$.

We now proceed to construct the bipartite caching graph $B(K,F,D)$. Construct a bipartite graph $B$ with left (user) vertex set \underline{$\mathcal{K}=\mathbb{X}$} and right (subfile) vertex set \underline{$\mathcal{F}=\mathbb{Y}$}. Define the edge set of $B$ as,
\[
E(B) \triangleq \{\{X,Y\} : X\in \mathbb{X}, Y\in \mathbb{Y} , X\cup Y \in \mathbb{Z}\}.
\]
We now find the values of $K,F$ and the degree of any user vertex (left degree) $D$.


\begin{lemma}
\label{lemma K,F,D,c expressions of scheme 2}
From the given construction, we have the following.
\begin{align*}
K = |\mathbb{X}| &= \frac{1}{n!} ~ q^{\frac{n(n-1)}{2}} ~ \prod\limits_{i=0}^{n-1}\gbinom{k-i}{1} ,\\
F = |\mathbb{Y}| &= \frac{1}{m!} ~ q^{\frac{m(m-1)}{2}} ~ \prod\limits_{i=0}^{m-1}\gbinom{k-i}{1} , \\
D \triangleq  |\mathcal{N}(X)| &=\dfrac{q^{nm}}{m!} ~ q^{\frac{m(m-1)}{2}} ~ \prod\limits_{i=0}^{m-1}\gbinom{k-n-i}{1}.
\end{align*}
where the last equation holds for any $X\in {\mathbb{X}}$.
\end{lemma}
\vspace{0.1cm}
\begin{IEEEproof}
See Appendix \ref{appendix proof of K,F,D,c in construction 2}.
\end{IEEEproof}



Note that, by Lemma \ref{lemma K,F,D,c expressions of scheme 2}, $B$ is a $D$-left regular bipartite graph with $K$ left vertices and $F$ right vertices. Therefore, by Definition \ref{definition bipartite caching scheme,graph}, $B(K,F,D)$ is a valid bipartite caching graph.

\begin{remark} \label{remark on bi-regularity of graph in scheme B}
It is also easy to show that the degree of any right (subfile) vertex $Y\in \mathbb{Y}$ is $|\mathcal{N}(Y)| = \dfrac{q^{nm}}{n!} ~ q^{\frac{n(n-1)}{2}} ~ \prod\limits_{i=0}^{n-1}\gbinom{k-m-i}{1}$ (the proof is similar to that of $D$ in Appendix \ref{appendix proof of K,F,D,c in construction 2} with an interchange of $n$ and $m$). Therefore, $B$ is a bi-regular bipartite caching graph.
\end{remark}

We now show that $B$ has an induced matching cover
$\{{\C C}_i : i\in [S]\}$ such that $|{\C C}_i|=g, \forall i \in [S]$, for some $g,S \in \mathbb{Z}^+$.

\textbf{\underline{Induced matching cover:}} We first describe an induced matching of $B$ and show that such equal-sized induced matchings partition $E(B)$. This will suffice to show the delivery scheme as per Theorem \ref{theorem bipartite coded caching}.

We now present an induced matching of size $\binom{n+m}{n}$ in $B$ (where $\binom{a}{b}$ represents binomial coefficient). Recall the definition of $\mathbb Z$ from (\ref{eqn Z}). 
\begin{lemma} 
\label{lemma transmission clique in scheme 2}
Consider $Z=\{T_1,T_2,\cdots,T_{n+m}\}\in \mathbb{Z}$. Then
$\mathcal{C}_Z \triangleq \left\{\{X,Z \setminus X\}: X \in \binom{Z}{n}\right\} \subseteq E(B)$ is an induced matching of size $\binom{n+m}{m}$ in $B$.
\end{lemma}


\begin{IEEEproof}
First note that $\mathcal{C}_Z$ is well-defined as $Z \in \mathbb{Z}$. Consider an arbitrary $X \subset Z$ such that $X\in \mathbb{X}$. It is clear that $\{X,Z\setminus X\}\in E(B)$. Consider two distinct edges $\{X_1,Z\setminus X_1\}, \{X_2,Z\setminus X_2\}\in \mathcal{C}_Z$ (where $X_1 \neq X_2$). It is clear that $Z\setminus X_1 \neq Z\setminus X_2$. WLOG, let $T_a \notin X_1$ and $T_a \in X_2$ (as $X_1\neq X_2$, such a $T_a$ exists). Then, $T_a \notin X_1 \cup (Z\setminus X_2)$ and hence $X_1 \cup (Z\setminus X_2) \subsetneq Z$. By the definition of $\mathbb{Z}$ in $(\ref{eqn Z})$, it is clear that $X_1 \cup (Z\setminus X_2) \notin \mathbb{Z}$. Therefore, we have $\{X_1,Z\setminus X_2\}\notin E(B)$, by the definition of $E(B)$. Similarly $\{X_2,Z\setminus X_1\}\notin E(B)$.
By invoking Definition \ref{defintion induced matching&cover}, it is clear that $\mathcal{C}_Z$ is an induced matching of $B$. It is easy to see that $|\mathcal{C}_Z|=\binom{n+m}{n}$. This completes the proof.
\end{IEEEproof}
We now show that the set of induced matchings $\{\mathcal{C}_Z : Z\in \mathbb{Z}\}$ partition $E(B)$.

\begin{lemma} \label{lemma transmission clique cover in scheme 2}
The induced matchings $\{\mathcal{C}_Z: Z\in \mathbb{Z}\}$ as defined in Lemma \ref{lemma transmission clique in scheme 2} partition the edge set ${E(B)}$.
\end{lemma}
\begin{IEEEproof}
Consider $Z, Z'\in \mathbb{Z}$ such that $Z\neq Z'$. By definition of $\mathcal{C}_Z,\mathcal{C}_{Z'}$, we have $\mathcal{C}_Z \cap \mathcal{C}_{Z'}=\phi$. Now consider an arbitrary edge $\big\{\{T_1,T_2,\cdots,T_n\},$ $\{T_{n+1},T_{n+2}\cdots,T_{n+m}\}\big\} \in E(B)$. By the construction of $B$,  $\{T_1,T_2,\cdots,T_n\}\cup \{T_{n+1},T_{n+2}\cdots,T_{n+m}\}\in \mathbb{Z}$. Therefore, the edge $\big\{\{T_1,T_2,\cdots,T_n\},\{T_{n+1},T_{n+2}\cdots,T_{n+m}\}\big\}$ lies in the unique induced matching, $ \mathcal{C}_{\{T_1,T_2,\cdots,T_{n+m}\}} $ (defined as in Lemma \ref{lemma transmission clique in scheme 2}). This completes the proof.
\end{IEEEproof}

Therefore, $\{\mathcal{C}_Z : Z\in \mathbb{Z}\}$ is an induced matching cover of $B$. We are now ready to present our coded caching scheme using the bipartite caching graph constructed above.

\begin{tcolorbox}
\begin{theorem}
\label{theorem construction 2}
\textbf{(Scheme B)} Let $k,n,m,q$ be positive integers such that $n+m\leq k$ and $q$ be some prime power. The bipartite graph $B$ constructed in Section \ref{subsection scheme 2} is a $B(K,F,D)$ bipartite caching graph with an induced matching cover having $S=\frac{1}{(n+m)!} ~ q^{\frac{(n+m)(n+m-1)}{2}} ~ \prod\limits_{i=0}^{n+m-1}\gbinom{k-i}{1}$ induced matchings, each having $g=\binom{n+m}{n}$ edges, and defines a coded caching scheme with,

\[K=\frac{1}{n!} ~ q^{\frac{n(n-1)}{2}} ~ \prod\limits_{i=0}^{n-1}\gbinom{k-i}{1},\] 
\[F=\frac{1}{m!} ~ q^{\frac{m(m-1)}{2}} ~ \prod\limits_{i=0}^{m-1}\gbinom{k-i}{1},\] 

\[\frac{M}{N}=1-q^{n m} ~ \prod\limits_{i=0}^{m-1}\dfrac{\gbinom{k-n-i}{1}}{\gbinom{k-i}{1}},\] 
 
\[R=\dfrac{m!~ q^{nm}}{(n+m)
!} ~ q^{\frac{n(n-1)}{2}} ~ \prod\limits_{i=0}^{n-1}\gbinom{k-m-i}{1},\]

\[\text{Global caching gain }\gamma = \binom{n+m}{n}.\]
\end{theorem}
\end{tcolorbox}


\begin{IEEEproof}
From Lemma \ref{lemma K,F,D,c expressions of scheme 2}, we get the expressions of $K, F$ and $D$. By Lemma
\ref{lemma transmission clique in scheme 2} and Lemma \ref{lemma transmission clique cover in scheme 2}, the size of the 
induced matchings of  $B$ is $g= |\mathcal{C}_Z|= \binom{n+m}{m}$ for any $Z\in \mathbb{Z}$ and they partition the edge set $E(B)$. Further, by Lemma \ref{lemma transmission clique in scheme 2}, the number of induced matchings in the induced matching cover is $S=|\mathbb{Z}|$. Finding $|\mathbb{Z}|$ is same as that of finding $K$ (replace $n$ with $n+m$ in the computation of $K$ in Appendix \ref{appendix proof of K,F,D,c in construction 2}). Therefore,
\begin{align}
    \label{eqn S in scheme 2}
    S &= \frac{1}{(n+m)!}\prod\limits_{i=0}^{n+m-1}(\theta(k)-\theta(i))\\
    \nonumber
    &= \frac{1}{(n+m)!} ~ q^{\frac{(n+m)(n+m-1)}{2}} ~ \prod\limits_{i=0}^{n+m-1}\gbinom{k-i}{1}.
\end{align}
Hence the bipartite graph $B$ satisfies all the conditions in Theorem \ref{theorem bipartite coded caching}. Therefore, there exists a coded caching scheme with $K$ users, subpacketization $F$,
\begin{align}
    \label{eqn 1-M/N in scheme 2}
    1-\frac{M}{N}&=\frac{D}{F}=\dfrac{\frac{1}{m!}\prod\limits_{i=0}^{m-1}(\theta(k)-\theta(n+i))}{\frac{1}{m!}\prod\limits_{i=0}^{m-1}(\theta(k)-\theta(i))}\\
    \nonumber
    &= \prod\limits_{i=0}^{m-1}\frac{q^k-q^{n+i}}{q^k-q^i}=\prod\limits_{i=0}^{m-1}q^n ~ \frac{q^{k-n-i}-1}{q^{k-i}-1},
\end{align}
(where $D,F$ expressions are from Appendix \ref{appendix proof of K,F,D,c in construction 2}).

Therefore,
\[\frac{M}{N}=1-q^{n m} ~ \prod\limits_{i=0}^{m-1}\dfrac{\gbinom{k-n-i}{1}}{\gbinom{k-i}{1}}.\]
\begin{align*}
R=\frac{S}{F} &= \dfrac{\frac{1}{(n+m)!}\prod\limits_{i=0}^{n+m-1}(\theta(k)-\theta(i))}{\frac{1}{m!}\prod\limits_{i=0}^{m-1}(\theta(k)-\theta(i))} \\ &= \frac{m!}{(n+m)!} \prod\limits_{i=m}^{n+m-1}(\theta(k)-\theta(i)) \\
&= \frac{m!}{(n+m)!} \prod\limits_{i=0}^{n-1}(\theta(k)-\theta(m+i)) \\ &= \frac{m! ~ q^{nm}}{(n+m)!} q^{\frac{n(n-1)}{2}} ~ \prod\limits_{i=0}^{n-1}\gbinom{k-m-i}{1}.
\end{align*}
(where $S$ expression is from (\ref{eqn S in scheme 2}) and $F$ expression is from Appendix \ref{appendix proof of K,F,D,c in construction 2}).
Finally, we have that the global caching gain  $\gamma = g,$ which completes the proof.
\end{IEEEproof}


We regard the coded caching scheme (Scheme B) presented in Theorem \ref{theorem construction 2} as the main scheme of this work. We present Algorithm \ref{algorithm} in which the caching and delivery scheme of Scheme B are captured. The coded caching scheme proposed in Theorem \ref{theorem construction 2}, does not exist for all $K$ (similar to most of the low subpacketization coded caching schemes in the literature). So based on the design parameters (desired number of users, cache size) we add some dummy users and treat some fraction of the available cache as unused cache. This is done as follows. For the given number of users $(K')$, cache size $(M')$ and number of files $(N)$, select the appropriate  parameters $k,n,m,q$ which give a coded caching scheme according to Theorem \ref{theorem construction 2} with parameters $K,\frac{M}{N},F,R$ such that $(K- K')$ and $(\frac{M'}{N}-\frac{M}{N})$ are non-negative and as small as possible (we treat the extra users $K-K'$ as dummy users and the extra cache $M'-M$ is left unused). Now construct a bipartite caching graph $B(K,F,D)$ as mentioned in Section \ref{subsection scheme 2} and find $\mathbb{X}$ (user indices), $\mathbb{Y}$ (subfile indices), $\mathbb{Z}$ (indices of induced matchings of $B$, equivalently indices of transmissions$)$ by using (\ref{eqn X}),(\ref{eqn Y}),(\ref{eqn Z}).

\begin{algorithm}
\caption{Coded caching scheme proposed in Theorem \ref{theorem construction 2}}
\label{algorithm}
\begin{algorithmic}[1]
\Procedure{Placement Phase}{}
    \For{each $i\in [N]$}
    \State Split $W_i$ into $\{W_{i,Y}:Y\in\mathbb{Y}\}$.
    \EndFor
    \For {each $X \in \mathbb{X}$}
    \State user $X$ caches the subfiles $W_{i,Y}, \forall i\in [N], \forall Y\in \mathbb{Y}$ such that $X \cup Y \notin \mathbb{Z}$.
    \EndFor
\EndProcedure
\Procedure{Delivery Phase}{ demand of user $X$ is represented as $W_{d_{X}}, \forall X\in \mathbb{X}$}
    \For {each $ Z=\{V_1,V_2,\cdots,V_{n+m}\} \in \mathbb{Z}$ }
    \State Server transmits $\bigoplus\limits_{X \in \binom{Z}{n}}  W_{d_{X},Z\backslash X}$.
    \EndFor
\EndProcedure
\end{algorithmic}
\end{algorithm}



\subsection{Asymptotic Analysis and Numerical Comparisons of Scheme B with the State of the Art}
\label{subsection asymptotics & numberical comparisons for scheme 2}

In Appendix \ref{appendix asymptotics of scheme 2}, we provide the asymptotic analysis for the scheme presented in Theorem \ref{theorem construction 2}. When $\frac{M}{N}$ is upper bounded by a constant $(\frac{M}{N}\leq constant)$ and as $K$ grows large, from Appendix \ref{appendix asymptotics of scheme 2} we see that our scheme has subexponential subpacketization i.e., $F=q^{O\left((\log_q{K})^2\right)}$ and rate $R = \Theta\left(\frac{K}{(\log_{q}{K})^{n}}\right)$. Hence, Scheme B overcomes the drawback of Scheme A (high cache requirement), but with higher rate. The asymptotics of Scheme B obtained in this section are summarized in the third row of Table \ref{table summary of schemes} in Section \ref{section summary of results}.

We now compare our scheme with some schemes from Table \ref{table known results}. We first discuss asymptotic comparison in subpacketization as the number of users $K$ increases. In Table \ref{table known results}, we see that the subpacketization levels of several schemes are exponential in $K^{\frac{1}{r}}$ for some positive integer $r$. Comparatively, our scheme achieves subpacketization exponential in $O((\log_{q}K)^2)$, which is an improvement. Matching our scheme's parameters with those from \cite{strongedgecoloringofbipartitegraphs} is hard. A special case of the general scheme in \cite{strongedgecoloringofbipartitegraphs} is discussed in that work, which achieves subpacketization exponential in $\sqrt{K}$. Our Scheme B thus improves over this special case. The scheme of \cite{STD} achieves linear subpacketization, but it works only for an extremely large number of users and hence we do not compare with this numerically. The PDA scheme $P_1$ from \cite{cheng2017coded} as given in Table \ref{table known results} achieves a subpacketization that is smaller than $K$, but uses a large cache fraction in general ($\geq 0.5$). Hence, we do not compare with this scheme also.



\begin{table*}[htbp]

\caption{Comparison of the coded caching scheme in Theorem \ref{theorem construction 2} (Scheme B) with the scheme in \cite{user_grouping_Shanmugam} (Ali-Niesen scheme with grouping). We match the cache fraction, global caching gain, and number of users, as closely as possible, and compare the subpacketization level.}
\label{table for construction 2 with Ali-Niesen scheme with grouping}

\setlength{\tabcolsep}{6pt} 
\renewcommand{\arraystretch}{1.1}
\centering

\begin{tabular}{|c|c||c||c||g|c|}
\hline

\multicolumn{2}{|c||}{\textbf{Number}} & \textbf{Cache} & \textbf{Global} & \multicolumn{2}{|c|}{\textbf{Subpacketization}} \\

\multicolumn{2}{|c||}{\textbf{of users}} & \textbf{fraction} & \textbf{caching gain} & \multicolumn{2}{|c|}{} \\

\hline

$(k,n,m,q)$ & $(K^{'},l)$ & & & &  \\

$K_1$ & $K_2$ & $\frac{M}{N}$ & $\gamma$ & $F_1$ & $F_2$ \\

(Theorem \ref{theorem construction 2}) & \cite{user_grouping_Shanmugam} & & & (Theorem \ref{theorem construction 2}) & \cite{user_grouping_Shanmugam}  \\

\hline

$(7,2,4,2)$ & $(56,143)$ & & & & \\

$8001$ & $8008$ & $\dfrac{125}{381}$ & $15$ & $10^{7}$ & $10^{12}$ \\

& & & & & \\
 
\hline

$(7,2,2,2)$ & $(75,107)$ & & & & \\

$8001$ & $8025$ & $\dfrac{187}{2667}$ & $6$ & $8001$ & $10^{7}$ \\

& & & & & \\

\hline

$(5,2,2,2)$ & (20,23) & & & &  \\

$465$ & $460$ & $\dfrac{43}{155}$ & $6$ & $465$ & $15504$ \\

& & & & & \\

\hline

$(4,2,1,2)$ & $(10,10)$ & & & & \\

$105$ & $100$ & $\dfrac{1}{5}$ & $3$ & $15$ & $45$ \\

& & & & & \\

\hline

$(4,1,2,3)$ & $(20,2)$ & & & & \\

$40$ & $40$ & $\dfrac{1}{10}$ & $3$ & $780$ & $190$ \\

& & & & & \\

\hline

\end{tabular}

\end{table*}




\begin{table*}[htbp]

\caption{Comparison of the coded caching scheme in Theorem \ref{theorem construction 2} (Scheme B) with the scheme in \cite{YCTCPDA}. (inf represents $> 10^{307}$).
(We try to match the $K$ and  $\frac{M}{N}$ values of Theorem \ref{theorem construction 2} with that of \cite{YCTCPDA} as closely as possible by choosing appropriate parameters, and compare the subpacketization and gain).}

\label{table for construction 2 with PDA basic paper}

\setlength{\tabcolsep}{6.5pt} 
\renewcommand{\arraystretch}{1.2}
\centering

\begin{tabular}{|g|c||g|c||g|c||g|c|}

\hline

\multicolumn{2}{|c||}{\textbf{Number of users}} & \multicolumn{2}{|c||}{\textbf{Cache fraction}} & \multicolumn{2}{|c||}{\textbf{Subpacketization}} & \multicolumn{2}{|c|}{\textbf{Global caching gain}}\\
\hline

$(k,n,m,q)$ & $(q,m)$ & & & & & & \\

$K_1$ & $K_2$ & $\left(\frac{M}{N}\right)_1$ & $\left(\frac{M}{N}\right)_2$ &$F_1$ & $F_2$ & $\gamma_1$ & $\gamma_2$\\

(Theorem \ref{theorem construction 2})&\cite{YCTCPDA} & (Theorem \ref{theorem construction 2}) & \cite{YCTCPDA} & (Theorem \ref{theorem construction 2}) & \cite{YCTCPDA} & (Theorem \ref{theorem construction 2}) &\cite{YCTCPDA}\\

\hline

$(4,2,2,2)$ & $(2,51)$ & & & & & & \\

105 & 104 & 0.54 & 0.50 & 105 & $10^{15}$ & 6 & 52\\

\hline

$(5,2,2,2)$ & $(4,116)$ & & & & & & \\

465 & 468 & 0.28 & 0.25 & 465 & $10^{69}$ & 6 & 117\\

\hline

$(4,2,2,3)$ & $(3,259)$ & & & & & & \\

780 & 780 & 0.38 &  0.33 & 780 & $10^{123}$ & 6 & 260\\

\hline

$(7,2,4,2)$ & $(3,2666)$ & & & & & & \\

8001 & 8001 & 0.33 & 0.33 & $10^7$ & inf & 15 & $2667$\\

\hline

$(7,2,2,2)$ & $(14,571)$ & & & & & & \\

8001 & 8008 & 0.07 & 0.07 & 8001 & inf & 6 & 572\\

\hline

\end{tabular}

\end{table*}




\begin{table*}[htbp]

\caption{Comparison of the coded caching scheme in Theorem \ref{theorem construction 2} (Scheme B) with the scheme in \cite{TaR}.
(We try to match the $K$ and $\frac{M}{N}$ values of Theorem \ref{theorem construction 2} with that of \cite{TaR} as closely as possible by choosing appropriate parameters,  and compare the subpacketization and gain).}
\label{table for construction 2 with ramamoorthy}

\setlength{\tabcolsep}{6.5pt} 
\renewcommand{\arraystretch}{1.2}
\centering

\begin{tabular}{|g|c||g|c||g|c||g|c|}

\hline

\multicolumn{2}{|c||}{\textbf{Number of users}} & \multicolumn{2}{|c||}{\textbf{Cache fraction}} & \multicolumn{2}{|c||}{\textbf{Subpacketization}} & \multicolumn{2}{|c|}{\textbf{Global caching gain}}\\
\hline

$(k,n,m,q)$ & $(k,n,z,q)$ & & & & & & \\

$K_1$ & $K_3$ & $\left(\frac{M}{N}\right)_1$ & $\left(\frac{M}{N}\right)_3$ &$F_1$ & $F_3$ & $\gamma_1$ & $\gamma_3$\\

(Theorem \ref{theorem construction 2})&\cite{TaR} & (Theorem \ref{theorem construction 2}) & \cite{TaR} & (Theorem \ref{theorem construction 2}) & \cite{TaR} & (Theorem \ref{theorem construction 2}) &\cite{TaR}\\

\hline

$(3,1,2,5)$ & $(7,12,2,3)$ & & & & & & \\

31 & 36 & 0.2 & 0.3 & 465 & $4374$ & 3 & 8\\

\hline

$(4,1,2,3)$ & $(8,12,3,5)$ & & & & & & \\

40 & 60 & 0.1 &  0.2 & 780 & $10^{6}$ & 3 & 9\\

\hline

$(4,2,1,2)$ & $(9,12,3,11)$ & & & & & & \\

105 & 132 & 0.2 & 0.1 & 15 & $10^9$ & 3 & 10\\

\hline

$(5,2,1,2)$ & $(8,12,3,29)$ & & & & & & \\

465 & 348 & 0.09 & 0.03 & 31 & $10^{12}$ & 3 & 9\\

\hline

\end{tabular}

\end{table*}



We now come to the numerical comparisons. In Table \ref{table for construction 2 with Ali-Niesen scheme with grouping}, we compare numerically the scheme in Theorem \ref{theorem construction 2} with a modified version of Ali-Niesen scheme with user grouping from \cite{user_grouping_Shanmugam} (Section V-A in \cite{user_grouping_Shanmugam}, which is also given in row 2 of Table \ref{table known results}). The scheme in \cite{user_grouping_Shanmugam} is parameterized by the cache fraction $\frac{M}{N}$, global caching gain $\gamma$ and number of user groups $l$, and gives a scheme with the number of users $K_2=K^{'}l$ and subpacketization $F_2=\binom{K^{'}}{\gamma -1}$, where $K^{'} =(\gamma -1) \lceil\frac{N}{M}\rceil$. The number of users and subpacketization corresponding to Theorem \ref{theorem construction 2}, are labelled as $K_1$ and $F_1$ respectively. From the table it is clear that Scheme B performs better than \cite{user_grouping_Shanmugam}, for most of the cases in terms of the subpacketization.

In Table \ref{table for construction 2 with PDA basic paper} and Table \ref{table for construction 2 with ramamoorthy}, we compare numerically the scheme in Theorem \ref{theorem construction 2} with the schemes in \cite{YCTCPDA} and \cite{TaR} respectively, for some choices of $K$ and $\frac{M}{N}$. We chose these two schemes for the reason that these two schemes show a large improvement in the subpacketization level without compromising much on the coding gain compared to the basic scheme of \cite{MaN}. For instance, the PDA based scheme in \cite{YCTCPDA} achieves lower subpacketization (though not in the asymptotic sense) over \cite{MaN}, while having a global caching gain only one less than the scheme of \cite{MaN}. 

We label the parameters of our scheme in Theorem \ref{theorem construction 2} as $K_1,\left(\frac{M}{N}\right)_1,F_1,\gamma_1$. The parameters of the scheme presented in \cite{YCTCPDA} are $K_2=q(m+1),\left(\frac{M}{N}\right)_2=\frac{1}{q},F_2=q^{m},\gamma_2=\frac{K(1-\frac{M}{N})}{q-1}=m+1$ where $q(\geq 2),m\in \mathbb{Z}^+$. 
The parameters of the scheme presented in \cite{TaR} are labeled as $K_3=nq,\left(\frac{M}{N}\right)_3 = \frac{1}{q}, F_3 = q^{k}z, \gamma_3 = k+1$ where $n,k,q,z$ parameters are defined as per \cite{TaR}. As it is difficult to match exact parameters, we try to match the $K$ and $\frac{M}{N}$ values as closely as possible between our scheme and these two schemes, while comparing the rate and subpacketization. Also, as the subpacketization can take large values, we approximate it to the nearest positive power of $10$. 

Throughout, we notice that our Scheme B has parameters for small cache sizes, offering large reductions in subpacketization compared to the other two existing schemes in the literature, while having smaller caching gains (and equivalently, higher rate). Since many of the subpacketization values are of the order of $10^2-10^4$, we consider our Scheme B to be of importance in practice.

We finally remark that we can also achieve subpacketization $F=K$ (we discuss this in the next subsection) and even $F<K$ by choosing the parameters appropriately. For instance, for $k=4,n=2,m=1,q=4$ (in Theorem \ref{theorem construction 2}), we have $K=3570$, $F=85$, with $\frac{M}{N}=0.0588$, and gain $3$.



\subsection{Scheme C: A Flexible Scheme with Linear Subpacketization (a special case of Scheme B)} \label{subsection linear scheme}
One of the most interesting regimes for the coded caching problem is that of linear subpacketization, i.e., the case when $F=O(K)$. It is known from \cite{SZG} via the theory of hypergraphs that linear subpacketization is not sufficient for achieving constant rate.  A well known result from \cite{STD} shows that there exist coded caching schemes which have $F=K$ and achieve a rate of $K^{\delta}$ (for small $\delta$), and for a small cache fraction. However, the number of users required for this construction is extremely high. The Ali-Niesen scheme \cite{MaN} itself achieves $F=K$ when $\frac{M}{N}=\frac{1}{K},$ but the rate is then $R=\frac{K-1}{2}$. A similar scheme with the same parameters is known from \cite{cheng2017coded} (see Section V-B in \cite{cheng2017coded}). Thus, most of the existing schemes with linear subpacketization either require an extremely large number of users to exist, or have a very small caching gain. 

The following corollary to Theorem \ref{theorem construction 2} gives a new linear subpacketization scheme (Scheme C) obtained using our projective geometry based technique. Note that the scheme is parametrized by the prime power $q$  (finite-field size) and the cache fraction $\lambda$, and hence is flexible for different numbers of users and cache fraction.

\begin{tcolorbox}
\begin{corollary}
\label{corollary linear scheme}
\textbf{(Scheme C)} For $q$ being some prime power and $\lambda\in (0,1)$ such that $\lambda q$ is a positive integer, then there exists a linear subpacketization coded caching scheme with $F=K\leq \dfrac{q^{2\lambda^2 q^2}}{(\lambda q)!}$, cache fraction $\frac{M}{N}\leq \lambda,$ and global caching gain $\gamma \geq \dfrac{4^{\lambda q}}{2\sqrt{\lambda q}}$ $\bigg($with the rate achieved being $\frac{K(1-\frac{M}{N})}{\gamma}\bigg)$.
\end{corollary}
\end{tcolorbox}


\begin{IEEEproof}
We choose some specific values for the parameters for our construction in Section \ref{subsection scheme 2} to prove this result. From (\ref{eqneqn1}) in Appendix \ref{appendix asymptotics of scheme 2}, we have that $\frac{M}{N}\leq \frac{n}{q^{\alpha-n+1}} \leq \frac{n}{q}.$ We choose  $n=\lambda q$ (note that $n$ must be an integer and $q$ is a prime power and hence we have our constraints on $\lambda$).

From the expressions in Theorem \ref{theorem construction 2}, we have that $K=F$ when $m=n$. We choose the least valid value of $k$, i.e., $k=n+m=2n$. We thus have with these parameters, 
\begin{align*}
   K=F\stackrel{(\ref{eqn K inequality})}{\leq}  \frac{q^{kn}}{n!}=\frac{q^{2n^2}}{n!}= \frac{q^{2\lambda^2q^2}}{(\lambda q)!},
\end{align*}
as stated in the statement of the corollary. We now come to the global caching gain. From Theorem \ref{theorem construction 2}, we have that the global caching gain of the scheme is 
\begin{align*}
    \gamma = \binom{2n}{n} \geq \frac{4^{n}}{\sqrt{4n}} = \frac{4^{\lambda q}}{2\sqrt{\lambda q}},
\end{align*}
where the above inequality is a well-known inequality for the middle binomial coefficient that holds for $n\geq 1.$ This completes the proof.
\end{IEEEproof}


\subsection{Generalization of Scheme B}
\label{subsection scheme 3}

We can further generalize the scheme presented in Section \ref{subsection scheme 2} (Scheme B). In Scheme B, the user vertices, the subfile vertices and the induced matchings are the sets of linearly independent $1$-dim subspaces, whereas in the generalized scheme which we present now, these are the sets of linearly independent $l$-dim subspaces.

Let $k,m,n,l,q \in \mathbb{Z}^+$ such that $nl+ml\leq k$ and $q$ is some prime power. Consider a $k$-dim vector space $\fq^{k}$ and the following sets of subspaces.
\begin{align*}
    \mathbb{L} &\triangleq  PG_q(k-1,l-1). \text{  (set of all $l$-dim subspaces)}\\
    \mathbb{R} &\triangleq  PG_q(k-1,nl-1). \text{  (set of all $nl$-dim subspaces)}\\
    \mathbb{S} & \triangleq  PG_q(k-1,ml-1). \text{  (set of all $ml$-dim subspaces)}\\
    \mathbb{U} & \triangleq  PG_q(k-1,(nl+ml)-1). \text{  \footnotesize{(set of all $(nl+ml)$-dim spaces)}}
\end{align*}

Similar to Scheme B, consider the following sets, which are used to present the new bipartite caching graph.

\begin{align}
    \label{eqn Xl}
    \mathbb{X} & \triangleq \left\{\{L_1,L_2,\cdots,L_{n}\}: \forall L_i \in \mathbb{L}, \sum\limits_{i=1}^{n} L_i\in \mathbb{R} \right\}. \\
    \label{eqn Yl}
    \mathbb{Y} & \triangleq \left\{\{L_1,L_2,\cdots,L_{m}\}: \forall L_i \in \mathbb{L}, \sum\limits_{i=1}^{m}L_i \in \mathbb{S} \right\}. \\
    \label{eqn Zl}
    \mathbb{Z} & \triangleq \left\{\{L_1,\cdots,L_{n+m}\}: \forall L_i \in \mathbb{L}, \sum\limits_{i=1}^{n+m}L_i \in \mathbb{U}\right\}.
\end{align}

i.e., $\mathbb{X}$ is the set of all $n$-sized sets of linearly independent $l$-dim  subspaces, hence their sum (or direct sum) is an $nl$-dim subspace. Similarly, we have $\mathbb{Y}$ and $\mathbb{Z}$.

Construct a bipartite caching graph $B$, similar to that of Scheme B, with the left (user) vertex set \underline{$\mathcal{K}=\mathbb{X}$} and right (subfile) vertex set \underline{$\mathcal{F}=\mathbb{Y}$}. Define the edge set of $B$ as,
\[
E(B) \triangleq \{\{X,Y\} : X\in \mathbb{X}, Y\in \mathbb{Y} , X\cup Y \in \mathbb{Z}\}.
\]

The parameters of $B$ such as the number of user vertices $(K)$, the number of subfile vertices $(F)$ and the degree of any user vertex (left degree) $D$ are given in the following lemma.

\begin{lemma}
\label{lemma K,F,D,c expressions of scheme 3}
From the given construction, we have the following.
\begin{align*}
K = |\mathbb{X}|&= \dfrac{(l!)^n~ q^{\frac{n(n-1)l^2}{2}}}{(nl)!~ (n!)}~ \dfrac{\prod\limits_{i=0}^{nl-1}\gbinom{k-i}{1}}{\left(\prod\limits_{i=0}^{l-1}\gbinom{l-i}{1}\right)^n}~ y_1 , \\
\end{align*}
\begin{align*}
F =|\mathbb{Y}| &= \dfrac{(l!)^m~ q^{\frac{m(m-1)l^2}{2}}}{(ml)!~ (m!)}~ \dfrac{\prod\limits_{i=0}^{ml-1}\gbinom{k-i}{1}}{\left(\prod\limits_{i=0}^{l-1}\gbinom{l-i}{1}\right)^m}~ y_2 , \\
D = |\mathcal{N}(X)| &=\dfrac{(l!)^m~ q^{ml^2\left(\frac{m-1}{2}+n\right)}}{(ml)!~ (m!)}~ \dfrac{\prod\limits_{i=0}^{ml-1}\gbinom{k-nl-i}{1}}{\left(\prod\limits_{i=0}^{l-1}\gbinom{l-i}{1}\right)^m}~ y_2 .
\end{align*}
where the last equation hold for any $X\in {\mathbb{X}}$, and $y_1=\prod\limits_{i=0}^{n-1}\binom{(n-i)l}{l}$, ~ $y_2=\prod\limits_{i=0}^{m-1}\binom{(m-i)l}{l}$.
\end{lemma}
\begin{IEEEproof}
See Appendix \ref{appendix proof of K,F,D,c in construction 3}.
\end{IEEEproof}

Note that, by Lemma \ref{lemma K,F,D,c expressions of scheme 3}, $B$ is a $D$-left regular bipartite graph with $K$ left vertices and $F$ right vertices. Therefore, by Definition \ref{definition bipartite caching scheme,graph}, $B(K,F,D)$ is a valid bipartite caching graph. We remark that, similar to scheme B, the graph $B(K,F,D)$ is a bi-regular bipartite graph. We now construct the induced matching cover similar to that of Scheme B. 

\textbf{Induced matching cover:}
$\mathcal{C}_Z \triangleq \Big\{\{X,Z \setminus X\}: X \in \binom{Z}{n}\Big\},$
where $Z \in \mathbb{Z}$ is a valid induced matching and $\{\mathcal{C}_Z :Z \in \mathbb{Z} \}$ is an induced matching cover of $B$. The proof of these statements is similar to Lemma \ref{lemma transmission clique in scheme 2} and Lemma \ref{lemma transmission clique cover in scheme 2}.
We now present the coded caching scheme corresponding to the bipartite caching graph constructed above.

\begin{tcolorbox}
\begin{theorem}
\label{theorem construction 3}
Let $k,m,n,l,q \in \mathbb{Z}^+$ such that $nl+ml\leq k$ and $q$ is some prime power. The bipartite caching graph $B$ constructed in Section \ref{subsection scheme 3} is a $B(K,F,D)$-bipartite caching graph with an induced matching cover having 
\[S= \dfrac{(l!)^{(n+m)}~ q^{\frac{(n+m)(n+m-1)l^2}{2}}}{((n+m)l)!~ ((n+m)!)}~ \dfrac{\prod\limits_{i=0}^{(n+m)l-1}\gbinom{k-i}{1}}{\left(\prod\limits_{i=0}^{l-1}\gbinom{l-i}{1}\right)^n} \times \]
\[~~~~\times \prod\limits_{i=0}^{n+m-1}\binom{(n+m-i)l}{l} \]
induced matchings, each having $g=\binom{n+m}{n}$ edges and defines a coded caching scheme with, number of users $K$ and subpacketization $F$ (where $K,F,D$ are as given in Lemma \ref{lemma K,F,D,c expressions of scheme 3}) and 
\begin{align*}
   \frac{M}{N} ~ & = 1-q^{m n l^2} ~ \prod\limits_{i=0}^{ml-1}\dfrac{\gbinom{k-nl-i}{1}}{\gbinom{k-i}{1}},
\end{align*}
\[ R  = \frac{S}{F}, ~~~~~~
    \text{Global caching gain } \gamma = g. \]

\end{theorem}
\end{tcolorbox}
\begin{IEEEproof}
Similar to the proof of Theorem \ref{theorem construction 2}.
\end{IEEEproof}

\begin{remark}
For the special case of $l=1$, the scheme in Theorem \ref{theorem construction 3}, reduces to Scheme B (Theorem \ref{theorem construction 2}). The asymptotic analysis for the scheme in Theorem \ref{theorem construction 3} is similar to that of Scheme B (Appendix \ref{appendix asymptotics of scheme 2}). When $\frac{M}{N}$ is upper bounded by a constant $\left(\frac{M}{N}\leq \dfrac{nl}{q^{k-ml-nl+1}}\right)$ and as $K$ grows large, we can show (by following the similar techniques as that of Appendix \ref{appendix asymptotics of scheme 2}) that the scheme in Theorem \ref{theorem construction 3} achieves  subexponential subpacketization i.e., $F=q^{O\left((\log_q{K})^2\right)}$ and rate $R = \Theta \left(\frac{K}{(\log_{q}{K})^{n}}\right)$. Thus, this generalized scheme has more flexibility than Scheme B, but unfortunately does not improve upon Scheme B asymptotically. 
\end{remark}



\section{Extension to Other Settings}
\label{section extensions}
In this section, we extend our main scheme, Scheme B, to the distributed computing setting \cite{distributed_computing_maddah_ali}, and the wireless interference channel setting \cite{interferencemanagement}. Our main motivation for this section is to present a low subpacketization scheme for each of these two settings which improves upon the currently existing schemes. Most of the existing schemes for these settings are adapted from existing broadcast coded caching schemes, and hence inherit the large subpacketization issue from the same. 

The connection from broadcast coded caching schemes to distributed computing schemes was established in \cite{PDAapplications}, through the concept of placement delivery array (PDA). In \cite{PDAapplications}, a method was shown to derive distributed computing schemes from a special class of PDAs known as $g$-PDAs. After recollecting a result from \cite{strongedgecoloringofbipartitegraphs}, which establishes the connection between bipartite caching graphs and PDA's, we use the result from \cite{PDAapplications} to adapt our Scheme B to the distributed computing setting. Towards that end, we recall the definition of PDA presented in \cite{YCTCPDA}.

\begin{definition}[Placement Delivery Array \cite{YCTCPDA}] 
\label{PDA definition}
For positive integers $K,F,Z$ and $S$, an $F\times K$ array $\boldsymbol{A}= [a_{j,k}],j\in [F],k\in [K]$, composed of a specific symbol ``$*$" and $S$ integers $1,\cdots,S,$ is called a $(K,F,Z,S)$ placement delivery array (PDA), if it satisfies the following conditions:
\begin{enumerate}
    \item[C1.] The symbol ``$*$" appears $Z$ times in each column.
    \item[C2.] Each integer occurs at least once in the array.
    \item[C3.] For any two distinct entries $a_{j_1,k_1}$ and $a_{j_2,k_2}$ we have $a_{j_1,k_1}=a_{j_2,k_2} =s$, an integer, only if
    \begin{enumerate}
        \item[1.] $j_1\neq j_2,k_1\neq k_2,$ i.e., they lie in distinct rows and distinct columns; and
        \item[2.] $a_{j_1,k_2}=a_{j_2,k_1} =*$.
    \end{enumerate}
\end{enumerate}


If each integer $s\in [S]$ occurs exactly $g$ times, then $\boldsymbol{A}$ is called a regular $g-(K,F,Z,S)$ PDA, or $g$-PDA for short.
\end{definition}

Most known coded caching schemes in the literature correspond to PDAs. 
The following result shows that any bipartite caching graph $B(K,F,D)$ with an induced matching cover with the size of each induced matching being at least $2$ is equivalent to a $g$-PDA. This is equivalent to the result in \cite{strongedgecoloringofbipartitegraphs} which relates a PDA with a strong edge coloring of the bipartite caching graph $B$, which as we have discussed in Section \ref{subsection bipartite model} is equivalent to an induced matching cover of $B$.
\begin{theorem}[equivalent to Theorem 1 in \cite{strongedgecoloringofbipartitegraphs}]
\label{theorem bipartite graph to pda connection}
$B$ is a $(K,F,D)$ bipartite caching graph with an induced matching cover consisting of $S$ induced matchings, each of size $g \geq 2$ if and only if there exist a $g-\left(K,F,Z=F-D,S\right) \text{ regular PDA.}$
\end{theorem}

By Theorem \ref{theorem bipartite graph to pda connection}, it is clear that the bipartite caching graph developed in Section \ref{subsection scheme 2}, which resulted in Scheme B, corresponds to a $g-(K,F,F-D,S)$ regular PDA, where the expressions of $K,F,S,g$ are given in Theorem \ref{theorem construction 2} and the expression for $D$ is given in Lemma \ref{lemma K,F,D,c expressions of scheme 2}.


\subsection{Extension  to Distributed Computing Systems}
\label{subsection distributed computing}
Using coded transmissions to reduce the communication load is a general technique that can potentially be used in any distributed communication system. One such system is the distributed computing framework called the Map-Reduce framework \cite{Mapreduce}. In this framework, one or more functions have to be evaluated on some given large amount of data. To do this in a distributed manner, subsets of the data are assigned to a number of distributed computing nodes, each of which compute the functions on their own data subset. After this, the intermediate values are accumulated at the nodes to give the total computed function values. In \cite{distributed_computing_maddah_ali}, for this framework presented in \cite{Mapreduce}, a technique to reduce the communication load was presented, which is inspired from the coded caching framework. 

We now recall the model from \cite{distributed_computing_maddah_ali}. Consider $N^{\mathcal{C}}$ files, $K^{\mathcal{C}}$ computing nodes and each node is assigned one or more functions to be computed finally (we use superscript $\mathcal{C}$ to represent the parameters of the distributing computing system). The total number of functions to be computed on the files is $Q$. In the \textit{map} phase, the set of $N^{\cal C}$ files are split into $F^{\mathcal{C}}$ batches, each containing $\frac{N^{\cal C}}{F^{\cal C}}$ files. Each node stores some batches of files and computes all the $Q$ functions on the files present in each batch. We denote the files present at node $k\in [K^{\cal C}]$ as ${\cal M}_k.$ These computed function values are referred as intermediate values, and we assume they are represented as $T$-length bit-vectors. Thus, there are $Q|{\cal M}_k|$ intermediate values computed at each node $k$ at the end of the map phase. In the \textit{shuffle} phase, each node $k \in \left[K^{\cal C}\right]$ computes a signal denoted by $X_k$ (of length $l_k$ bits) from the intermediate values, it computed in the map phase, and communicates to other nodes. This is known as \textit{data shuffling}. In the final \textit{reduce} phase, each node is assigned to reduce a set of $\frac{Q}{K^{\mathcal{C}}}$ functions (we assume $\frac{Q}{K^{\mathcal{C}}}$ is an integer). Using the signals received from the other nodes via data shuffling, and the intermediate values computed at itself during the map phase, each node decodes all the intermediate values of its assigned output functions, to finally compute the complete value of the output functions.

The \textit{computation load}, denoted by $r^{\mathcal{C}}$, is the number of map functions computed at all the nodes, normalized by $N$. The communication load, denoted by $L^{\mathcal{C}}$, is the ratio of the total number of bits transmitted to the total number of bits in all the intermediary values $QNT$. Therefore,
\[
r^{\cal C}=\dfrac{\sum\limits_{k\in \left[K^{\cal C}\right]}|{\cal M}_k|}{N}, ~~~~~~~~~~~
L^{\cal C}=\dfrac{\sum\limits_{i=1}^{K^{\cal C}}l_i}{QNT}.
\]

Similar to the coded caching problem, it is a practical necessity to obtain distributed computing schemes for the above model with smaller number of batches $F^{\mathcal{C}}$ for low computation and communication loads. We adapt our Scheme B to this setting in order to obtain a distributed computing scheme which has lower number of batches for similar computation load requirements compared to other major schemes in the literature, at the cost of having larger communication loads. For this purpose, we rely on the literature, which relates the PDAs to the distributed computing schemes \cite{PDAapplications,Q.Yan_CDC_storage_computation_communication,Q.Yan_PDA_CDC_stragglers}. In particular, we are interested in the following result from \cite{PDAapplications}.
\begin{lemma}(Corollary 3 in \cite{PDAapplications}) 
\label{lemma PDA to distributed computing}. For a given $g-(K,F,Z,S)$ regular PDA  with $g\geq 2$, there exists a scheme for distributed computing system with $K^{\mathcal{C}}=K$ nodes, achieving the computation load $r^{\mathcal{C}}=\frac{KZ}{F}$ and communication load $L^{\mathcal{C}}=\frac{g}{g-1}\frac{S}{KF}$, which can be implemented with the minimum number of batches requirement $F^{\mathcal{C}}=F$. 
\end{lemma}


Now, by applying Lemma \ref{lemma PDA to distributed computing}, we can get the corresponding distributed computing system which is presented in Theorem \ref{theorem distributed computing scheme} (the proof follows from Theorem \ref{theorem bipartite graph to pda connection} and Lemma \ref{lemma PDA to distributed computing}).

\begin{tcolorbox}
\begin{theorem}\label{theorem distributed computing scheme}
Let $k,n,m,q$ be positive integers such that $n+m\leq k$ and $q$ be some prime power. The bipartite caching graph given in Section \ref{subsection scheme 2} corresponds to a distributed computing scheme with,

\[K^{\mathcal{C}}=\frac{1}{n!} ~ q^{\frac{n(n-1)}{2}} ~ \prod\limits_{i=0}^{n-1}\gbinom{k-i}{1},\] 

\[r^{\mathcal{C}} =K^{\mathcal{C}}\left(1-q^{nm} ~ \prod\limits_{i=0}^{m-1}\dfrac{\gbinom{k-n-i}{1}}{\gbinom{k-i}{1}}\right),\]

\[F^{\mathcal{C}}=\frac{1}{m!} ~ q^{\frac{m(m-1)}{2}} ~ \prod\limits_{i=0}^{m-1}\gbinom{k-i}{1},\] 

\[L^{\mathcal{C}}= \dfrac{q^{nm}}{\binom{n+m}{n}-1}~\prod\limits_{i=0}^{n-1}\dfrac{\gbinom{k-m-i}{1}}{\gbinom{k-i}{1}}.\]
\end{theorem}
\end{tcolorbox}


In Table \ref{table distributed computing}, we compare our distributed computing system presented in Theorem \ref{theorem distributed computing scheme} with that of \cite{distributed_computing_maddah_ali}. The scheme in \cite{distributed_computing_maddah_ali} is the original scheme in the distributed computing literature for the given setting, and can be viewed as an adaptation of the fundamental coded caching scheme \cite{MaN} into a distributed computing scheme. This scheme is optimal in terms of the communication load for a given computation load, but uses a large batch size, as shown in Table \ref{table distributed computing}. The parameters of the scheme presented in \cite{distributed_computing_maddah_ali} are  $K^\mathcal{C}_2 =K, F^\mathcal{C}_2 =\binom{K}{r}, r^\mathcal{C}_2 =r, L^\mathcal{C}_2 =\frac{1}{r}\left(1-\frac{r}{K}\right)$. where $K$ is a positive integer and $r\in [K]$. The parameters corresponding to Theorem \ref{theorem distributed computing scheme} are labelled as $K^\mathcal{C}_1, F^\mathcal{C}_1, r^\mathcal{C}_1, L^\mathcal{C}_1 $. We choose matching values for $K^\mathcal{C}$ and $r^\mathcal{C}$ for the purpose of comparison.



\begin{table*}[htbp]

\caption{Comparison of the distributed computing scheme in Theorem \ref{theorem distributed computing scheme} with the scheme in \cite{distributed_computing_maddah_ali}. (inf represents $> 10^{307}$).
}

\label{table distributed computing}

\setlength{\tabcolsep}{6.3pt} 
\renewcommand{\arraystretch}{1.4}

\centering
\begin{tabular}{|g|c||g|c||g|c||g|c|}

\hline

\multicolumn{2}{|c||}{\textbf{Number of computing nodes}} & \multicolumn{2}{|c||}{\textbf{Computation load}} & \multicolumn{2}{|c||}{\textbf{Number of batches}} & \multicolumn{2}{|c|}{\textbf{Communication load}}\\
\hline

$(k,n,m,q)$ & $(K,r)$ & & & & & & \\

$K_1^\mathcal{C}$ & $K_2^\mathcal{C}$ & 
$r_{1}^\mathcal{C}$ & 
$r_{2}^\mathcal{C}$ &
$F_1^\mathcal{C}$ & $F_2^\mathcal{C}$ &  $L_1^\mathcal{C}$ & $L_2^\mathcal{C}$ 

\\

(Theorem \ref{theorem distributed computing scheme})&\cite{distributed_computing_maddah_ali} &  (Theorem \ref{theorem distributed computing scheme}) & \cite{distributed_computing_maddah_ali} & (Theorem \ref{theorem distributed computing scheme})
& \cite{distributed_computing_maddah_ali} &  (Theorem \ref{theorem distributed computing scheme}) &\cite{distributed_computing_maddah_ali}

\\
\hline

$(6,2,2,2)$ & $(1953,273)$ & & & & & & \\

1953 & 1953  & 273 & 273  & 1953 & inf & 0.1720 & 0.0032 
\\

\hline

$(4,2,1,3)$ & $(780,78)$ & & & & & & \\

780 & 780 & 78 & 78 & 40 & $10^{108}$ & 0.45 & 0.0115 
\\

\hline

$(5,2,2,2)$ & $(465,129)$ & & & & & & \\

465 & 465 & 129 & 129 & 465 & $10^{117}$ & 0.1445 & 0.0056
\\

\hline

$(5,1,2,3)$ & $(121,4)$ & & & & & & \\

121 & 121 & 4 & 4 & 7260 & $10^{6}$ & 0.4835 & 0.2417 
\\

\hline

$(4,2,1,2)$ & $(105,21)$ & & & & & & \\

105 & 105 & 21 & 21 & 15 & $10^{21}$ & 0.40 & 0.0381 
\\

\hline

$(6,1,2,2)$ & $(63,3)$ & & & & & & \\

63 & 63 & 3 & 3 & 1953 & 39711 & 0.4762 & 0.3175 
\\

\hline

\end{tabular}

\end{table*}


We see from Table \ref{table distributed computing} that for a given number of computing nodes and computation load, our scheme presented in Theorem \ref{theorem distributed computing scheme} performs much better than the scheme of \cite{distributed_computing_maddah_ali} in terms of the number of file batches required (which makes our scheme more practical for distributed systems with tens of nodes or more) for the same computation load. However, the communication load is higher in general. However, for some examples, for instance the entries in Table \ref{table distributed computing} with number of nodes equal to $63$ or $121$, for the communication load at most twice of what the scheme of \cite{distributed_computing_maddah_ali} achieves, we achieve order-of-magnitude gains in the batch-number $F^{\cal C}.$

\begin{remark}
\label{remark D2D scheme}
The coded caching technique has also been utilized in the Device-to-Device (D2D) setting in recent work, for instance, \cite{D2D}. The underlying communication model is very similar to the distributed computing model. Hence, similar to Theorem \ref{theorem distributed computing scheme}, we can also extend the proposed Scheme B to the D2D coded caching system \cite{D2D}, using Theorem \ref{theorem bipartite graph to pda connection} and \cite{PDAapplications} (Corollary 1 in \cite{PDAapplications}) to get low subpacketization schemes for D2D systems. For more details, the reader is referred to \cite{Hari_GLOBECOM_2019}.
\end{remark}

\vspace{-0.2cm}



\begin{figure}
  \centering
        \includegraphics[height=2in]{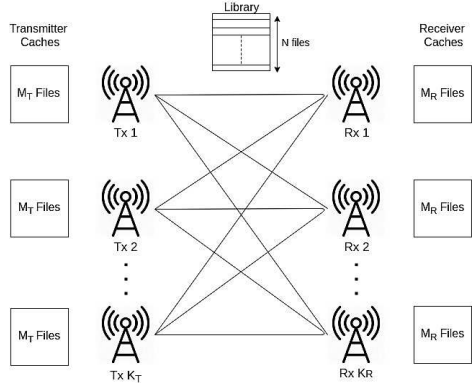}
\caption{Channel model for coded caching in the wireless interference channel.}
\label{fig:interference channel}
\hrule
\end{figure}



\subsection{Extension to Coded Caching in a Wireless Interference Channel}
\label{subsection interference channel}
We now present another setting in which we can apply our low subpacketization scheme, namely the interference channel. Coded caching for the interference channel setting with caches at both transmitters and receivers was considered in \cite{interferencemanagement}, where the fundamental coded caching scheme of Ali-Niesen \cite{MaN} was adapted to this setting. In \cite{addingtransmitters}, this scheme was further refined to present a scheme with lower subpacketization requirement (particularly, no subpacketization at the transmitter end). We leave the details of these schemes to the reader and refer them to the respective papers. In this subsection, we present the adaptation of our low-subpacketization Scheme B for the setting presented in \cite{interferencemanagement}, also motivated by the techniques from \cite{addingtransmitters}. We then compare via numerical examples our scheme's performance with that of \cite{addingtransmitters} 
which itself is an improvement over \cite{interferencemanagement} in terms of the subpacketization, with all other parameters being the same.

We first review the model given in \cite{interferencemanagement} as shown in Fig. \ref{fig:interference channel}. Consider a wireless channel with $K_T$ transmitters and $K_R$ receivers with $N$ files (denoted as $W_i:i\in[N]$) such that each transmitter can store $M_T$ files and each receiver can store $M_R$ files. We assume $L \triangleq \frac{K_TM_T}{N}$ is an integer, and also that $L$ divides $K_R$. Each transmitter and receiver has one antenna, and the channel coefficient between any particular transmitter-to-receiver pair is a complex number that is a realization of an independent continuous probability distribution, and assumed to be constant over the time of communication. The coded caching scheme in this setting involves two phases as before, the caching phase where the transmitter and receiver caches are populated, and then the delivery phase in which the receiver demands (each receiver demands a particular file as before) are served collaboratively by the transmitters. In every round of transmission, some subset of $K_T$ transmitters send signals to the receivers. In a valid scheme, at the end of a finite number of rounds, the decoding of demands at the respective receivers should be complete. The sum-$\mathsf{DoF}$ in this setting refers intuitively to the number of receivers served per transmission. For a precise definition of this parameter, we refer to the reader to \cite{interferencemanagement}. 

For this setting, under the condition that $L+\frac{K_RM_R}{N}\leq K_R$, the scheme in \cite{interferencemanagement} achieves a sum-$\mathsf{DoF}$ of $L+\frac{K_RM_R}{N}$, with a subpacketization level of $\binom{K_T}{L}\binom{K_R}{\frac{K_RM_R}{N}}$ which arises as a result of subpacketization for caching at both the transmitter-side and the receiver-side. The work \cite{addingtransmitters} achieves the same sum-$\mathsf{DoF}$ with a subpacketization level of $\binom{{K_R}/{L}}{{K_RM_R}/{(LN)}}$, which avoids subpacketization at the transmitter-side completely. For more details, we refer the reader to \cite{interferencemanagement} and \cite{addingtransmitters}. In the rest of this subsection, we shall adapt our low subpacketization scheme, Scheme B, to this interference channel setting. For the sake of notational convenience, we focus on adapting a particular special class of Scheme B (which has parameters $(k,m,n,q)$) with parameter $n=1$. However, the more general scheme can also be adapted, which gives more flexibility in terms of the parameters.

The principle we use is similar to the grouping scheme for the multi-transmitter coded caching scheme given in \cite{addingtransmitters}. The idea behind this is to divide the receivers into a number of $K\triangleq \frac{K_R}{L}$ groups, each containing $L$ receivers. Then the broadcast-channel coded caching scheme (in our case, Scheme B) is applied to the groups considering them to be super-users, while also utilizing the presence of transmitter caches to zero-force some non-demanded and un-cached subfiles. We now give the detailed scheme.

Let $k,m,q$ be positive integers such that $m\leq k$ and $q$ is some prime power. We assume that $K_R=KL$, where $K=\gbinom{k}{1}$ for some $k$. Let $\mathbb{V} \triangleq \{V \in PG_q(k-1,0)\}$ (Note that in Section \ref{subsection scheme 2}, we used the notation $\mathbb T$ for the same). Consider $\frac{M_R}{N}=1-\frac{q^{m}{\gbinom{k-m}{1}}}{{\gbinom{k}{1}}}$ (note that this is equal to the $M/N$ value we get from Scheme $B$ with $n=1$ in Theorem \ref{theorem construction 2}). We group the $K_R$ users into $K$ groups, such that in each group there are $L$ users. 
The groups are indexed by distinct 1-dim subspaces of $\mathbb{F}^k_q$, i.e., by the elements of $\mathbb{V}$. 
Let the users in a group $V \in \mathbb{V}$ be denoted as $V(1),...,V(L)$.
We now describe the placement and delivery phase.


\subsubsection{\textbf{Placement Phase}}

\underline{\textit{Transmitter Caching Strategy:}} We follow the caching strategy described in \cite{addingtransmitters} (Appendix A in \cite{addingtransmitters}) for the transmitter side. We require each subfile of each file to be cached at precisely $L$ transmitters. 
\begin{itemize}
    \item 
For $i\in[K_T]$, the cache content $C_{T_i}$ of the $i^{th}$ transmitter denoted by $T_i$ is given by
$$\hspace{-0.4cm}C_{T_i}=\{W_{1+(p-1)(mod~N)}:p \in \{1+(i-1)M_T,...,M_Ti\}\}.$$ 
\end{itemize}
Thus, the transmitters cache successive $M_T$ files into their caches. Because of this caching strategy, it should also be clear that each file cached at $L=\frac{K_TM_T}{N}$ transmitters. This property of the transmitter-side caching will be used to obtain a sum-$\mathsf{DoF}=L+\frac{K_RM_R}{N}$ via the zero-forcing technique.

\textit{\underline{Receiver Caching Strategy:}} On the receiver side, the caching strategy is based on the projective geometry as described in Section \ref{subsection scheme 2} (Scheme B, with the parameter $n=1$). Consider $m \in \mathbb{Z}^+$ such that $m\leq k$. Let 

\begin{align*}
\mathbb{S}&\triangleq\{S \in PG_q(k-1,m-1)\}.\\
\mathbb{Y}&\triangleq\bigg\{\{V_1,V_2,...,V_m\}: \forall V_i \in \mathbb{V},  \sum\limits_{i=1}^{m}V_i \in \mathbb{S}\bigg\}.
\end{align*}

Following Scheme B, the subfiles are indexed by the elements of $\mathbb{Y}$. The subfiles of file $W_i$ are denoted as $W_i=\{{W_i^Y}: Y \in \mathbb{Y}\}$. The subpacketization $F$ is thus equal to $|\mathbb{Y}|$, which was shown in Lemma \ref{lemma K,F,D,c expressions of scheme 2} in Section \ref{section scheme 2} to be $ \frac{1}{m!} ~ q^{\frac{m(m-1)}{2}} ~ \prod\limits_{i=0}^{m-1}\gbinom{k-i}{1}$. 

 \begin{itemize}
     \item 
The caches of the receivers $V(r):r\in[L]$ in the group $V\in {\mathbb V}$ are populated with the same content, given as 
$$
Z_{V}\triangleq \bigg\{W_i^Y: Y \in \mathbb{Y}, V \not\subseteq \sum\limits_{V_j \in Y} V_j\bigg \}_{i=1}^N.
$$
 \end{itemize}
Thus we have the $K=\gbinom{k}{1}$ groups of our current setting in the place of $K$ users in the original Scheme B. We now describe the delivery phase corresponding to this caching strategy.
\subsubsection{\textbf{Delivery Phase}}
In the delivery phase, each receiver demands for a file. We use a projective geometry based delivery scheme developed in Section \ref{subsection scheme 2} over the groups of users.
Let the demand of receiver $V(i)$ be denoted as $W_{d_{V(i)}}$. The delivery scheme serves $m+1$ groups of receivers in each transmission. For this purpose, we develop some notations following the description of Scheme B. 

Let $$\mathbb{Z}=\bigg\{\{V_1,V_2,...,V_{m+1}\}\subseteq {\mathbb V}: \sum\limits_{j=1}^{m+1}V_j \in PG_q(k-1,m) \bigg\}.$$ 
Consider $Z=\{V_1,V_2,...,V_{m+1}\} \in \mathbb{Z}$, denoting a set of $(m+1)$ groups of receivers, the $(m+1)L$ receivers which will be served in one round of transmission. Let $Y_j=Z\backslash V_j:j\in[m+1]$. In this round, the subfiles to be transmitted to the receivers in the groups given in $Z$ are $\{W_{d_{V_j(i)}}^{Y_j}:j\in[m+1],i\in[L]\}$. Clearly, the subfile ${W}_{d_{V_j(i)}}^{Y_j}$ which is desired at the user $V_j(i)$, is not cached at any receiver in the group $V_j$ but available at all the receivers in the groups $V_{j'}: j'\in [m+1]\backslash j.$ For the purpose of transmitting on a wireless channel, let $\widetilde{W}_{d_{V_j(i)}}^{Y_j}$ denote the signal (from some complex constellation) denoting the mapping of the subfile ${W}_{d_{V_j(i)}}^{Y_j}$; this mapping is known to all receivers and transmitters. We now construct the idea behind the delivery scheme, by showing the round corresponding to $Z\in {\mathbb Z}$.

We denote by $\boldsymbol{s}_i \in \mathbb{C}^{1 \times 1}: i \in [K_T]$ the signal transmitted by $i^{th}$ transmitter during this round.  We have to design our transmission signals such that the mapped subfile $\widetilde{W}_{d_{V_j(i)}}^{Y_j}$ can be obtained at the user $V_j(i)$, while (a) the same subfile can be zero-forced at the receivers in $V_j\backslash V_j(i)$ by utilizing the presence of the $L$ transmitters in which the subfile ${W}_{d_{V_j(i)}}^{Y_j}$ is available, and (b) the same subfile can be cancelled using the cache content at each receiver in the group $V_{j'}:j'\in\left[m+1\right]\backslash{j}$.  

Let $\boldsymbol{\overline{w}}_j=[\widetilde{W}_{d_{V_j(1)}}^{Y_j},...,\widetilde{W}_{d_{V_j(L)}}^{Y_j}]^T$ denote the mapped subfiles involved in this round of transmissions corresponding to the group $V_j$. The round of transmissions is described as follows:

\begin{align}
\label{tx}
{\begin{bmatrix}
\boldsymbol{s}_1\\ \vdots \\\boldsymbol{s}_{K_T}
\end{bmatrix}}={\begin{bmatrix}
\overline{A}_1 & \cdots & \overline{A}_{m+1}
\end{bmatrix}}{\begin{bmatrix}
\boldsymbol{\overline{w}}_1\\ \vdots \\\boldsymbol{\overline{w}}_{m+1}
\end{bmatrix}},
\end{align}
where $\overline{A}_i \in \mathbb{C}^{K_T \times L} : i \in [m+1]$ contains as its columns the $L$ precoding vectors of length $K_T$ corresponding to the mapped subfiles $\widetilde{W}_{d_{V_j(i)}}^{Y_j}:i\in[L]$ generated  by the $K_T$ transmitters. Note that any such precoding vector can have only $L$ non-zeros, as the subfile ${W}_{d_{V_j(i)}}^{Y_j}$ is available only at some $L$ transmitters. Thus, every column of the matrices $\overline{A}_i:i\in[m+1]$ contains $K_T-L$ zeros, with the remaining entries to be chosen to satisfy the zero-forcing requirements. In the remainder of this subsection, we show that such precoding vectors can indeed be chosen, and show that decoding of the mapped subfiles (and thereby, via the inverse mapping, the uncached subfiles) can be decoded at the respective receivers. The delivery scheme can therefore be completed by constructing transmissions as in (\ref{tx}) for every $Z\in{\mathbb Z},$ with appropriately chosen precoding vectors to effect successful decoding.

We now show that the precoding vectors of (\ref{tx}) in the matrices $\overline{A}_i:i\in[m+1]$ can be chosen to ensure decoding of the desired subfiles at the respective users. The received signals at the receivers of groups $Z=(V_1,...,V_{m+1})$ is given as,
\begin{align}
\label{rx}
{\begin{bmatrix}
\boldsymbol{y}_{1}\\ \vdots \\\boldsymbol{y}_{{m+1}}
\end{bmatrix}}={\begin{bmatrix}
H_{1}\\ \vdots \\H_{{m+1}}
\end{bmatrix}}{\begin{bmatrix}
\boldsymbol{s}_1\\ \vdots \\\boldsymbol{s}_{K_T}
\end{bmatrix}}+{\B z},
\end{align}
where $\boldsymbol{y_j}=[y_{j,1},...,y_{j,L}]^T:j \in [m+1]$, with $y_{j,i}:i\in[L]$ being the received signal at the user $V_j(i)$, the matrix $H_j:j \in [m+1]$ is the ${L \times K_T}$ channel matrix from the $K_T$ transmitters to the $L$ receivers of the group $V_j$ and ${\B z}=({\B z}_1,\hdots,{\B z}_{m+1})^T$ denotes the additive white Gaussian noise, each component of which is distributed as a circular symmetric complex Gaussian.

We now have from (\ref{tx}) and (\ref{rx}), the received signal at the group $V_j$ as
\begin{align}
\label{rx_sig}
    \boldsymbol{y_j}={\begin{bmatrix}
    H_{j}A_1 \hdots H_{j}A_{m+1}
    \end{bmatrix}}{\begin{bmatrix}
    \overline{\B{w}}_1 \\ \vdots \\ \overline{\B{w}}_{m+1}
    \end{bmatrix}}_{}+{z}_j
\end{align}
Because the receiver $V_j(i)$ has the subfiles in (\ref{rx_sig}) except $W_{d_{V_j(i)}}^{Z\setminus V_j}:i \in [L]$, from (\ref{rx_sig}), the receivers of the group $V_j$ can obtain, 

$$\tilde{\boldsymbol{y}}_{j}=H_{j}A_{j}\overline{\boldsymbol{w}}_{j}+{\B z}_j.$$

Because of the fact that the entries of $H_j$ are picked from a continuous distribution, any $L$ columns of $H_j$ are linearly independent almost surely, there exists a vector, with the condition that it is non-zero only in the positions corresponding to the transmitters caching $W_{d_{V_j(i)}}$ and $0$ everywhere else, which can be fixed as the  $i^{th}$ column of $A_j$, such that $H_{j}A_j=I$ (the identity matrix of size $L$). Hence, we have found the desired solution for the precoding vectors. This ensures decoding (as $\mathsf{SNR}$ grows large, in the $\mathsf{DoF}$ sense) of the mapped subfiles $\{\widetilde{W}_{d_{V_j(i)}}^{Y_j}:j\in[m+1],i\in[L]\}$ at the respective receivers, and thus by inverse mapping the original desired subfiles can be decoded. The delivery scheme, which constructs the transmissions as in (\ref{tx}) for every $Z\in {\mathbb Z}$ thus ensures decoding of all uncached and desired subfiles at the respective receivers. As in each round of transmissions, the number of users served is $L(m+1)$, which is the sum-$\mathsf{DoF}$ for the presented scheme. The results of the coded caching scheme constructed above is presented in the following theorem.

\begin{tcolorbox}
\begin{theorem}
\label{theorem interference channel scheme}
In a wireless channel consisting of $N$ files, $K_T$ transmitters each with cache that can contain $M_T$ files such that $L=\frac{K_TM_T}{N}$ is a positive integer, and 
consisting of $K_R=L\gbinom{k}{1}$ receivers each of cache size $M_R$ files where $\frac{M_R}{N}=1-\frac{q^{m}{\gbinom{k-m}{1}}}{{\gbinom{k}{1}}}$, we can achieve with high SNR, a sum-$\mathsf{DoF}=L(m+1)$ with subpacketization 
$F= \frac{1}{m!} ~ q^{\frac{m(m-1)}{2}} ~ \prod\limits_{i=0}^{m-1}\gbinom{k-i}{1},$
where $k,m \in \mathbb{Z}^+$ with $m\leq k$ and $q$ is some prime power. 
\end{theorem}
\end{tcolorbox}


\begin{table}[htbp]

\captionof{table}{Comparison of the coded caching scheme for the interference channel presented in Theorem \ref{theorem interference channel scheme} with the scheme in \cite{addingtransmitters}, for some specific values of $K_R,L,\frac{M_R}{N}$. }
\label{table interferen channel}
\centering

\setlength{\tabcolsep}{6.3pt} 
\renewcommand{\arraystretch}{1.4}
\resizebox{8.9cm}{!} {
\begin{tabular}{|c|c|c|g|c|g|c|}

\hline

$\mathbf{(k,m,q) }$ & $\mathbf{L}$ & $\mathbf{\frac{M_R}{N}}$ & $\mathbf{F_1}$ & $\mathbf{F_2}$ & \textbf{sum-}$\mathbf{\mathsf{DoF}_1}$ & \textbf{sum-}$\mathbf{\mathsf{DoF}_2}$\\

$\mathbf{{K_{R}}}$& & & (Theorem \ref{theorem interference channel scheme}) &{\cite{addingtransmitters}}& (Theorem \ref{theorem interference channel scheme}) &{\cite{addingtransmitters}}\\
\hline

$(4,3,2)$ & & & & & &  \\

$30$ & $2$ & $0.4667$ & $420$ & $6435$ & $8$ & $16$ \\

\hline

$(5,3,2)$ & & & & & & \\

$62$ & $2$ & $0.2258$ & $4340$ & $7.3\times10^5$ & $8$ & $15$ \\

\hline

$(4,3,3)$ & & & & & & \\

$80$ & $2$ & $0.3250$ & $9360$ & $1.2\times10^{10}$ & $8$ & $28$ \\

\hline

$(5,4,2)$ & & & & & & \\

$124$ & $4$ & $0.4839$ & $2.6 \times 10^4$ & $3\times10^8$ & $20$ & $64$ \\

\hline

$(6,4,2)$ & & & & & & \\

$252$ & $4$ & $0.2381$ & $5.4 \times 10^5$ & $1.22\times10^{14}$ & $20$ & $64$ \\

\hline

\end{tabular}
}

\end{table}


\vspace{-0.2cm}
Table \ref{table interferen channel} gives a numerical comparison of our scheme's parameters, subpacketization and sum-$\mathsf{DoF}$, with that of \cite{addingtransmitters} by choosing matching values for $K_R, L$ and $\frac{M_R}{N}$. The comparison parameters corresponding to Theorem \ref{theorem interference channel scheme} are labelled as $F_1$ and sum-$\mathsf{DoF}_1$. The comparison parameters corresponding to the scheme in \cite{addingtransmitters} are $F_2=\binom{K_R/L}{\lfloor (K_R M_R)/(NL) \rfloor}$ and sum-$\mathsf{DoF}_2 = L+ \lfloor\frac{K_R M_R}{N}\rfloor$. As can be seen, our scheme performs better in terms of the subpacketization, while having lesser sum-$\mathsf{DoF}$. The quantities are rounded off to a few decimal places wherever applicable.




\section{Lower Bounds on Rate of Delivery Scheme for Symmetric Caching}
\label{section lowerbound}
In this section, we present two information theoretic lower bounds on the rate of the transmission scheme associated with a $(K,F,D=F(1-\frac{M}{N}))$ bipartite caching scheme (which is associated with the bipartite caching graph $B(K,F,D)$) and numerically compare with the existing lower bounds from \cite{WTP,cheng2017coded}. We obtain two bounds for the rate of coded caching with the fixed parameters $K,F,M$ and $N$. The first bound holds for all symmetric caching schemes, but is quite loose. The second one holds for a special class of symmetric caching schemes, in which each subfile is stored in the same number of users (equivalently the bipartite caching scheme associated with a bi-regular bipartite graph).  These bi-regular schemes include all the symmetric caching schemes in the literature to the best of our knowledge, including those in Table \ref{table known results}. In this special class of caching schemes, the second bound is shown to be better, for a number of parameter choices, than the existing bounds from the prior work via numerical comparisons.

We first give some preliminary ideas and definitions before we present our bounds. 
As the subfile $W_{i,f}$ of the file $W_i$ takes values from the finite set ${\cal A}$ with uniform distribution, taking the base of logarithm as $|{\cal A}|$, we have the Shannon entropy of $W_{i,f}$ as $H(W_{i,f})=1,\forall i,f$. Thus $H(W_i)=F, \forall i\in[N]$.  

\begin{definition}
For the given parameters $K, \frac{M}{N}$, and $F$ such that $\frac{FM}{N} \in \mathbb{Z}^{+}$, a rate $R$ is said to be \textit{achievable} if there exists some $(K,F,D=F(1-\frac{M}{N}))$ bipartite caching scheme, with a delivery scheme with the rate $R$ that satisfies all the client demands. For the given parameters $K,F,D$, we define the optimal rate $R^*(K,F,D)$ as follows:
\[
R^*(K,F,D) \triangleq \inf \{R : R \text{ is achievable}\}.
\]
\end{definition}
\begin{remark}
We abbreviate $R^*(K,F,D)$ as simply $R^*$, as the parameters involved will be clear from the context.
\end{remark}
It is known from \cite{WTP} that $R^*\geq \frac{K(1-\frac{M}{N})}{1+\frac{MK}{N}}$ for any value of $F$, and this is achieved by the scheme in \cite{MaN} with $F=\binom{K}{\frac{MK}{N}}$. Further, for the coded caching schemes with given parameters derived from the PDAs (which correspond to the bipartite caching schemes with the induced matching based delivery schemes), it was shown in \cite{cheng2017coded} that 
\begin{align}
\nonumber
R^{*}F &\geq  \left\lceil{\frac{DK}{F}}\right\rceil +  \left\lceil{\frac{D-1}{F-1}\left\lceil{\frac{DK}{F}}\right\rceil}\right\rceil +\cdots \\
\label{eqnearlierlowerbound}
\cdots &
+\left\lceil{\frac{1}{\frac{FM}{N}+1}\left\lceil {\frac{2}{\frac{FM}{N}+2}\left\lceil\cdots\left\lceil\frac{D-1}{F-1}\left\lceil{\frac{DK}{F}}\right\rceil\right\rceil\cdots\right\rceil}\right\rceil}\right\rceil.
\end{align} 
Note that in (\ref{eqnearlierlowerbound}), the notation $R^*$ is abused to correspond to the optimal rate only among the PDA based delivery schemes (which corresponds to the induced matching based delivery schemes), which is a restricted class of delivery schemes among all the delivery schemes obtainable for any symmetric caching scheme. We compare our new lower bounds with these two bounds.

\begin{figure}
  \centering
        \includegraphics[height=1.9in]{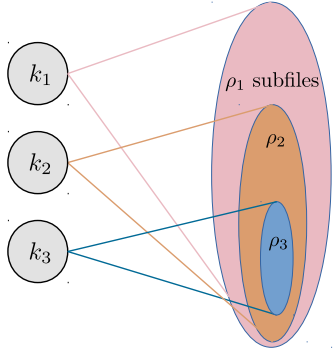}
\caption{Intuition behind the lower bound presented in Theorem \ref{genericlowerbound}. The nesting of the $\rho_3$ neighbours of the vertex $k_3$ within the $\rho_2$ neighbours of $k_2$, and these further within the $\rho_1$ neighbours of $k_1$, results in the number of transmissions $R^*F\geq \rho_1+\rho_2+\rho_3$.} 
\label{fig:lowerbound}
\hrule
\vspace{-0.4cm}
\end{figure}

We will first prove a generic lower bound on the rate, in Theorem \ref{genericlowerbound}, for the coded caching schemes whose caching scheme is based on the bipartite caching model, about which we briefly give some intuition. Consider the induced-matching based delivery scheme in Section \ref{subsection bipartite model} of the bipartite caching graph. The cardinality of any subset of edges of the bipartite caching graph such that no two among the subset can appear in any single induced matching, gives us a lower bound on the number of transmissions in any induced-matching based delivery scheme. This is illustrated in Fig. \ref{fig:lowerbound}. Consider a bipartite caching graph $B$, of which a subgraph is shown in Fig. \ref{fig:lowerbound}. Let $k_1$ be an arbitrary user vertex. Let $ \rho_1  = |\mathcal{N}(k_1)|$. By Definition \ref{defintion induced matching&cover}, it is clear that no two of these $\rho_1$ edges (from $k_1$ to $\mathcal{N}(k_1)$) must lie in the same induced matching. Now consider an arbitrary user vertex $k_2 (\neq k_1)$. Let $ \rho_2 = |\mathcal{N}(k_1) \cap \mathcal{N}(k_2)|$. We can see that no two of the $\rho_1 +\rho_2$ edges $\big($from $k_1$ to $\mathcal{N}(k_1)$ and from $k_2$ to $\mathcal{N}(k_1) \cap \mathcal{N}(k_2) \big)$ lie in the same induced matching, by Definition \ref{defintion induced matching&cover}. Following this idea, consider $N^{'}$ user vertices $\left\{k_1,k_2, \cdots, k_{N^{'}}\right\}$. Let $\rho_{j} =  \Big| \bigcap\limits_{i=1}^{j} \mathcal{N}(k_i) \Big|$ where $j \in \big[N^{'}\big]$. Now consider the set of edges of $B$ from $k_j$ to $\bigcap\limits_{i=1}^{j} \mathcal{N}(k_i), \forall j \in \big[N^{'}\big]$. There are exactly $\sum\limits_{j=1}^{N^{'}} \rho_j$ edges in this set and no two edges lie in the same induced matching, by Definition \ref{defintion induced matching&cover}. Therefore, $\sum\limits_{j=1}^{N^{'}} \rho_j$ is a lower bound on the number of transmissions of any induced-matching based delivery scheme.

Interestingly, we show in  Theorem \ref{genericlowerbound} that the bound $\sum\limits_{j=1}^{N^{'}} \rho_j$ derived in the above discussion applies in the information theoretic sense also, for a given bipartite caching scheme. We then use this result to show our two lower bounds on the rate of the coded caching schemes with the fixed parameters $K, F$ and $D=F(1-M/N)$ in Corollary \ref{corollaryfirstlowerbound} and Theorem \ref{theorem lower bound 2}. Thus, these bounds hold for non-linear schemes as well, in contrast with (\ref{eqnearlierlowerbound}), which was shown to be true for induced matching based delivery schemes. The lower bound in Theorem \ref{genericlowerbound} can be viewed as the maximum acylic induced subgraph (MAIS) lower bound \cite{MAIS_lowerbound} for the index coding problem induced by the coded caching setup. The proof of this bound uses a technique from \cite{exactratememorytradeoff} (Appendix A in \cite{exactratememorytradeoff}). 

\begin{tcolorbox}
\begin{theorem}
\label{genericlowerbound}
Let $B$ be a bipartite caching graph representing a caching scheme on a broadcast network with $N$ files at the server. For some $N^{'} \leq N$, let $U=\{k_j:j\in[N']\}$ be an arbitrary subset of $N'$ user vertices of $B$. For $j\in[N']$, let $\rho_j$ be the number of right vertices (subfiles) in $B$ which are adjacent to each vertex in $\{k_i:i\in[j]\}.$ Let $\tilde{R}^*$ be the infimum of all achievable rates for the bipartite caching scheme defined by $B$. Then, \vspace{-0.5cm}
\[ \tilde{R}^{*} F \geq \sum_{j=1}^{N'}\rho_j.\]
\end{theorem}
\end{tcolorbox}
\begin{IEEEproof}
We are given a valid caching scheme associated with $B$. As $N' \leq N$, we can assume a demand scenario in which the $N'$ users all demand different files. Let $\boldsymbol{Y}$ denote the set of all transmissions in a valid transmission scheme. Let $W_{d_j}$ be the demand and $Z_j$ be the cache content of the user $k_j\in U$. Let $S_j$ denote the set of subfiles of $W_{d_j}$ not cached in any of the users $k_i:i\in[j]$. This corresponds to the subfile vertices adjacent to all the users $k_i:i\in[j]$. In our notation, $|S_j|=\rho_j.$ Since $W_{d_j}$s are distinct, thus each subfile in $S_j:j\in[N']$ is distinct. We then follow an idea similar to \cite{exactratememorytradeoff}. We construct a virtual receiver which contains an empty cache at first. In the $j^{th}$ step, the cache of this virtual user is populated with all the cache contents of the user $k_j$ except those pertaining to the files demanded by $k_i:i\in[j-1].$ Let $\tilde{Z}_j=Z_j \backslash \{W_{d_i,f}:i<j,\forall f\}.$ Then $\{\tilde{Z}_j:j\in[N']\}$ is the final cache content of this virtual user. By the given transmission scheme, the receivers can decode their demands. Hence, we must have
\begin{equation}
\label{eqn201}
H\left(\{W_{d_j}:j\in [N']\}\mid\{\tilde{Z}_j:j\in[N']\},\boldsymbol{Y}\right)=0,
\end{equation}
as the virtual user must be successively able to decode all the demands of the $N'$ users. Since $RF$ denotes the size of transmissions (assuming each subfile to be of unit size) in a code of rate $R$, we must have the following inequalities. 
\begin{align*}
\tilde{R}^*F&\geq H(\boldsymbol{Y})\\
&\geq I\left(\boldsymbol{Y};\{W_{d_j}:j\in[N']\}\mid\{\tilde{Z}_j:j\in[N']\}\right)\\
&= H\left(\{W_{d_j}:j\in[N']\}\mid\{\tilde{Z}_j:j\in[N']\}\right) ~~(\text{by}~ (\ref{eqn201}))\\
& \geq H\left(\{S_j:j\in[N']\}\right)=\sum_{j=1}^{N'}\rho_j,
\end{align*}
where $I(;)$ denotes the mutual information, and the last inequality is obtained by noting the missing subfiles in $\{\tilde{Z}_j:j\in[N']\}$. This completes the proof.
\end{IEEEproof}
Once again, we note that as the bound in Theorem \ref{genericlowerbound} applies to the rate of any delivery scheme for the given bipartite caching scheme, not just the induced-matching based scheme. The following example illustrates the lower bound presented in Theorem \ref{genericlowerbound}.

\begin{example}(Continuation of Example \ref{example of bipartite caching}, Example \ref{example of induced matching})
\label{example of lower bound}
Label the user vertices of the bipartite caching graph $B(4,5,3)$ presented in Fig. \ref{fig: bipartite caching graph} as $k_1 = 1, k_2 = 3, k_3 = 2, k_4 = 4$. Assume that the number of files $N\geq 4$. Following Theorem \ref{genericlowerbound}, we get $\rho_1=\big|\mathcal{N}(k_1)\big| =\big|\{f_1,f_2,f_3\}\big| =3$, $\rho_2= \bigg|\bigcap\limits_{i=1}^{2}\mathcal{N}(k_i)\bigg|= \big|\{f_2,f_3\}\big|=2$, $\rho_3=\bigg|\bigcap\limits_{i=1}^{3}\mathcal{N}(k_i)\bigg|=\big|\{f_2\}\big| =1$, $\rho_4=\bigg|\bigcap\limits_{i=1}^{4}\mathcal{N}(k_i)\bigg|=|\phi| =0$. We then get the lower bound on the rate as, $R^*\geq \frac{\sum\limits_{j=1}^4\rho_j}{5}=\frac{6}{5}$. The induced matching based delivery scheme presented in Example \ref{example of induced matching} achieves the rate $R=\frac{6}{5}$. Therefore, for the symmetric caching scheme defined by $B(4,5,3)$, the rate $R=\frac{6}{5}$ is information theoretically optimal. 
\end{example}

Using the  result in Theorem \ref{genericlowerbound}, we obtain our first lower bound in Corollary \ref{corollaryfirstlowerbound} for any symmetric caching scheme (left-regular bipartite caching schemes), and then the second bound in Theorem \ref{theorem lower bound 2} for all bi-regular bipartite caching schemes. 
\begin{tcolorbox}
\begin{corollary}
\label{corollaryfirstlowerbound}
For any symmetric caching scheme with $K$ users, cache fraction $\frac{M}{N}$, subpacketization $F$, and the number of files $N\geq K(1-\frac{M}{N});$ the optimal rate $R^*$ satisfies
\begin{equation}
\label{eqn110}
R^{*} F \geq (K+F)\left(1-\frac{M}{N}\right)-1.
\end{equation}
\end{corollary}
\end{tcolorbox}
\begin{IEEEproof}
A symmetric caching scheme, with the given parameters, is equivalently represented by a bipartite caching graph $B(K,F,D)$ as in Section \ref{subsection bipartite model}. In any such graph, by the pigeon-holing argument, it is easy to see that there is a subfile vertex $f\in {\cal F}$ in $B$ having at least $K\left(1-\frac{M}{N}\right)$ adjacent user vertices, which we refer to as $U=\{k_i:i\in [K\left(1-\frac{M}{N}\right)]$. Clearly, by definition of $B$, the vertex $k_1$ has $D=F(1-\frac{M}{N})$ neighbours. Thus, following the notations in Theorem \ref{genericlowerbound}, we have $\rho_1= D$. Further, for $ 2 \leq i \leq K(1-\frac{M}{N})$, we have $\rho_i\geq 1$, as $f$ is adjacent to each vertex in $U$. As the bounds on $\rho_i:\forall i\in\left[K(1-\frac{M}{N})\right]$ are independent of the bipartite graph chosen and depend only on the given parameters $K,F,\frac{M}{N}$, by applying Theorem \ref{genericlowerbound} for the bipartite graph corresponding to the caching scheme which gives the delivery scheme with optimal rate $R^*$, we get (\ref{eqn110}).
\end{IEEEproof}

In the following bound, we abuse the notation $R^*$ to denote the optimal rate (given $K,F,\frac{M}{N}$) among all delivery (including non-linear) schemes defined for a \textit{bi-regular} bipartite caching graph $B(K,F,D)$. It is clear that for such a graph the right degree is $d \triangleq K\left(1-\frac{M}{N}\right)$ which means that, each subfile is stored in a constant $\frac{KM}{N}$ number of clients. All known symmetric caching schemes belong to this class of graphs, to the best of our knowledge, and hence this bound is applicable to each of them. This bound also applies to the schemes which we have developed in this work, as they are obtained from bi-regular bipartite graphs (see Remarks \ref{remark on bi-regularity of graph in scheme A} and \ref{remark on bi-regularity of graph in scheme B}). We show in Table \ref{table lower bound} that this second lower bound outperforms prior bounds for many chosen values of parameters. 
\begin{tcolorbox}
\begin{theorem} 
\label{theorem lower bound 2}
Let $R^{*}$ be the infimum of all achievable rates for the coded caching problem defined by a bi-regular bipartite caching graph $B(K,F,D)$, with the right degree $d = K\left(1-\frac{M}{N}\right)$ and assume that the number of files $N\geq K\left(1-\frac{M}{N}\right)$. Then $R^*$ satisfies
\small
\begin{align*} \nonumber
R^*F ~ & \geq D+ \left\lceil{\tiny\frac{(d-1)D}{K-1}}\right\rceil +  
\cdots
\\ & \hspace{-1.15cm}\cdots 
+
\left\lceil{\frac{1}{\frac{KM}{N}+1}\left\lceil {\frac{2}{\frac{KM}{N}+2}\left\lceil\cdots\left\lceil \frac{d-2}{K-2}\left\lceil{\frac{(d-1)D}{K-1}}\right\rceil\right\rceil\cdots\right\rceil}\right\rceil}\right\rceil
\end{align*}
\end{theorem}
\end{tcolorbox}


\begin{table*}[htbp]
\caption{\small Comparison of the
two proposed information theoretic lower bounds on the rate of the transmission schemes associated with the special class of caching schemes defined by the bi-regular bipartite caching graphs with parameters $K,F,D$, with that of \cite{cheng2017coded}, \cite{WTP}. The last column gives the number of transmissions in the scheme constructed in Section \ref{subsection scheme 2} for whatever values are applicable. 
(NA represents not applicable).}
\label{table lower bound}
\setlength{\tabcolsep}{9pt} 
\renewcommand{\arraystretch}{1.7}
\centering
\begin{tabular}{|c|c|c|c|c|c|c|c|c|}
\hline
$K$ & $F$ &$D$ & $\frac{M}{N}$ & (Corollary \ref{corollaryfirstlowerbound}) & (Theorem \ref{theorem lower bound 2}) & \cite{cheng2017coded}  & \cite{WTP} & Scheme B
\\ 
& & & & & & & & [Section \ref{subsection scheme 2}]\\
& & & $=1-\frac{D}{F}$ & $R^{*} \geq $ & $R^{*} \geq $ & $ R^{*} \geq $ & $ R^{*} \geq $ & $R$
\\
\hline
15 & 50 & 30 & 0.4 & 0.76 & 1.42 & 1.08 & 1.2857 & NA\\
\hline
24 & 54 & 36 & 0.3333 & 0.9444 & 2.0185 & 1.6667 & 1.7778 & NA\\
\hline
15 & 20 & 12 & 0.4 & 1 & 1.5 & 1.55 & 1.2857 & NA\\
\hline
7 & 21 & 12 & 0.4286 & 0.7143 & 1.0476 & 0.8571 & 1 & 1.33\\
\hline
13 & 78 & 54 & 0.3077 & 0.7949 & 1.8333 & 1.3205 & 1.8 & 3\\
\hline
15 & 105 & 84 & 0.2 & 0.9048 & 3.0952 & 2.2286 & 3 & 4\\
\hline
21 & 210 & 160 & 0.2381 & 0.8333 & 2.7381 & 1.7952 & 2.6667 & 105\\
\hline
31 & 465 & 420 & 0.0968 & 0.9613 & 7.0839 & 4.9247 & 7 & 9.3333\\
\hline
40 & 780 & 702 & 0.1 & 0.9449 & 7.2936 & 4.7397 & 7.2 & 12 \\
\hline
105 & 105 & 48 & 0.5429 & 0.9048 & 1.2571 & 1.2571 & 0.8276 & 8 \\
\hline
465 & 4340 & 1792 & 0.5871 & 0.4569 & 0.7435 & 0.4880 & 0.7007 & 19.2 \\
\hline
4340 & 465 & 192 & 0.5871 & 4.2645 & 4.5548 & 6.9398 & 0.7030 & 179.2 \\
\hline
465 & 465 & 335 & 0.28 & 1.4387 & 3.8215 & 3.8215 & 2.5573 & 56 \\
\hline
8001 & 9,921,240 & 6,666,081 & 0.3281 & 0.6724 & 2.0479 & 1.0330 & 2.0471 & 358.4 \\
\hline
\end{tabular}
\end{table*}


\begin{IEEEproof}
Consider a bi-regular bipartite caching graph $B(K,F,D)$. We know that $D=F\left(1-\frac{M}{N}\right)$ and $d = K\left(1-\frac{M}{N}\right)$. Consider an arbitrary user vertex and call it $k_1$. We know that $|\mathcal{N}(k_1)|=D$. From the notations of Theorem \ref{genericlowerbound}, we have $\rho_1 = D$. WLOG, let $\mathcal{N}(k_1)= \{f_1,f_2,\cdots, f_{\rho_1}\}$. Now, consider the graph induced by $\mathcal{K}\cup \mathcal{N}(k_1)$ of $B$ and call it $B^{'}$.

\textbf{\underline{Finding lower bound on $\rho_2$:}} Since the degree of each $f\in \mathcal{N}(k_1)$ is $d$, there are exactly $d\rho_1$ edges in $B^{'}$. It is easy to see that the number of edges in $B^{'}$ between $\mathcal{K}\setminus \{k_1\}$ and $\mathcal{N}(k_1)$ is $(d-1)\rho_1$. By the pigeon-holing argument, there exists a user vertex in $\mathcal{K}\setminus \{k_1\}$ with degree at least $\big\lceil \frac{(d-1)\rho_1}{K-1}\big\rceil$, in $B^{'}$. Consider such a user vertex and call it $k_2$.
Therefore, 
\[\rho_2 \geq \bigg\lceil \frac{(d-1) \rho_1}{K-1}\bigg\rceil=\bigg\lceil \frac{\left(d-1 \right)D}{K-1}\bigg\rceil.\]
WLOG, let $\mathcal{N}(k_2)= \{f_1,f_2,\cdots, f_{\rho_2}\}$ in $B^{'}$. It is clear that $\mathcal{N}(k_2) \subseteq \mathcal{N}(k_1)$. Therefore, from each subfile vertex in $\mathcal{N}(k_2)$ two edges will go to $k_1,k_2$ each.

\textbf{\underline{Finding lower bound on $\rho_3$:}} Since the degree of each $f\in \mathcal{N}(k_2)$ is $d$, there are exactly $d\rho_2$ edges incident at the edges in $\mathcal{N}(k_2)$. Now, it should be clear that the number of edges in $B'$ between $\mathcal{K}\setminus \{k_1,k_2\}$ and $\mathcal{N}(k_2)$ is $(d-2)\rho_2$. Again by the pigeon-holing argument, there exists a user vertex in $\mathcal{K}\setminus \{k_1,k_2\}$ with degree at least $\big\lceil \frac{(d-2)\rho_2}{K-2}\big\rceil$, in $B^{'}$. Consider such a user vertex and call it $k_3$.
Therefore, 
\[\rho_3 \geq \bigg\lceil \frac{(d-2 )\rho_2}{K-2}\bigg\rceil.\]
By using the lower bound on $\rho_2$ we can write,
\[
\rho_3 \geq \bigg\lceil \frac{d-2}{K-2}\bigg\lceil \frac{(d-1 )\rho_1}{K-1}\bigg\rceil\bigg\rceil.
\]
Note that these $\rho_3$ edges are incident on $\rho_3$ subfile vertices in $B'$, which we denote as ${\cal N}(k_3)$. By construction of $k_3$ we note that ${\cal N}(k_3)\subseteq {\cal N}(k_2).$ 
Iterating this procedure for $j=4,\hdots,d$, we can identify vertices $k_j,$ with neighbouring vertices ${\cal N}(k_j)$ in $B'$ numbering $\rho_j$ respectively, such that ${\cal N}(k_j)\subseteq {\cal N}(k_{j-1}),\forall j\leq d$, satisfying the following inequality for each $j$ by the pigeon-hole principle
\[\rho_j \geq \Bigg\lceil \frac{\left(d-(j-1) \right)\rho_{j-1}}{K-(j-1)}\Bigg\rceil.\]

Note that $d=K\left(1-\frac{M}{N}\right)$. As $N\geq K\left(1-\frac{M}{N}\right),$ we have in Theorem \ref{genericlowerbound}, $N^{'}=K\left(1-\frac{M}{N}\right)$. Now, applying Theorem \ref{genericlowerbound} upon noticing that the bounds on $\rho_j$s depend only on the parameters $K,F,\frac{M}{N}$, completes the proof.
\end{IEEEproof}
%

We now present the numerical comparisons (in Table \ref{table lower bound}) between the information theoretic bounds we have obtained in this section, with earlier results. Throughout this numerical comparison, we assume that the caching schemes are defined by the bi-regular bipartite caching graphs. For a number of choices of parameters $\left(K,F\text{ and } \frac{M}{N} \right)$,  in Table \ref{table lower bound}, we compare numerically the new subpacketization-dependent lower bounds based on Corollary \ref{corollaryfirstlowerbound} and Theorem \ref{theorem lower bound 2} on the optimal rate (column 5 and column 6 of Table \ref{table lower bound}), with the lower bound (\ref{eqnearlierlowerbound}) of \cite{cheng2017coded} (given in column 7), as well as the lower bound of \cite{WTP} $\left(R^*\geq \dfrac{K(1-\frac{M}{N})}{1+\frac{MK}{N}}~\text{, calculated in column 8}\right)$. Note that the bound in \cite{cheng2017coded} holds for PDA based delivery schemes with given subpacketization level, while the bound in \cite{WTP} holds for non-linear schemes as well and is subpacketization-independent.  

It can be seen that for many of the chosen parameters, our bound of Theorem \ref{theorem lower bound 2} is better than those in \cite{cheng2017coded},\cite{WTP} (when applied to the special class of bi-regular bipartite caching graphs). However, the bound in Corollary \ref{corollaryfirstlowerbound} is quite loose. Further, the last column of Table \ref{table lower bound} denotes the rate achieved (for whichever parameters are applicable) by our new coded caching scheme, titled Scheme B, whose construction we have presented in Section \ref{subsection scheme 2} (from Remark \ref{remark on bi-regularity of graph in scheme B},  recall that Scheme B is defined on a bi-regular bipartite caching graph). Also, seeing the table, we remark that there is in general a wide gap between the lower bounds and achievable rates for small subpacketization levels, which indicate scope for future work.



\section{Conclusion and Discussion}
\label{section conclusion}
In this work, we have presented the coded caching schemes which achieve low subpacketization compared to a number of existing schemes in the literature. Our coded caching schemes are constructed over the foundations of the bipartite graphs and the projective geometry. The literature in this area is now extensive, therefore we have presented comparisons with only a few important existing coded caching schemes, which are considered state-of-the-art to the best of our knowledge. We have also extended our scheme to other channel settings, thereby showing similar gains in the subpacketization in those settings also. There are a number of questions and open problems, which are yet to be answered in terms of the subpacketization-rate trade-off, some of which are listed here.
\begin{itemize}
    \item Is there a closed-form expression for the optimal rate achievable  for a given subpacketization level? Can we obtain impossibility results, for certain regimes of (asymptotic) coded caching gains given a certain level of subpacketization, or vice-versa? (For instance, the work \cite{SZG} gives an impossibility of the subpacketization being linear in $K$, for constant rate and cache-fraction).  
    
    \item The current paper can be said to subsume the construction of \cite{MaN} in the following sense. Suppose we substitute $n=1$ in Theorem \ref{theorem construction 2}, and let $q\rightarrow 0$, then it is easy to see that we obtain the scheme from \cite{MaN}. Further, retaining $n$ as a parameter, just letting $q\rightarrow 0$ gives us a scheme from \cite{strongedgecoloringofbipartitegraphs} (which is based on the set-theoretic principles). The likely common theory which underlies these types of combinatorial constructions, which are based on set-containment and subspace-containment principles, is the notion of geometric lattices. Exploring such connections may lead to further interesting and useful constructions. 
    
    \item One drawback of our constructions in this work is the number of parameters and the lack of flexibility in designing for all possible values for the number of users, and for cache sizes, without introducing dummy users or wasting existing user cache memory. It is possible that using a similar grouping scheme as in \cite{user_grouping_Shanmugam}, we could achieve some flexibility in the number of users. Exploring this is a direction for future work. 
    
    \item The single server with multi-antenna and multi-receiver wireless scenario can be looked at as a multi-transmitter scenario (which we discuss in this current work) with all the transmitters having access to the entire library. Recently, the works \cite{subpackmultiantenna1,subpackmultiantenna2} carry forward the discussion of subpacketization in the multi-antenna wireless communication. In particular, in \cite{subpackmultiantenna2}, a simple linear subpacketization scheme is presented for multi-antenna scenario provided some parameter conditions are satisfied. This raises the question of whether there are schemes for other scenarios, which are of low theoretical complexity, low subpacketization, and offer good caching gain. 
\end{itemize}

\appendices
\section{Proof of Lemma \ref{lemma perfectmatching}}
\label{appendix proof perfect matching}
Construct a bipartite graph with left vertices as $\left\{ V_i, i=1,\hdots,\gbinom{m+t}{t}\right\}$ (the $t$-dim subspaces of $X$) and right vertices as $\left\{T	 : T~\text{is a}~m\text{-dim subspace of}~X\right\}$. By the lemma statement, we know that the number of right vertices is $\gbinom{m+t}{m}$. By A1 of Lemma \ref{lemma gaussiancoeff}, we have $\gbinom{m+t}{m} = \gbinom{m+t}{t}$. Therefore, the number of right vertices is equal to the number of left vertices.

We now define the edges of the bipartite graph. For a left vertex $V$, let the adjacent right-vertices in the bipartite graph be $\{T: V\cap T=\{\mathbf{0}\} \}$. 

Thus, the left-degree is $\gbinom{m+t}{t}-|\{T: V\cap T\neq \{\mathbf{0}\} \}|$. Now, $V\cap T$ is a subspace. By A3 of Lemma \ref{lemma gaussiancoeff}, the number of $m$-dim subspaces of $X$ intersecting a fixed $t$-dim subspace in some $i$-dim subspace ($1\leq i\leq min(t,m)$) is 
\begin{align*}
q^{(m-i)(t-i)}\gbinom{(m+t)-(t)}{m-i}\gbinom{t}{i}
= q^{(m-i)(t-i)}\gbinom{m}{i}\gbinom{t}{i}.
\end{align*}
Thus the left-degree in this bipartite graph is 
\[
\gbinom{m+t}{t}-\sum_{i=1}^{min(m,t)}q^{(m-i)(t-i)}\gbinom{m}{i}\gbinom{t}{i},
\]
where the second term above is precisely $|\{T: V\cap T\neq \{\mathbf{0}\} \}|$.

Similarly, by Lemma \ref{lemma gaussiancoeff}, the number of $t$-dim subspaces of $X$ intersecting a fixed $m$-dim subspace in some $i$-dim subspace is
\begin{align*}
q^{(t-i)(m-i)}\gbinom{(m+t)-(m)}{t-i}\gbinom{m}{i}=q^{(t-i)(m-i)}\gbinom{t}{i}\gbinom{m}{i}.
\end{align*}
And hence the right degree is equal to the left-degree. Hence, the bipartite graph we have constructed is regular. 

A perfect matching of a graph $G$ is a matching of $G$ such that every vertex of $G$ is incident on some edge of the matching. It should be clear that what we are looking for is a perfect matching of the regular bipartite graph we have constructed. The reason is as follows. Define $T_{V_i,X}$ as the $m$-dim subspace  adjacent to $V_i$ in the perfect matching. Since for given $V_i$, any $T$ adjacent to $V_i$ in our bipartite graph is such that $T \oplus V_i = X$, thus we have $T_{V_i,X}\oplus V_i=X$. Thus, a perfect matching gives us the collection of $T_{V_i,X},\forall V_i$ as we desire. 
 
Now, for a regular bipartite graph with $n$ left-vertices, algorithms are known to find a perfect matching with complexity as small as $O(n \log{n})$ \cite{perfect_matchings}. This completes the proof.

\section{Asymptotic Analysis of Scheme A (Theorem \ref{theorem construction 1})}
\label{appendix asymptotics of scheme 1}

We analyse the asymptotic behaviour of our scheme as $K$ grows large in two cases. In the first case, we bound $\frac{M}{N}$ from above by a constant and analyse the asymptotic behaviour of $F$ and $R$. In the second case, we bound $R$ from above by a constant and analyse the asymptotic behaviour of $F$ and $\frac{M}{N}$. 

We first give the equivalent expressions of $\frac{M}{N}$ and $R$ (which can be easily verified using the definition of the Gaussian binomial coefficient). We do this because we can apply the bounds in Lemma \ref{lemma bounds on gaussian binomial coefficients} conveniently to these expressions in order to obtain our asymptotics.

\begin{align*}
\frac{M}{N} &= 1-\frac{\gbinom{k-t}{m}}{\gbinom{k}{m+t}} = 1-\frac{\gbinom{m+t}{t}}{\gbinom{k}{t}} \text{   ~~~~   and  } \\
R &= \frac{\gbinom{k}{m}}{\gbinom{k}{m+t}}=\frac{\gbinom{m+t}{t}}{\gbinom{k-m}{t}}.
\end{align*}

Throughout we assume $q$ is constant. We have $K=\gbinom{k}{t}$. We analyse our scheme as $k$ grows large. 

By using (\ref{eqn31}) we have

\begin{equation}
\label{eqn41}
    q^{(k-t)t} \leq K \leq q^{(k-t+1)t}.
\end{equation}

We can write this as,

\begin{equation}
\label{eqn42}
\frac{1}{\sqrt{K}}q^{\frac{-t^2}{2}} \leq q^{\frac{-kt}{2}} \leq \frac{1}{\sqrt{K}}q^{\frac{-t^2 +t}{2}}.
\end{equation}

\underline{\textbf{Case 1:} If $\frac{M}{N}$ is upper bounded by a constant}

\vspace{0.2cm}

From Theorem \ref{theorem construction 1} we have,
\begin{align}
\label{eqn34}
1-\frac{M}{N}=\frac{\gbinom{m+t}{t}}{\gbinom{k}{t}} \stackrel{(\ref{eqn32})}{\geq} q^{(m+t-k-1)t}.
\end{align}

To lower bound $1-\frac{M}{N}$ (or upper bound $\frac{M}{N}$) by a constant, assume $t$ and $k-m$ as constants. Note that $m+t \leq k$. 

\textbf{Asymptotics of $F$:} We now analyse the asymptotics for $F$. Consider,

\begin{align}
\nonumber
\frac{F}{K}=\frac{\gbinom{k}{m+t}}{\gbinom{k}{t}}&\stackrel{(\ref{eqn33})}{\leq}q^{(k-t-m-t+1)m} \leq q^{(k-2t-m+1)(k-t)} \\
\nonumber
&\stackrel{(\ref{eqn41})}{\leq} K^{\frac{k-2t-m+1}{t}}
\\ 
\nonumber
\text{Hence } ~~F&\leq K^{\frac{k-t-m+1}{t}}.
\end{align} 
Therefore $F = O(poly(K))$. (since $t$ and $k-m$ are constants)

\vspace{0.2cm}

\textbf{Asymptotics of $R$:} 

We now analyse the asymptotics for $R = \frac{\gbinom{m+t}{t}}{\gbinom{k-m}{t}}$. 

By using (\ref{eqn32}) we get,

\begin{equation*}
    q^{(m+t-k+m-1)t} \leq R \leq q^{(m+t-k+m+1)t}.
\end{equation*}
Now by using (\ref{eqn41}) we can write,
\begin{equation*}
    q^{2(t+m-k-1)t} \leq \frac{R}{K} \leq q^{2(t+m-k)t+t}.
\end{equation*}

Therefore $R = \Theta(K)$. (since $t$ and $k-m$ are constants)

\vspace{0.3cm}

\underline{\textbf{Case 2:} If $R$ is upper bounded by a constant}

\vspace{0.2cm}

We have, $R\stackrel{(\ref{eqn32})}{\leq} q^{(2m-k+t+1)t}.$

\vspace{0.2cm}

To bound $R$ from above by a constant, assume $t$ and $k-2m$ as constants. Note that $m+t \leq k$. 
\vspace{0.2cm}

\textbf{Asymptotics of $F$:} We now analyse the asymptotics for $F$. Consider,

\begin{align}
\nonumber
\frac{F}{K}&\stackrel{(\ref{eqn33})}{\leq}q^{(k-t-m-t+1)m} \leq q^{(k-2t-m+1)(k-t)} \\
\nonumber
F&\leq K ~ q^{\left(\frac{k}{2}+\frac{k-2m}{2}-2t+1\right)(k-t)} \\
\label{eqn43}
&\leq q^{\log_{q}K} ~ q^{\frac{k^2-kt}{2}+ \alpha_1 (k-t)},
\end{align} 

where $\alpha_1 =\frac{k-2m}{2}-2t+1$ is a constant.
From (\ref{eqn41}) we have $k \leq \frac{1}{t} \log_{q}{K}+t$.

\vspace{0.2cm}

By using this in the inequality in (\ref{eqn43}), it is easy to see that $F = q^{O\left((log_qK)^2\right)}$. (since $t$ and $k-2m$ are constants)

\vspace{0.2cm}

\textbf{Asymptotics of $\frac{M}{N}$:} We now analyse the asymptotics for $\frac{M}{N}$. By using (\ref{eqn32}) we have,

\begin{equation*}
q^{(m+t-k-1)t} \leq 1-\frac{M}{N} \leq q^{(m+t-k+1)t}
\end{equation*}
\begin{equation*}
q^{\left(\frac{2m-k}{2}+t-1\right)t} q^{\frac{-kt}{2}} \leq 1-\frac{M}{N} \leq q^{\left(\frac{2m-k}{2}+t+1\right)t} q^{\frac{-kt}{2}}.
\end{equation*}

By using (\ref{eqn42}) we get,
\begin{equation*}
\frac{1}{\sqrt{K}} q^{\left(\frac{2m-k}{2}+t-1\right)t - \frac{t^2}{2}} \leq 1-\frac{M}{N} \leq \frac{1}{\sqrt{K}} q^{\left(\frac{2m-k}{2}+t+1\right)t- \frac{t^2 +t}{2}}.
\end{equation*}
Therefore $\frac{M}{N}=1-\Theta\left(\frac{1}{\sqrt{K}}\right)$. (since $t$ and $k-2m$ are constants)


\section{Proof of Lemma \ref{lemma K,F,D,c expressions of scheme 2}}
\label{appendix proof of K,F,D,c in construction 2}
\underline{\textbf{Finding the number of user vertices $K\left(=|\mathbb{X}|\right)$:}} 

\vspace{0.2cm}

Finding $K$ means finding the number of distinct sets $\{T_1,T_2,\cdots,T_n\}$ such that $ T_i\in \mathbb{T},~\forall i\in [n]$ and $\sum\limits_{i=1}^{n}T_i \in \mathbb{R}$. By invoking Lemma \ref{lemma no of sets of LI 1D spaces} (with $a=0$), we have,

\begin{align}
\label{eqn K}
K&=\frac{1}{n!}\prod\limits_{i=0}^{n-1}(\theta(k)-\theta(i)) \\
\nonumber
&=\frac{1}{n!}\prod\limits_{i=0}^{n-1}\left(\frac{q^{k} -1}{q-1}-\frac{q^{i}-1}{q-1}\right) \\
\nonumber
&=\frac{1}{n!}\prod_{i=0}^{n-1}\frac{q^k-q^i}{q-1} = \frac{1}{n!}\left(\prod\limits_{i=0}^{n-1}q^i\right)\left(\prod_{i=0}^{n-1}\frac{q^{k-i}-1}{q-1}\right) \\
\nonumber
&= \frac{1}{n!} ~ q^{\frac{n(n-1)}{2}} ~ \prod\limits_{i=0}^{n-1}\gbinom{k-i}{1}.
\end{align}

\underline{\textbf{Finding the number subfile vertices $F\left(=|\mathbb{Y}|\right)$:}}

\vspace{0.2cm}

Proof is similar to that of $K$ (replace $n$ with $m$). 

\vspace{0.4cm}

\underline{\textbf{Finding the degree of user vertex $D\left(=|\mathcal{N}(X)|\right)$:}} 

\vspace{0.2cm}

Consider an arbitrary $X=\{T_1,T_2,\cdots,T_n\}\in \mathbb{X}$. We have $\sum\limits_{i=1}^{n}T_i =R$, for some $R \in \mathbb{R}$. We know that $dim(R)=n$. 
Now, finding $|\mathcal{C}_X|$ is equivalent to counting the number of distinct sets $ Y=\{T_1',T_2',\cdots, T_{m}'\} \in \mathbb{Y}$ such that $X\cup Y \in  \mathbb{Z} $, that is $dim\left(\sum\limits_{i=1}^{n}T_i +\sum\limits_{i=1}^{m}T_i'\right)=n+m$.  
By Lemma \ref{lemma no of sets of LI 1D spaces} we have,

\begin{align*}
|\mathcal{N}(X)|&=\frac{1}{m!}\prod\limits_{i=0}^{m-1}(\theta(k)-\theta(n+i)) \\
\nonumber
&=\frac{1}{m!}\prod_{i=0}^{m-1}\frac{q^k-q^{n+i}}{q-1} \\
\nonumber
&= \frac{1}{m!}\left(\prod\limits_{i=0}^{m-1}q^{n+i}\right)\left(\prod_{i=0}^{m-1}\frac{q^{k-n-i}-1}{q-1}\right) \\
\nonumber
&= \dfrac{q^{nm}}{m!} ~ q^{\frac{m(m-1)}{2}} ~ \prod\limits_{i=0}^{m-1}\gbinom{k-n-i}{1}.
\end{align*}

\vspace{0.4cm}




This completes the proof.


\section{Asymptotic Analysis of Scheme B (Theorem \ref{theorem construction 2})}
\label{appendix asymptotics of scheme 2}

In this appendix, we analyse the asymptotic behaviour of $F,R$ for our coded caching scheme presented in Theorem \ref{theorem construction 2} (Scheme B) as $\frac{M}{N}$ is upper bounded by a constant and $K\rightarrow \infty$. We show that $F=q^{O\left((\log_{q}{K})^2\right)}$, while $R=\Theta\left(\frac{K}{(\log_{q}{K})^{n}}\right)$. Throughout our analysis, we assume $q$ is a constant and some prime power, and $n$ is some constant. We now upper bound $\frac{M}{N}$ by a constant.
From (\ref{eqn 1-M/N in scheme 2}) we have,

\begin{align*}
1-\frac{M}{N} &=  \dfrac{\prod\limits_{i=0}^{m-1}(\theta(k)-\theta(n+i))}{\prod\limits_{i=0}^{m-1}(\theta(k)-\theta(i))}= \dfrac{\prod\limits_{i=n}^{n+m-1}(\theta(k)-\theta(i))}{\prod\limits_{i=0}^{m-1}(\theta(k)-\theta(i))} 
\end{align*}

\begin{align*}
&=\dfrac{\prod\limits_{i=0}^{n+m-1}(\theta(k)-\theta(i))}{\left(\prod\limits_{i=0}^{n-1}(\theta(k)-\theta(i))\right)\left(\prod\limits_{i=0}^{m-1}(\theta(k)-\theta(i))\right)} \\
&= \dfrac{\prod\limits_{i=m}^{n+m-1}(\theta(k)-\theta(i))}{\prod\limits_{i=0}^{n-1}(\theta(k)-\theta(i))} = \dfrac{\prod\limits_{i=0}^{n-1}(\theta(k)-\theta(m+i))}{\prod\limits_{i=0}^{n-1}(\theta(k)-\theta(i))} \\
&= \prod\limits_{i=0}^{n-1} \frac{q^{k}-q^{m+i}}{q^{k}-q^{i}}  =\prod\limits_{i=0}^{n-1} \frac{q^{k-i}-q^m}{q^{k-i}-1} \\
&\geq \prod\limits_{i=0}^{n-1} \frac{q^{k-i}-q^{m}}{q^{k-i}}
\\
1-\frac{M}{N} & \geq \prod\limits_{i=0}^{n-1} \left(1-\frac{q^i}{q^{k-m}}\right).
\end{align*}
Let $\alpha = k-m$. ($\alpha \geq n$, since $k\geq n+m$)
\begin{align}
\nonumber
    1-\frac{M}{N} & \geq \prod\limits_{i=0}^{n-1} \left(1-\frac{q^i}{q^{\alpha}}\right) \geq \prod\limits_{i=0}^{n-1} \left(1-\frac{q^{n-1}}{q^{\alpha}}\right) 
    \\ 
    \label{eqneqn1}
    &\geq \left(1-\frac{1}{q^{\alpha-n+1}}\right)^{n} \geq 1-\frac{n}{q^{\alpha-n+1}}.
\end{align}

Therefore, the upper bound on $\frac{M}{N}$ is given as
$\frac{M}{N} \leq \frac{n}{q^{\alpha -n+1}}$, where $\alpha=k-m$ and $n$ are constants.

\vspace{0.2cm}

We have $K=\frac{1}{n!} ~ q^{\frac{n(n-1)}{2}}  \prod\limits_{i=0}^{n-1}\gbinom{k-i}{1}$. We analyse our scheme as $k$ grows large (thus $K$ grows large). By Lemma \ref{lemma bounds on gaussian binomial coefficients} we have,
\begin{align}
\nonumber
\frac{1}{n!} ~ q^{\frac{n(n-1)}{2}} \prod\limits_{i=0}^{n-1}q^{k-i-1} &\leq K \leq \frac{1}{n!} ~ q^{\frac{n(n-1)}{2}}  \prod\limits_{i=0}^{n-1}q^{k-i}  \\
\nonumber
\frac{\prod\limits_{i=0}^{n-1}q^i}{n!}  ~ q^{(k-1)n} ~ \prod\limits_{i=0}^{n-1}q^{-i}&\leq  K \leq \frac{\prod\limits_{i=0}^{n-1}q^i}{n!} ~ q^{kn} ~ \prod\limits_{i=0}^{n-1}q^{-i} \\
\label{eqn K inequality}
\frac{1}{n!} ~ q^{(k-1)n} & \leq  K \leq \frac{1}{n!} ~ q^{kn}  \\
\nonumber
(k-1)n &\leq \log_q{\left(n! K\right)} \leq kn.
\end{align}

Hence, we have,
\begin{equation}
\label{eqn k-t bounds}
    \frac{1}{n}\log_q{(n! K)} \leq k \leq \frac{1}{n}\log_q{(n!K)} +1.
\end{equation}

\vspace{0.2cm}

\underline{\textbf{Asymptotics of $R$:}}

\vspace{0.2cm}

We now get the asymptotics for the rate. We have,  
$R=\dfrac{K(1-\frac{M}{N})}{\gamma}$. From Theorem \ref{theorem construction 2}, we have $\gamma= \binom{n+m}{n}$. Since $\alpha= k-m$ we can write,
$\gamma=\binom{k-\alpha+n}{n}$.
We have the following well known bounds on the binomial coefficient ($e$ being the base of the natural logarithm),
$$\left(\frac{a}{b}\right)^b \leq \binom{a}{b} \leq e^b \left(\frac{a}{b}\right)^b.$$ 

By using this result, the bounds on $\gamma$ can be written as,


$$\tiny \left(\frac{k-\alpha+n}{n}\right)^{n} \leq \gamma \leq  \left(\frac{e(k-\alpha+n)}{n}\right)^{n}.$$   

By using (\ref{eqn k-t bounds}) the lower bound on $\gamma$ can be written as,

\[\left(\frac{\frac{1}{n}\log_q{\left(n! K\right)}-\alpha+n}{n}\right)^{n} \leq \gamma ~  ~,\]

and the upper bound on $\gamma$ can be written as,

\[\gamma \leq \left(\frac{e\left(\frac{1}{n}\log_q{\left(n! K\right)}-\alpha+n+1\right)}{n}\right)^{n}.\]

After some simple manipulations we get $\gamma= \Theta\left((\log_q{n!K})^{n}\right) = \Theta\left((\log_q{K})^{n}\right)$.
(since $n$ is a constant).
Therefore, we get 
$R = \Theta\left(\frac{K}{(\log_q{K})^{n}}\right)$. 

\vspace{0.2cm}

\underline{\textbf{Asymptotics of $F$:}}

\vspace{0.2cm}

We now obtain the asymptotics for the subpacketization $F$. 
By using $K,F$ expressions in Theorem  \ref{theorem construction 2} we get,
\begin{align*}
\frac{F}{K} &= \frac{n!}{m!}\frac{q^{\frac{m(m-1)}{2}}}{q^{\frac{n(n-1)}{2}}} \frac{\prod\limits_{i=0}^{m-1}\gbinom{k-i}{1}}{\prod\limits_{i=0}^{n-1}\gbinom{k-i}{1}}.
\end{align*}
By Lemma \ref{lemma bounds on gaussian binomial coefficients} we have,
\begin{align*}
\frac{F}{K} &\leq \frac{n!}{m!}~\frac{q^{\frac{m(m-1)}{2}}}{q^{\frac{n(n-1)}{2}}}~\frac{\prod\limits_{i=0}^{m-1}q^{k-i}}{\prod\limits_{i=0}^{n-1}q^{k-i-1}}  \\
&= \frac{n!}{m!} ~ \frac{q^{\frac{m(m-1)}{2}}}{q^{\frac{n(n-1)}{2}}} ~  \frac{\prod\limits_{i=0}^{m}q^{-i}}{\prod\limits_{i=0}^{n}q^{-i}} ~ \frac{q^{km}}{q^{(k-1)n}} \\
F&\leq \frac{n!~K}{m!}~q^{km-kn+n}.
\end{align*}

By using $m=k-\alpha$ we get,

\begin{align*}
F &\leq \frac{q^{\log_q{(n!K)}}~q^{\left(k^2+(\alpha+n)(-k)+n\right)}}{(k-\alpha)!}.
\end{align*}

By (\ref{eqn k-t bounds}) we have,
\begin{align*}
&k^2+(\alpha+n)(-k)+n \\
& \small \leq \left(\frac{1}{n}\log_q{(n!K)} +1\right)^2 + (\alpha+n)\left(\frac{-1}{n}\log_q{(n!K)}\right)+n \\
& = \left(\frac{1}{n}\log_q{(n!K)} \right)^2 + \left(\frac{2-\alpha-n}{n}\log_q{(n!K)}\right)+n+1.
\end{align*}

By the lower bound of (\ref{eqn k-t bounds}) we have,

\[\frac{1}{(k-\alpha)!} \leq \frac{1}{\big\lfloor\frac{1}{n}\log_q{\left(n! K\right)}-\alpha \big\rfloor!}.\]

By using these bounds, we get,
\[F\leq \frac{q^{\left(\frac{1}{n}\log_q{(n!K)} \right)^2 + \left(\frac{2-\alpha}{n}\log_q{(n!K)}\right)+n+1}}{\big\lfloor\frac{1}{n}\log_q{\left(n! K\right)}-\alpha\big\rfloor!}.\]

Using Stirling's approximation for $x!$ as $\sqrt{2\pi x}\left(\frac{x}{e}\right)^x$ for large $x$, and after some simple manipulations, we see that $F=q^{O\left((\log_q{(n!K
)})^2\right)}=q^{O\left((\log_q{K})^2\right)}$ (Since $n$ is a constant).


\section{Proof of Lemma \ref{lemma K,F,D,c expressions of scheme 3}}
\label{appendix proof of K,F,D,c in construction 3}

\underline{\textbf{Finding the number of user vertices $K\left(=|\mathbb{X}|\right)$:}} 

\vspace{0.2cm}

Let $\mathbb{T} \triangleq  PG_q(k-1,0) \text{  (set of all $1$-dim subspaces)}$. Finding $K$ means finding the number of distinct sets $\{L_1,L_2,\cdots,L_n\}$ such that $ L_i\in \mathbb{L},~\forall i\in [n]$ and $\sum\limits_{i=1}^{n}L_i \in \mathbb{R}$. 
Now to find $K$, we prove the following smaller claims.

\vspace{0.2cm}

\underline{\textit{Claim 1:}} The number (say $x_1$) of distinct $nl$ sized sets $\{T_1,T_2,\cdots,T_{nl}\}$ such that $T_i\in \mathbb{T}, \forall i\in [nl]$ and $\sum\limits_{i=1}^{n}T_i \in \mathbb{R}$ is 
$x_1= \frac{1}{(nl)!}\prod\limits_{i=0}^{nl-1}(\theta(k)-\theta(i))$.

\vspace{0.2cm}

\textit{Proof of Claim 1:}
By invoking Lemma \ref{lemma no of sets of LI 1D spaces} with $a=0$ and $b=nl$ we see the result.

\vspace{0.2cm}

\underline{\textit{Claim 2:}} Consider a set $\mathcal{T}=\{T_1,T_2,\cdots,T_{nl}\}$ such that $T_i\in \mathbb{T}, \forall i\in [nl]$ and $\sum\limits_{i=1}^{n}T_i \in \mathbb{R}$. The number (say $x_2$) of distinct $X\in \mathbb{X}$ generated from $\mathcal{T}$ is $x_2= \frac{1}{n!} \prod\limits_{i=0}^{n-1}\binom{(n-i)l}{l}$.

\vspace{0.2cm}

\textit{Proof of Claim 2:} It is clear that $\mathcal{T}$ contains $nl$ number of linearly independent $1$-dim subspaces. So the addition of any $l$ subspaces from $\mathcal{T}$ will generate a unique $l$-dim subspace. Hence, finding $x_2$ is equivalent to counting the number of distinct $n$-sized sets of $l$-sized sets that can be formed from $\mathcal{T}$. It is clear that each such $n$-sized set will generate a unique $X$. From the basic combinatorics, the expression for $x_2$ can be inferred.

Therefore, the total number of $X\in \mathbb{X}$ which can be generated from $\fq^k$ is $x_1x_2$. But there are some repetitions in $x_1x_2$. We identify them in the following claim.

\vspace{0.2cm}

\underline{\textit{Claim 3:}} Consider an arbitrary $X=\{L_1,L_2,\cdots,L_n\}\in \mathbb{X}$. The number (say $x_3$) of distinct $\{T_1,T_2,\cdots,T_{nl}\}$ (such that $T_i\in \mathbb{T}, \forall i\in [nl]$ and $\sum\limits_{i=1}^{nl}T_i \in \mathbb{R}$) which generate $X$ is $x_3= \left(\frac{1}{(l!)}\prod\limits_{i=0}^{l-1}(\theta(l)-\theta(i))\right)^n$.

\vspace{0.2cm}

\textit{Proof of Claim 3:} Consider an arbitrary $L_j \in X$. By Lemma \ref{lemma no of sets of LI 1D spaces}, the number of distinct sets $\{T_1^{'} ,T_2^{'} ,\cdots,T_{l}^{'}\}$ ($\forall T_i^{'} \in \mathbb{T}, i\in [l]$) such that $\sum\limits_{i=1}^{l}T_i^{'} = L_j$ is (substitute $a=0,b=l, k=l$ in Lemma \ref{lemma no of sets of LI 1D spaces})
$\frac{1}{(l!)}\prod\limits_{i=0}^{l-1}(\theta(l)-\theta(i))$. There are $n$ such $L_i$'s in $X$ (all are linearly independent subspaces).
Therefore $x_3= \left(\frac{1}{(l!)}\prod\limits_{i=0}^{l-1}(\theta(l)-\theta(i))\right)^n$. 

\vspace{0.1cm}

Hence $K=\frac{x_1x_2}{x_3}$. Now by using $\theta(k)=\frac{q^k-1}{q-1}$ and by doing some simple manipulations we see the expression of $K$ as per the lemma statement. The proofs of $F$ and $D$ follows similarly.




\bibliographystyle{IEEEtran}
\bibliography{IEEEabrv,main.bbl}

\end{document}